\documentclass[submission, Phys]{SciPost}
\hypersetup{
colorlinks,
linkcolor={red!50!black},
citecolor={blue!50!black},
urlcolor={blue!80!black}
}

\usepackage[bitstream-charter]{mathdesign}
\DeclareSymbolFont{usualmathcal}{OMS}{cmsy}{m}{n}
\DeclareSymbolFontAlphabet{\mathcal}{usualmathcal}
\usepackage{times}
\usepackage{color}
\usepackage{caption}
\usepackage{graphicx}
\usepackage{subcaption}
\usepackage{setspace}
\usepackage{multirow}
\usepackage{comment}
\usepackage{subcaption}
\usepackage{stackengine}
\usepackage{bm} 
\usepackage{hyperref}
\newcommand\bea{\begin{eqnarray}}
\newcommand\eea{\end{eqnarray}}
\newcommand\beq{\begin{equation}}
\newcommand\eeq{\end{equation}}
\newcommand\bib{\bibitem}

\newcommand{\noi}{\noindent}
\newcommand{\non}{\nonumber}
\newcommand{\al}{\alpha}
\newcommand{\de}{\delta}
\newcommand{\De}{\Delta}
\newcommand{\ga}{\gamma}
\newcommand{\ep}{\epsilon}

\newcommand{\om}{\omega}

\newcommand{\la}{\langle}
\newcommand{\ra}{\rangle}

\newcommand{\bra}[1]{\langle #1|}
\newcommand{\ket}[1]{|#1\rangle}

\begin{document}
\begin{center}{\Large \textbf{Dynamical localization and slow thermalization in a class 
of disorder-free periodically driven one-dimensional interacting systems}}\end{center}
\begin{center}
{Sreemayee Aditya\textsuperscript{1,$^\star$} and Diptiman Sen\textsuperscript{1,2,$^{\star\star}$}}
\end{center}
\begin{center}
{\bf 1} Center for High Energy Physics, Indian Institute of Science, Bengaluru 560012, India
\\
{\bf 2} Department of Physics, Indian Institute of Science, Bengaluru 560012, India

${}^\star$ {\small \sf sreemayeea@iisc.ac.in}\\
${}^{\star\star}$ {\small \sf diptiman@iisc.ac.in}
\end{center}

\section*{Abstract}
{\bf
We study if the interplay between dynamical localization and interactions in periodically driven quantum systems can give rise to anomalous thermalization 
behavior. Specifically, we consider one-dimensional models with interacting spinless fermions with nearest-neighbor hopping and density-density interactions, and 
a periodically driven 
on-site potential with spatial periodicity $\boldsymbol{m=2}$ and $\boldsymbol{m=4}$. 
At a dynamical localization point, these models 
evade thermalization either due to the presence of an extensive number of conserved quantities (for weak interactions) or due to the kinetic 
constraints caused by drive-induced resonances (for strong interactions). 
Our models therefore illustrate interesting mechanisms for generating 
constrained dynamics in Floquet systems which are difficult to realize in an undriven system.}

\vspace{10pt}
\noindent\rule{\textwidth}{1pt}
\tableofcontents\thispagestyle{fancy}
\noindent\rule{\textwidth}{1pt}
\vspace{10pt}

\section{Introduction}

The non-equilibrium dynamics of quantum systems has been extensively studied 
in recent years~\cite{rev1,rev2,rev3,rev4,rev5,rev6,rev7,rev8,rev9,rev10,
rev11,rev12,rev13}. Various kinds of time-dependent protocols have been 
considered such as quenching and ramping~\cite{rev1,rev2,rev3,rev4,quench1,quench2,quench3,ramp1,ramp2,ramp3,
ramp4,ramp5,ramp61,ramp62,ramp71,ramp72,ramp81,ramp82,ramp83,ramp9}, periodic driving~\cite{rev5,rev6,rev7,rev8,rev9,rev10,rev11,rev12,rev13}, 
and quasiperiodic and aperiodic driving~\cite{quasi11,quasi12,quasi2,quasi3,
quasi4,quasi5,quasi6,ap1,ap2}. There have been several experimental 
studies of non-equilibrium dynamics in systems of cold atoms trapped 
in optical lattices~\cite{exp1,exp2,exp3,exp4,exp5,exp6,exp7,exp8,exp9,exp10}.

Periodic driving of quantum systems can give rise to a host of interesting 
phenomena which have no equilibrium counterparts, such as the generation of 
drive-induced topological phases~\cite{topo1,topo2,topo3,topo4,topo5,topo6}, 
Floquet time crystals~\cite{tc1,tc2,tc3}, dynamical localization~
\cite{dynloc1,dynloc2,dynloc3,dynloc4,dynloc5,
dynloc6,dynloc7}, dynamical freezing~\cite{dynfreez1,dynfreez2,dynfreez3,
dynfreez4,dynfreez5,dynfreez6}, tuning between ergodic and non-ergodic
behaviors~\cite{erg1,pscar1,erg3}, and dynamical transitions~\cite{dyntran1,
dyntran2,dyntran3,dyntran4,dyntran5,dyntran6}. The out-of-equilibrium dynamics 
of a wide class of closes quantum systems 
is believed to be governed by the eigenstate thermalization 
hypothesis (ETH)~\cite{ETH1,ETH2,ETH3,ETH4,ETH5}. According to ETH, all the 
eigenstates near the middle of the energy spectrum of a closed, non-integrable and
disorder-free quantum system are thermal; the thermal nature of such 
states guarantees the ergodicity of the system. However,
some instances are known where ETH is violated, for example, in integrable quantum 
systems and in many-body localized phases in one dimension in the presence of
disorder and interactions~\cite{MBL1,MBL2}. 
In recent years, it has been found that ETH can be broken in some quantum systems 
which are not integrable and have no disorder. The breaking of ETH may be weak or 
strong. In the case of weak ergodicity breaking, systems evade
ergodicity due to the presence of quantum many-body scars~\cite{scar1,scar2,
scar3,scar4,scar5,scar6,scar7,scar8,scar9,scar10,scar11,scar12}. Quantum 
many-body scars are states which lie near the middle of the spectrum and have
anomalously low entanglement entropy between two halves of
the system. The number of scar states is typically much smaller than the
full Hilbert space dimension. Moreover, the scar states form a subspace which is
almost decoupled from the thermal subspace. Hence they are 
protected from thermalization for a long period of time and show
persistent long-time coherent oscillations in their dynamics; this
has been observed recently in Rydberg atoms simulators. Furthermore, the interplay 
between quantum many-body scar states and periodic driving can
generate rich dynamical phase diagrams which have been studied in a
number of papers~\cite{pscar1,pscar2,pscar3,pscar4,pscar5,pscar6,pscar7}. 
One possible mechanism for strong ergodicity breaking is Hilbert space 
fragmentation (HSF)~\cite{tomasi,yang,HSF1,HSF2,HSF3} which occurs due to the 
presence of certain kinetic constraints in the 
dynamics. These kinetic constraints lead to the fragmentation of the 
Hilbert space into many disconnected sectors which can give rise to 
non-ergodic behavior in such systems. The HSF
in Floquet systems has been examined recently~\cite{prefrag}.

It has been shown in a series of theoretical works that quantum 
many-body scar states can appear in systems hosting flat bands 
supported by compact localization~\cite{scar8,scar9,scar10}. 
Motivated by this idea, we will pose a similar question in the context 
of Floquet systems undergoing dynamical localization
(DL)~\cite{dynloc1,dynloc2,dynloc3,dynloc4,dynloc5,
dynloc6,dynloc7}. 
DL can be achieved in Floquet systems by tuning some of the system parameters
which makes the effective hopping amplitudes zero or very small. DL
can thus be a powerful tool for generating flat bands. The effects of interactions 
then become predominant which makes such systems highly
promising platforms for investigating correlated out-of-equilibrium phases of matter. DL is special to periodically driven systems since this phenomenon has no equilibrium analog. 
Furthermore, the interplay between DL and quasiperiodic driving can prevent thermalization in a quantum system; this
has been studied recently~\cite{dynqua}. Having understood the rich possibilities
that DL can offer in Floquet systems, the natural question that motivates our work is as follows. Can the interplay between DL and interactions in a 
one-dimensional disorder-free periodically driven closed quantum system induce
anomalous thermalization behavior?

\begin{center}
\begin{table}[!ht]
\begin{tabular}{|p{3.2cm}|p{3.8cm}|p{3.2cm}|p{3.2cm}|} 
\hline
Class of periodic potential & Condition for dynamical localization & Dynamical localization and $\mu ~\gg~ J,~V$ & Dynamical localization and $\mu=V ~\gg ~J$ \\
\hline
$~~~~~~~~~~~m=2$&$~~~~\mu=n\om~(n=1,2,\cdots)$ &cMany-body flat bands, slow thermalization due to an emergent integrability& Model of Hilbert space fragmentation \\
\hline
~~~~~$m=4~(\phi=0)$&$~~~~\mu=2n\om~(n=1,2,\cdots)$ & Same as period-2 model &
Same as the period-2 model 
\\
\hline
$~~~~m=4~(\phi=7\pi/4)$&$~~~~~\mu=n\om~(n=1,2,\cdots)$ &Many low-entanglement states near the middle of the spectrum due to the presence of an extensive number of conserved quantities & Model of Hilbert space fragmentation but different from the period-2 case \\
\hline
\end{tabular}
\caption{\label{main results}Schematic of main results obtained for the period-2 and period-4 models. $\mu$ and $\om$ denote the driving amplitude and frequency
respectively.}
\end{table}
\end{center}

Another motivation for our work is as follows. As mentioned above, it has been
known for many years that periodic driving of non-interacting systems can be
used to generate quantum systems with a wide variety of band structures. 
It is therefore natural to 
ask if periodic driving of interacting systems can produce new kinds of 
quantum many-body systems whose parameters can be readily tuned.

The plan of this paper is as follows. In Sec.~\ref{sec2}, we introduce our general model
which consists of a one-dimensional system of spinless fermions with nearest-neighbor
hopping, an on-site potential which is periodic in space~\cite{periodic} and is also driven
periodically in time, and a density-density interaction between nearest-neighbor
sites. In Sec.~\ref{sec3}, we study in detail a model in which the potential has a 
periodicity of 2 sites. We first use first-order Floquet perturbation theory to 
derive an effective Floquet Hamiltonian for a non-interacting system when the 
driving amplitude and frequency are much larger than the hopping amplitude. We
find that the system shows DL for certain values of the system parameters.
We look at a two-point correlation function as a function of the
stroboscopic time $t=nT$, where $T = 2 \pi/\om$ is the time period. We find that the
correlator can decay as a power-law, where the power depends on the structure
of the quasienergy dispersion around zero momentum. An interesting dynamical
phase transition is found to occur when the dispersion changes, and a 
crossover between different powers occurs. Next, we look at the effects
of interactions on DL. Exactly at a DL point, we find that there is an emergent integrability with a large number of conserved quantities; these quantities
are just the fermion occupation numbers (0 or 1) at the different sites. 
An exact numerical calculation of the time evolution of the system 
confirms this result from the Floquet Hamiltonian, namely, we find that
the two-point correlation function and Loschmidt
echo (which is the overlap between an initial state and its time-evolved state)
oscillate in time and show almost perfect revivals with a frequency which depends
on the interaction strength. The integrability disappears as we go away from
a DL point, and the Loschmidt echo then decays rapidly with time. In Sec.~\ref{sec4},
we consider the combined effects of resonances~\cite{reso} and DL for the period-2 model.
The first-order Floquet Hamiltonian then has a remarkable structure in which
we have both the large number of conserved quantities as well as a density-dependent
hopping in which the hopping between sites $j+1$ and $j+2$ depends on the occupation
numbers on their neighboring sites $j$ and $j+3$. We thus obtain dynamical constraints
on the hopping. This leads to the appearance of an exponentially large number
of zero quasienergy states (an expression for this number is derived in Appendix A
using a transfer matrix method), a highly fragmented Hilbert space, and several states
with low entanglement entropy which lie near the middle of the quasienergy spectrum.

In Sec.~\ref{sec5}, we study a model in which the potential has a periodicity of 4 sites.
The potential has an amplitude which is driven periodically in time and a phase $\phi$.
The system has a mirror symmetry for two values of the phase, 0 and $7 \pi/4$.
At a DL point, a period-4 model with $\phi = 0$ behaves similarly as the period-2 model. 
But a period-4 model with $\phi = 7 \pi/4$ exhibits a different and
remarkably rich set of behaviors. 
First, in the absence of interactions, the Floquet Hamiltonian has the form of the 
Su-Schrieffer-Heeger (SSH) model in which the nearest-neighbor hoppings have a staggered
structure; this leads to the appearance of modes near the ends of an open system.
At a DL point, we obtain an extreme limit of the SSH model in which hoppings
on alternate bonds vanish. This leads to a large number of conserved quantities
which is the total fermion occupation number on two sites between which the hopping
is non-zero. Labeling the unit cell with two such sites as $j$, the conserved occupation
number $n_j$ can take the values 0, 1 and 2. It is convenient to map the two possible 
states of a unit cell with $n_j = 1$ to the states of a spin-1/2 object. We then 
discover that when interactions are introduced, a set of consecutive unit cells all of 
which have $n_j = 1$ is described by the transverse field Ising model, in which 
neighboring spin-1/2's have $\sigma^x_j \sigma^x_{j+1}$ interactions and there is
a transverse magnetic field term $\sigma_j^z$. In addition, the two boundary sites
of this model have a longitudinal magnetic field term $\sigma_j^x$. The exact spectrum
of this model can be found by mapping the spin-1/2 model to a model of fermions using the
Jordan-Wigner transformation. Once again we examine the spectrum of the entanglement 
entropy versus the quasienergy and the time evolution of the Loschmidt echo. We find
a clear fragmentation of the Hilbert space in terms of the quasienergy spectrum, and the
Loschmidt echo shows oscillations for a long period of time. Both of these are
consequences of the conserved quantities. We then study the effects of a staggered
on-site potential at a DL point; we find that the fragments of the Hilbert space 
further break up into secondary fragments. Finally, we study what happens in the 
period-4 model when both resonances and DL are simultaneously present. The Floquet 
Hamiltonian again consists of a density-dependent nearest-neighbor hopping. 
We summarize our main results in Sec.~\ref{sec6}. 

\section{Hamiltonian of period-m model}
\label{sec2} 

In this paper we will discuss a class of periodically driven one-dimensional models 
of interacting spinless fermions. The general form of the Hamiltonian is
\bea H(t) ~=~ \sum_{j=1}^{N} ~[ J ~(c_{j}^{\dagger}c_{j+1}+ {\rm H.c.}) ~+~ \mu(t)~
\cos(2\pi j/m+\phi) ~c_j^\dagger c_j ~+~ V~ n_{j} n_{j+1}], \label{gen} \eea
where $J$ is a uniform time-independent nearest-neighbor hopping, $\mu(t)$ is the
strength of an on-site potential which varies in space with period $m$ (we will call
this a period-$m$ model), $\phi$ is a phase whose significance will 
be discussed later, $V$ is the strength of a nearest-neighbor density-density 
interaction, $n_j = c_j^\dagger c_j$, and $N$ denotes the number of sites 
(we will use periodic boundary conditions unless otherwise specified). We will take
$\mu(t)$ to be a periodic function of time with a time period $T$ and the form
\bea \mu(t) &=& \mu ~f(t), \non \\
{\rm where} ~~~f(t)&=& 1 ~~~{\rm for}~~~ 0~\le~t~ <~ T/2 \non\\
&=&- ~1 ~~~{\rm for}~~~ T/2~\le~t~ <~ T, \label{protocol} \eea
and $f (t+T) = f (t)$. Since $H(t)=H(t+T)$, we will use the Floquet formalism to 
examine
this model. The model can be analytically studied by performing a Floquet-Magnus expansion~\cite{rev1,rev2,rev3,rev4,rev5,rev6,rev7,rev8,rev9,rev10,rev11,rev12,rev13} which 
is valid in the large $\omega$ limit, where
$\om = 2\pi /T$ is the driving frequency. However, we will analytically study the 
model by finding an effective Floquet Hamiltonian $H_F$ using Floquet perturbation 
theory (FPT)~\cite{FPT1,FPT2,pscar1,haldar,rev11}; this
approach is valid when both $\om$ and $\mu$ are much larger than all the other
parameters of the model. We will examine in detail two classes 
of models, period-2 and period-4, both analytically using the Hamiltonian $H_F$ and 
numerically by computing the Floquet operator $U$ which evolves the system 
through one time period. We will set $\hbar = 1$ in this paper.

\section{Period-2 model}
\label{sec3} 

In this section we will consider the period-2 model, whose form can be obtained by 
putting $m=2$ and setting $\phi=0$ in Eq.~\eqref{gen},
\beq H(t)=\sum_{j=1}^{N} ~\left[J ~(c_{j}^{\dagger}c_{j+1}+{\rm H.c.}) ~+~ \mu(t)
~\cos(\pi j) ~+~ V~ n_{j}n_{j+1}\right]. \label{per21} \eeq
This Hamiltonian can be written in the following form in the language of 
a unit cell with two sites,
\bea H(t)&~=~&\sum_{j=1}^{N/2} ~\left[J ~(a_{j}^{\dagger}b_{j}+a_{j}^{\dagger}b_{j-1}
+ {\rm H.c.}) ~+~ \mu(t) ~(a_{j}^{\dagger}a_{j}-b_{j}^{\dagger}b_{j})
~+~ V (n_{j,a} n_{j,b} + n_{j,a} n_{j-1,b})\right],\non\\ \label{per22} \eea
where $N/2$ denotes the number of unit cells (we assume $N$ is even), and each unit cell consists of two sites labeled $a$ and $b$ with $a_{j}^{\dagger}$ and $b_{j}^{\dagger}$ 
denoting the creation operator for a particle at odd- and even-numbered sites, respectively. We will assume periodic boundary conditions. 

We will first study the non-interacting case with $V=0$. The Hamiltonian in 
momentum space is then given by
\bea H(t)~=~ \sum_{k} ~[J ~((1+e^{-2ik})a_{k}^{\dagger}b_{k}+ {\rm H.c.}) ~+~ \mu(t)~
(a_{k}^{\dagger}a_{k}-b_{k}^{\dagger}b_{k})], \label{per23} \eea
where $k$ takes $N/2$ equally spaced values lying in the range of $(-\pi/2$, $\pi/2]$. For each value of $k$, we define the Floquet operator
\bea U_{k}=\mathcal{T} \exp [-i\int_{0}^{T} ~dt ~H_{k}(t)], \label{per24} \eea
where $\mathcal{T}$ denotes time ordering. The Floquet operator can be 
written in the form
\bea U_{k}=e^{-iTH_{Fk}},\label{per25} \eea
where $H_{Fk}$ is the time-independent effective Floquet Hamiltonian. Assuming 
$\mu \gg J$, we can write
\bea H_{k}(t) &=& H_{0}(t) ~+~ H_{1},\non\\
H_{0}&=& \mu(t) ~(a_{k}^{\dagger}a_{k}-b_{k}^{\dagger}b_{k}),\non\\
H_{1} &=& \sum_{k} ~[J ~(1+e^{-2ik})a_{k}^{\dagger}b_{k}+ {\rm H.c.}]. \label{matper2}\eea 
We will now calculate $H_{Fk}^{(1)}$ using FPT to first order in the hopping $J$. 
We see from Eq.~\eqref{protocol} that the two instantaneous 
eigenvalues of $H_{0}$ given by $E_{k}^{\pm} (t) =\pm \mu (t)$ satisfy the condition
\bea e^{i\int_{0}^{T} ~dt ~[E_{k}^{+} (t) -E_{k}^{-} (t)]} ~=~ 1. \label{per26} \eea
Hence we need to use degenerate FPT~\cite{FPT1,pscar1,haldar,rev11}. 

The eigenfunctions corresponding to $E_{k}^{\pm}$ are given by
\bea \ket{+}_{k}=\left(\begin{array}{cc}
1 \\ 0 \end{array}\right) ~~~~{\rm and}~~~~
\ket{-}_{k}=\left(\begin{array}{cc}
0 \\ 1 \end{array}\right). \label{per27} \eea
To construct the first-order Floquet Hamiltonian, we start with the Schr\"odinger equation
\bea i\frac{d\ket{\psi(t)}}{dt}=(H_{0}+H_{1})\ket{\psi(t)},\label{per28} \eea
where we assume that $\ket{\psi(t)}$ has the form
\bea \ket{\psi(t)}=\sum_{n}c_{n}(t)e^{-i\int_{0}^{t} ~dt' E_{n}(t')}\ket{n},\label{per29} \eea
and $\ket{n}=\ket{\pm}_{k}$ in our case. Using Eq. \eqref{per28} and keeping terms up to first order in $H_{1}$, we find that 
\bea c_{m}(T)~=~ c_{m}(0) ~-~ i ~\sum_{n}\int_{0}^{T} ~dt ~\bra{m}H_{1}\ket{n}~ 
e^{i\int_{0}^{t} ~dt' [E_{m}(t')-E_{n}(t')]} ~c_{n}(0). \eea
This can be re-written as follows
\bea c_{m}(T)&=&\sum_{n} ~(I-iH_{Fk}^{(1)}T)_{mn} ~c_{n}(0), \label{per30} \eea
where $I$ denotes the identity matrix and $H_{Fk}^{(1)}$ refers to the first-order effective Hamiltonian. We then find that
\bea \bra{+}H_{Fk}^{(1)}\ket{+} &=& 0,~~~~~ \bra{-}H_{Fk}^{(1)}\ket{-}~=~ 0,\non\\
\bra{+}H_{Fk}^{(1)}\ket{-} &=& J~(1+e^{-2ik}) e^{iA}\left(\frac{\sin A}{A}\right),~~~~~
\bra{-}H_{Fk}^{(1)}\ket{+}=J(1+e^{2ik}) e^{-iA}\left(\frac{\sin A}{A}\right),\label{per31}
\eea
where 
\beq A ~=~ \frac{\mu T}{2} ~=~ \frac{\pi \mu}{\om}. \eeq
To first order in $H_1$, therefore, $H_{F}^{(1)}$ is given by
\bea H_{F}^{(1)}=e^{iA}\left(\frac{\sin A}{A}\right) ~\sum_{k} ~[J~(1+e^{-2ik}) a_{k}^{\dagger}b_{k}+ {\rm H.c.}]. \label{per32} \eea

Before discussing further, we note a symmetry of the Floquet operator,
\beq [U_{k}(\mu,J)]^{-1} ~=~ U_{k}(\mu,-J), \label{sym} \eeq 
which follows from the expression in Eq.~\eqref{per25} and the fact that the 
driving protocol satisfies $\mu (T-t) = - \mu (t)$. Eq.~\eqref{sym} implies\cite{dyntran4} that
\beq H_{Fk}(\mu,J) ~=~ - ~H_{Fk} (\mu,-J). \label{per33} \eeq
Hence $H_{F}$ can only have terms with odd powers of $J$. This implies that the 
next order term after the first order will be third order since there cannot be 
a term of second order in $J$. Hence, the first-order effective Hamiltonian will 
be a very good approximation to the exact Hamiltonian in the limit $\mu \gg J$. 

We also note that the Floquet quasienergy $E_{Fk}$ must be an even function of $k$ 
if we hold $J,~\mu$ fixed. To prove this, we do the unitary transformation
\bea H_{k}(t)~\rightarrow~V_{k} ~H_{k}(t) ~V_{k}^{-1},\non\\
\text{where}~~~~V_{k}~=~ \left(\begin{array}{cc}
1 & 0\\ 0&e^{-ik}\end{array}\right). \eea
Thus we obtain
\bea H_{k}(t)=\left(\begin{array}{cc}
\mu (t) & 2J\cos k\\ 2J\cos k&-\mu (t)\end{array}\right). \eea
With this Hamiltonian, it is clear that
\bea U_{k}(\mu,J) ~=~ U_{-k}(\mu,J) ~~~{\rm which ~implies ~that}~~~ H_{Fk}(\mu,J) ~=~
H_{F,-k}(\mu,J). \label{uksym} \eea
Hence the Floquet quasienergy must be an even function of $k$.

\subsection{Dynamical localization for a single-particle system}

\begin{figure}[!tbp]
\footnotesize
\stackunder[5pt]{\includegraphics[width=0.48\hsize]{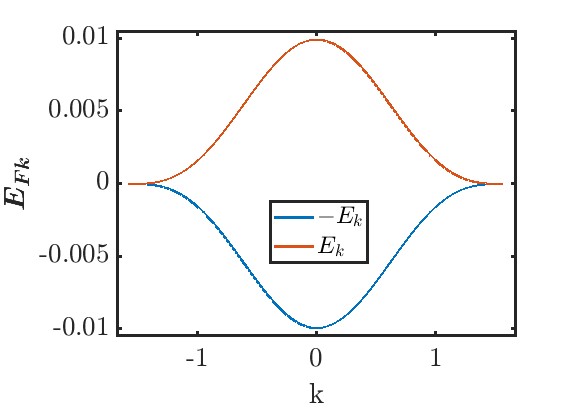}}{\large(a)}
\stackunder[5pt]{\includegraphics[width=0.48\hsize]{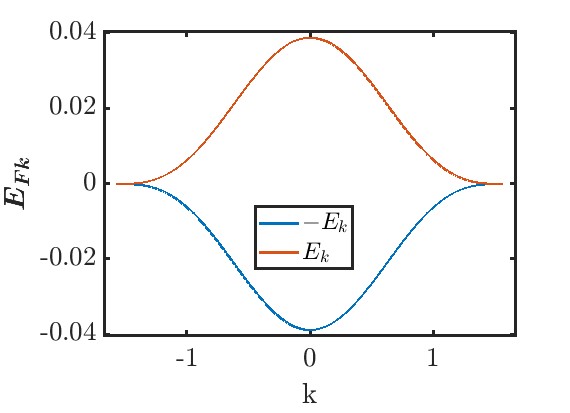}}{\large(b)}%
\caption{{\bf Dispersion of quasienergies $\mathbf{E_{Fk}}$ at DL 
points:} (a) Plot of $E_{Fk}$ versus $k$ obtained from the exact numerical 
calculation for $J=1$, $\mu=20$, and $\om=20$. We find that $\De ~\sim~0.02$, 
which implies an almost 
perfectly flat band at zero energy. (b) The same plot for $J=1$, $\mu=10$,
and $\om=10$. In this case, the bandwidth $\De = 0.08$ which is four times
larger than in the first case.}\label{fig01} \end{figure}

It is evident from the form of the Hamiltonian obtained by first-order FPT that the system will (approximately) exhibit DL~\cite{dynloc1,dynloc2,dynloc3,dynloc4,dynloc5,dynloc6,dynloc7} when
\beq A ~=~ n\pi, ~~~{\rm i.e.,}~~~ \mu ~=~ n\om, \label{per35} \eeq
where $n$ is a non-zero integer. We note that this condition for DL 
becomes more and more exact as $\mu /J \to \infty$ and the 
higher order corrections to the first-order effective Hamiltonian become 
negligible. In this limit, $H_{Fk}^{(1)}$ vanishes for all values of $k$ which 
produces a flat band with zero quasienergy. We can see in Fig.~\ref{fig01}
(a) that for a system with $J=1$, $\mu=20$, and $\om=20$, the quasienergy band is almost flat with a bandwidth $\De ~\sim~0.02$.
In this case, we have taken $\mu \gg J$, and the higher order corrections to the 
first-order effective Hamiltonian are therefore very small.
However, if we decrease the value of $\mu$ holding $J$ fixed, the first-order
Floquet Hamiltonian becomes less and lass accurate, and the DL
begins to fail as can be seen in Fig.~\ref{fig01} (b), for a system with $J=1$,
and $\mu=\om=10$. In this case, the bandwidth, $\De~\sim~0.08$, which is four times larger than in Fig.~\ref{fig01} (a). This is due to the fact that the 
third-order effective Hamiltonian scales as $J^{3}/\mu^2$ at the dynamical localization points ($\mu=n\om$) obtained from the first-order effective 
Hamiltonian as seen in \hyperlink{Appendix C}{Appendix C}. Interestingly, the third-order effective Hamiltonian does not explicitly depend on the value of 
$\om$ if one keeps $J$ 
and $\mu$ fixed at the DL points. Hence the bandwidth remains unaffected 
if $\om$ is changed but $\mu = n \om$ is kept fixed. For instance, we find 
numerically that the bandwidth is the same for the parameter values 
($J=1$, $\mu=20$, $\om=20$) and ($J=1$, $\mu=20$, $\om=10$).

\subsection{Dynamical phase transition}

Motivated by our previous work~\cite{dyntran5}, we now study the relaxation behavior of some correlators for the non-interacting model ($V=0$). We will examine the correlation 
function $a_{j}^{\dagger}b_{j}$ (where $j$ denotes a particular unit cell) at stroboscopic times $t=nT$,
\bea C_{n} ~=~ \bra{\Psi_{0}}a_{j}^{\dagger}(nT)b_{j}(nT)\ket{\Psi_{0}}, \eea
where $\ket{\Psi_{0}}$ is an initial state. 
For simplicity, we will take $\ket{\Psi_{0}}$ to be a product state in 
momentum space with the form
\bea \ket{\Psi_{0}} ~=~ \prod_{k}a_{k}^{\dagger} ~\ket{0}. \eea
Since this state is translation invariant, the correlator $C_n$ will not depend
on the unit cell index $j$.
We will now investigate the relaxation behavior of $C_n$ for large values of $n$, particularly to see if there is any crossover behavior. We observe 
numerically that generally in the $\mu\gg J$ limit (with a few exceptions),
the correlation function exhibits a $n^{-1/2}$ decay with oscillations and there is no crossover behavior. To explain this, we first note that the Floquet quasienergy obtained from the first-order Floquet Hamiltonian has the form
\bea E_{Fk}^{\pm}&=&\pm ~E_{k},\non\\
\text{where}~~~~ E_{k} &=& 2J\left(\frac{\sin A}{A}\right)\cos k. \eea
The stationary point of $E_{k}$ within the first Brillouin zone lies at $k=0$.
Next, we can show that the time-dependent part of the correlation 
function can be expressed as
\bea \delta C_{n} ~\sim~ \frac{2}{N}~\sum_{k} ~[f(k)e^{-i2nTE_{k}}+ {\rm H.c.}], 
\eea
where $f_{k}=\bra{\Psi_{0}}a_{j}^{\dagger}(0)b_{j}(0)\ket{\Psi_{0}}$.
For $N\rightarrow \infty$ and assuming $f(k)$ to be real, the above equation takes the integral form
\bea \delta C_{n}~ \sim~ \frac{1}{\pi} ~\int_{-\pi/2}^{\pi/2} dk ~f(k)~
\cos (2nT E_{k}). \label{cnint} \eea
For large values of $n$, this integral gets a dominant contribution from the regions close to the stationary points of $E_{k}$\cite{dyntran2,dyntran4,dyntran5}. Therefore, 
expanding Eq.~\eqref{cnint} around the stationary point at $k=0$, we find that 
\bea \delta C_{n} ~\sim~ \frac{1}{\pi} ~\int 
dk ~f(k=0)~ e^{i\zeta n(1-k^{2}/2)}, \label{eq29} \eea
where $\zeta=4JT (\sin A /A)$. Assuming $f(k)\neq 0$ for a generic 
initial state, we find that the correlation function for large values of 
$n$ will decay as a power $n^{-1/2}$, and there will be oscillations due 
to the term $\cos (\zeta n)$. This is the usual decay behavior unless 
there are competing terms which come from higher-order corrections. 

\begin{figure}[!tbp]
\begin{center}
\footnotesize
\stackunder[5pt]{\includegraphics[width=0.49\hsize]{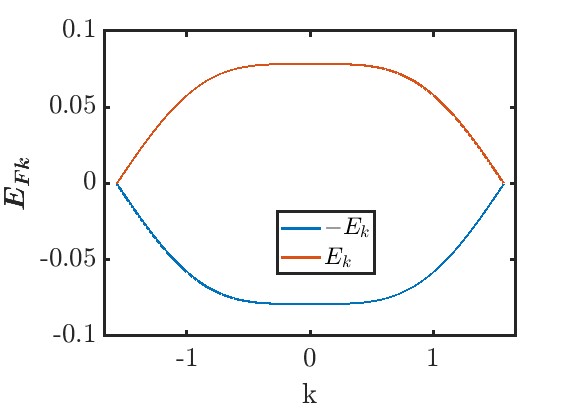}}{\large(a)}
\stackunder[5pt]{\includegraphics[width=0.49\hsize]{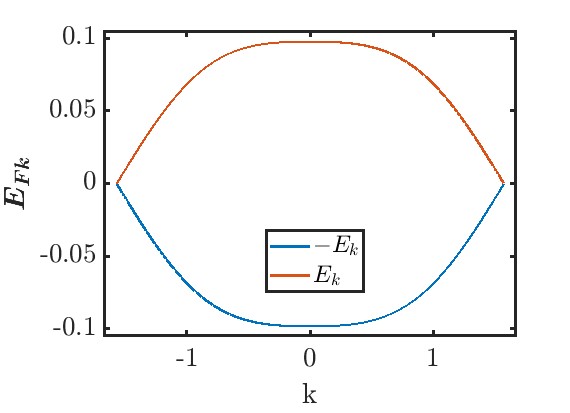}}{\large(b)}\\
\stackunder[5pt]{\includegraphics[width=0.49\hsize]{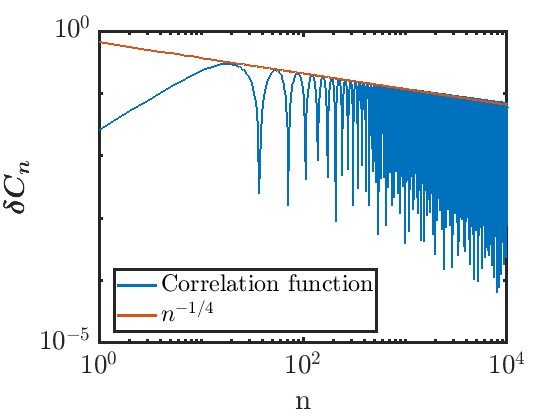}}{\large(c)}%
\stackunder[5pt]{\includegraphics[width=0.49\hsize]{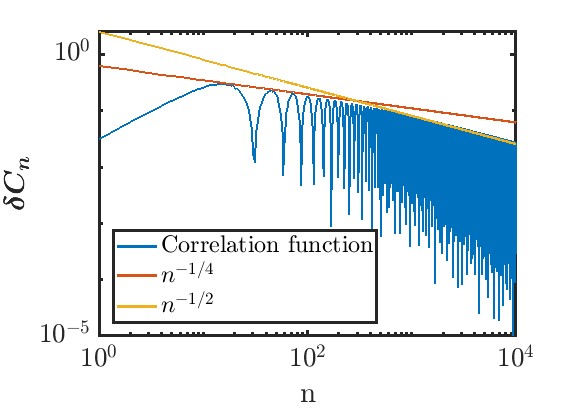}}{\large(d)}\\
\stackunder[5pt]{\includegraphics[width=0.49\hsize]{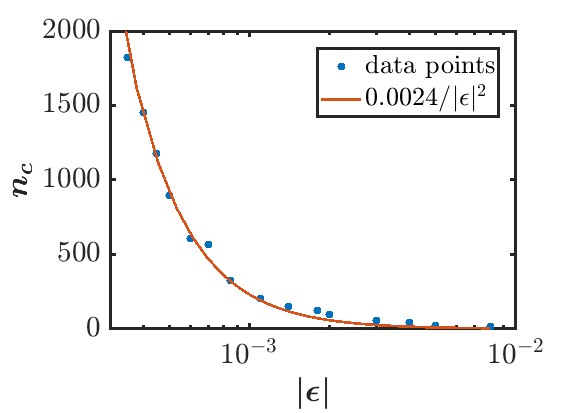}}{\large(e)}%
\end{center}
\caption{{\bf Quasienergies and crossover behaviors of correlation functions:} 
Plots showing $E_{Fk}$ as a function of $k$ obtained from the exact numerical 
calculation for $J=1$, $\mu=10$, and (a) $\om=10.6$ and (b) $\om=10.7$.
In plot (a), $E_{k} ~\sim~ k^{4}$ around $k=0$ as can be seen in Eq.~\eqref{eq31}.
In plot (b), $E_{k}~\sim~ k^{2}+\ep k^{4}$, where $|\ep|\ll1$, as seen in 
Eq.~\eqref{eq33}. Log-log plots of the absolute value of the $n$-dependent part 
of the correlation function $\delta C_{n}$ as a function of the time $nT$, 
for $J=1$, $\mu=10$, and (c) $\om=10.6$ and (d) $\om=10.7$.
(c) The correlation function decays as $n^{-1/4}$ along with oscillations. 
(d) The plot shows a crossover between an oscillatory term times $n^{-1/4}$ 
and an oscillatory term times $n^{-1/2}$. (e) Plot showing the variation of $n_c$ with $|\epsilon|$ as we approach the critical frequency from the 
$\om>\om_{c}$ side, where $\om_c \simeq 10.6$ is the frequency where the correlation function decays as $n^{-1/4}$. The numerically obtained fitting indicates that 
$n_{c}\sim 1/|\epsilon|^2$.} \label{fig02} \end{figure}

We encounter such higher-order terms slightly away from the DL points (we recall
that a DL point is where $A$ is an integer multiple of $\pi$, i.e., $\mu/\om$ 
is an integer). Since an analytical calculation of the third-order effective 
Hamiltonian is a tedious task, we analyze this regime numerically. We 
first find the Floquet quasienergy from the numerically exact calculation 
and then do a fitting analysis of it. The results we obtain from such an 
analysis are as follows. We consider the 
parameter values $J=1$, $\mu=10$, and $\om=10.6$. 
Taking into account the structure of the stationary point obtained from the 
first-order effective Hamiltonian, we fit the
numerically computed Floquet quasienergy around $k=0$ as a function of $k$ up to 
sixth order in $k$. We find the following functional form 
\bea E(k)~=~p_{4} k^{4}+ p_{0},~~~~ {\rm where} ~~~~p_{4}~=~ -0.02882,~
~~~p_{0}~=~ 0.07865. \label{eq30} \eea
All terms with odd powers of $k$ are found to be zero which is expected by
the symmetry discussed in Eq.~\eqref{uksym}. Further, the coefficients of $k^{2}$ and $k^{6}$ are also found to be zero at these particular parameter values. Hence
Eq.~\eqref{eq30} shows that $E_{k}$ goes as $k^{4}$ near $k=0$. An analysis 
similar to the one following Eq.~\eqref{eq29} then shows that 
\bea \delta C_{n}\sim\frac{1}{\pi}\int dk ~f(k=0) 
~e^{i2nT(p_{0}+p_{4}k^{4})}. \label{eq31} \eea
Assuming $f(k)\neq 0$, we see that for large values of $n$, the correlation
function will decay as $n^{-1/4}$ with oscillations due to the $\cos(2 nT p_{0})$
term. This is what we see in the Fig.~\ref{fig02} (c):
for a system with the parameter values mentioned above, the correlation function function decays as an oscillatory term times $n^{-1/4}$. If we plot $|\de C_n|$ (rather than $\de C_n$) versus $n$, the period of oscillations $\De n$ will be
given by the condition, $2 \De n p_{0}T=\pi$, which implies that
$\De n=\om/(4p_{0})$. Putting $\om=10.6$, and $p_{0}=0.07865$, we find that $\De n~ \sim~ 34)$, 
which agrees very well with the oscillation period seen in Fig.~\ref{fig02}.
Interestingly, we observe a crossover from $n^{-1/4}$ to $n^{-1/2}$ as we move slightly away from $\om=10.6$. As mentioned earlier, the correlation function in general decays as a power $n^{-1/2}$ for large values of $n$ in this class of systems. However, we observe a different power law decay behavior ($n^{-1/4})$ emerging at $\om \simeq 10.6$, and we, therefore, call $\om \simeq 10.6$ the critical frequency $\om_c$. To see the crossover, we consider $J=1$,
$\mu=10$, and $\om=10.7$ as an example. We again follow the same fitting procedure, and find the following functional form
\bea E(k) ~=~ p_{4} k^{4} + p_{2} k^{2}+p_{0},~~ {\rm where}~~ p_{4}~=
-0.02823 ,~~p_{2}~= -0.009116,~ ~p_{0}~=~ 0.0981. \label{eq32} \eea
The terms with odd powers of $k$ vanish as before.
In contrast to the previous case, however, there is now a competition between 
the $k^{2}$ and $k^{4}$ terms, which can be seen in Eq.~\eqref{eq32}. Close
to $k=0$, we see that $E_{k}$ has a leading contribution coming from the
$k^{4}$ term, followed by a subleading correction due to the $k^{2}$ term very 
close to $k=0$. Expanding the integrand in Eq.~\eqref{cnint} around $k=0$ then gives
\bea \delta C_{n} ~\sim~ \frac{1}{\pi} ~\int dk ~f(k=0) ~e^{i2nT(p_{0}+p_{2}k^{2}+p_{4}k^{4})}. \label{eq33} \eea

Defining a scaled variable $k'=kn^{1/4}$, and assuming $f(k)\neq0$, we find
\bea \delta C_{n} ~\sim~ \frac{1}{\pi n^{1/4}} ~e^{i2nT p_{0}} ~f(k=0)~
\int dk' ~e^{iT(p_{2}k'^{2}n^{1/2}+p_{4}k'^{4})}. \eea
Since $|p_{2}|\ll|p_{4}|$, it is clear from Eq.~\eqref{eq33} that there will
be a $n^{-1/4}$ scaling (along with oscillations) when $|\epsilon|n^{1/2}\ll1$, where $\epsilon'=p_{2}/p_{4}$. However, when 
$|\epsilon'|n^{1/2}\gg 1$, the $n^{-1/4}$ scaling breaks down and we then 
encounter a different scaling law, namely, $n^{-1/2}$ (along with oscillations 
due to the $\cos (2nT p_{0})$ term). We can extract the crossover scale 
$n_{c}$ from this analysis; a crossover between the $n^{-1/4}$ and $n^{-1/2}$ power-laws occurs when $|\epsilon'|n_{c}^{1/2} \sim 1$, which implies that 
$n_{c} \sim 1/|\epsilon'|^{2}$. To see this behavior, we define $\epsilon$ 
as $\om=1/(1/\om_c-\ep)$~\cite{dyntran5}, where $\ep\propto\ep'$, and we look at
the divergence behavior near $\om_c$. We plot the crossover scale $n_c$ with 
$\epsilon$ in Fig.~\ref{fig02} (e), and then numerically fit the plot 
of $n_c$ versus $\ep$. We find that $n_c \sim 1/|\epsilon|^2$ which agrees 
with the analytically derived result.

\subsection{Effects of interactions on dynamical localization}

In this section, we will look at the effects of density-density interactions~\cite{dynloc4,dynloc6}, $H_{I}=\sum_{j}Vn_{j}n_{j+1}$ on DL. 
For this case, we take the system to be at 
half-filling with periodic or antiperiodic boundary conditions ($c_{N+1} =
c_1$ or $- ~c_1$) depending on whether the particle number is even 
or odd, respectively, to avoid any degeneracy in the spectrum. To get an analytical 
understanding of this system, we will first compute the effective Hamiltonian up 
to first order in $V$. Since $H_{I}$ is diagonal in the position
basis and commutes with the unperturbed Hamiltonian, $H_{0}(t)$, the
effective Hamiltonian to first order in $V$ simply reads as
\bea H_{FI}^{(1)}=\sum_{j} ~V n_{j}n_{j+1}. \eea
Hence the full effective Hamiltonian to first order in $J$ and $V$ is as follows
\bea H_{F}^{(1)}&=&H_{F,J}^{(1)} ~+~ H_{FI}^{(1)},\non\\
H_{F,J}^{(1)}&=&e^{iA} ~\left(\frac{\sin A}{A} ~\right) ~\sum_{k} ~\left[J(1+e^{-2ik}) a_{k}^{\dagger}b_{k}+ {\rm H.c.}\right],\non\\
H_{FI}^{(1)}&=& \sum_{j} ~V n_{j}n_{j+1}. \label{heff1} \eea
We note that the Floquet evolution operator $U(T)$ satisfies the same condition
as mentioned in Eq.~\eqref{per33}\cite{dyntran4}, and therefore $H_{F}$ possesses the symmetry
\bea H_{F}(\mu,J,V) ~=~-~ H_{F}(\mu,-J,-V). \label{hamsym} \eea
Hence the first-order effective Hamiltonian will be a very good approximation to the exact Hamiltonian for $\mu\gg J,~V$, since the higher order 
corrections will be negligible compared to the first-order term. We see from
Eq.~\eqref{heff1} that $H_{F,~J}^{(1)}=0$ at the DL points where $A$ is
an integer multiple of $\pi$, and then
$H_{F}^{(1)}$ just reduces to $H_{FI}^{(1)}$. Consequently, the spectrum of the Floquet quasienergies becomes easy to compute at any filling due to the diagonal form of the effective Hamiltonian in the position basis.

\begin{figure}[!tbp]
\footnotesize
\begin{center}
\stackunder[5pt]{\includegraphics[width=0.45\hsize]{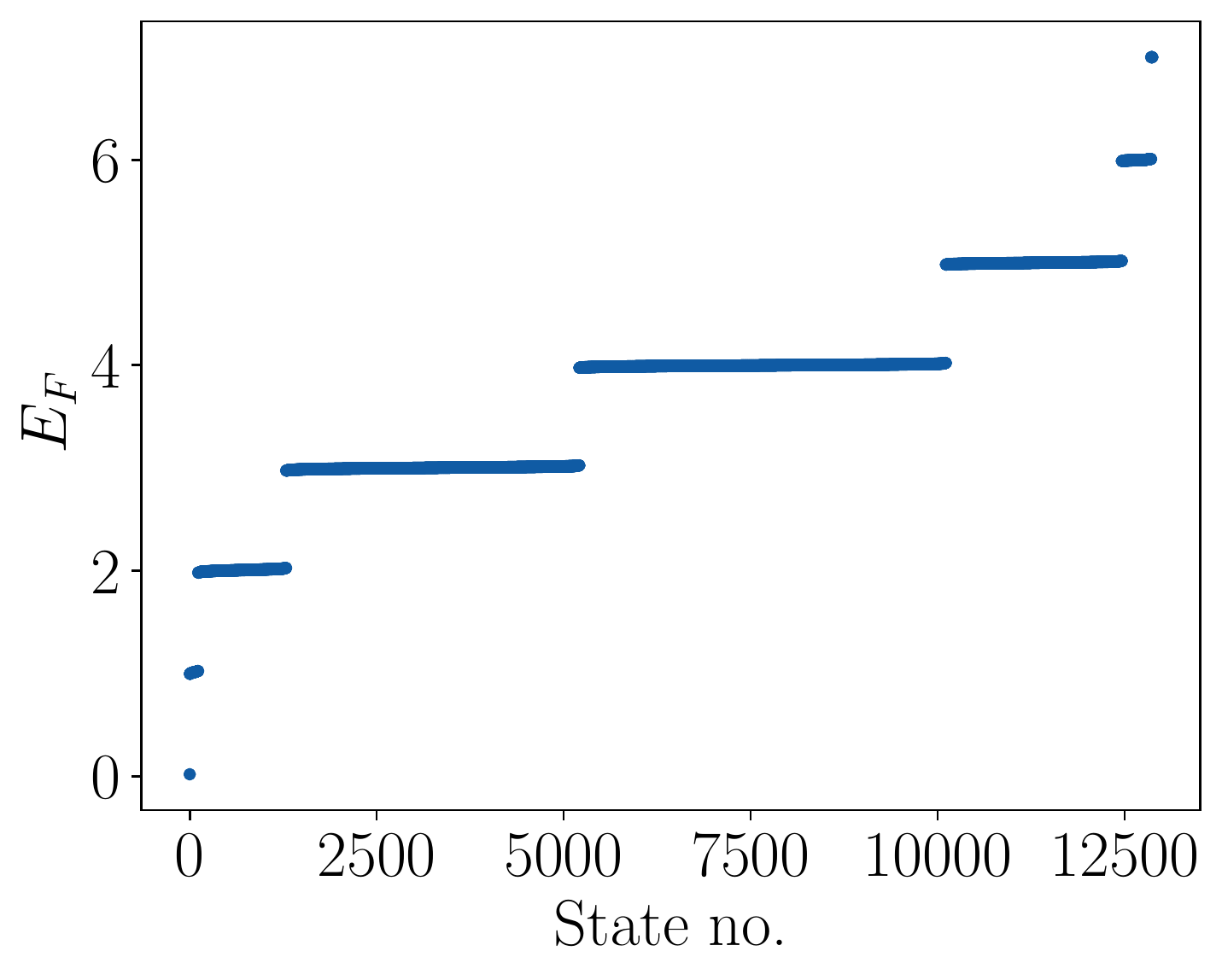}}{\large(a)}
\stackunder[5pt]{\includegraphics[width=0.45\hsize]{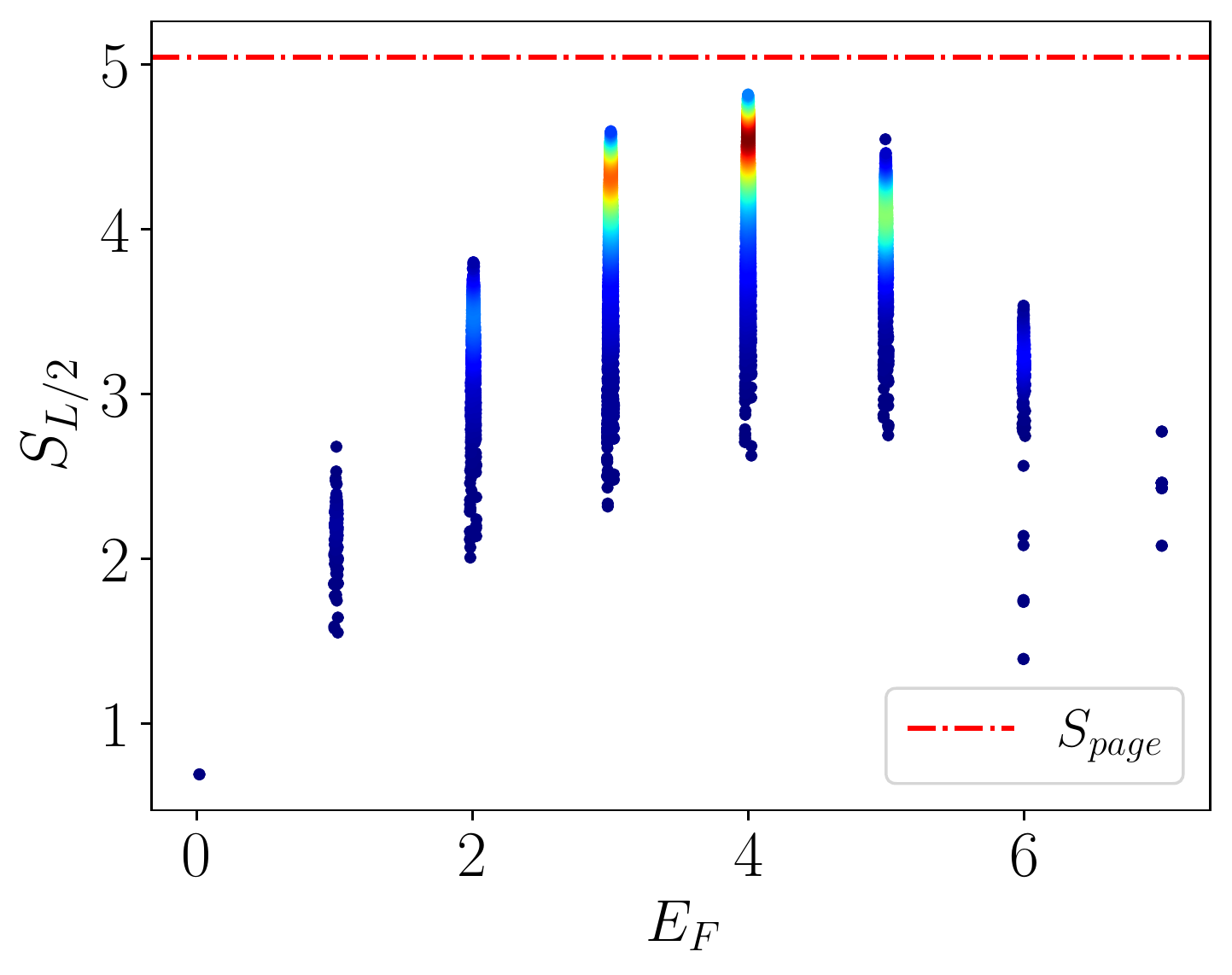}}{\large(b)}\\
\stackunder[5pt]{\includegraphics[width=0.45\hsize]{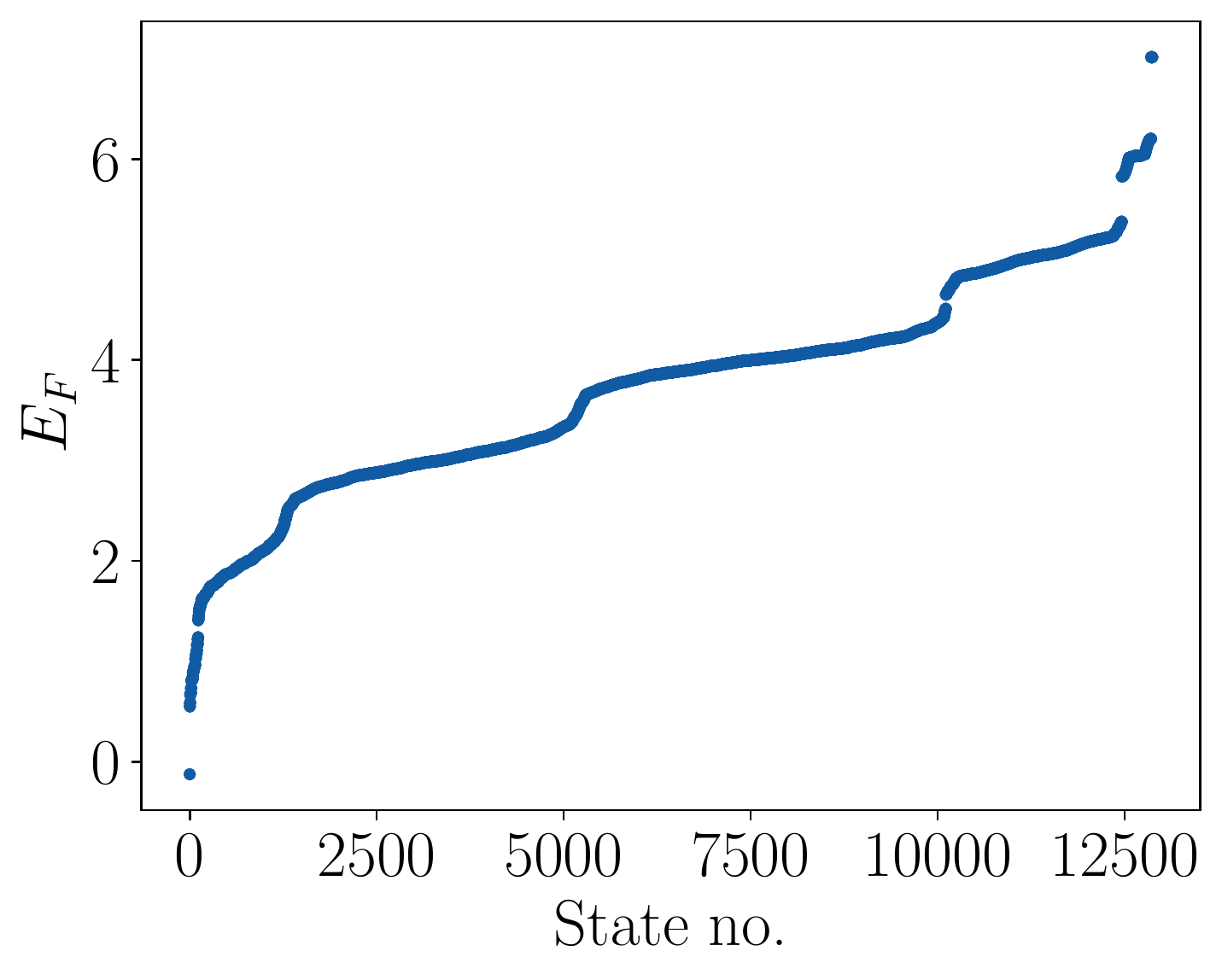}}{\large(c)}
\stackunder[5pt]{\includegraphics[width=0.45\hsize]{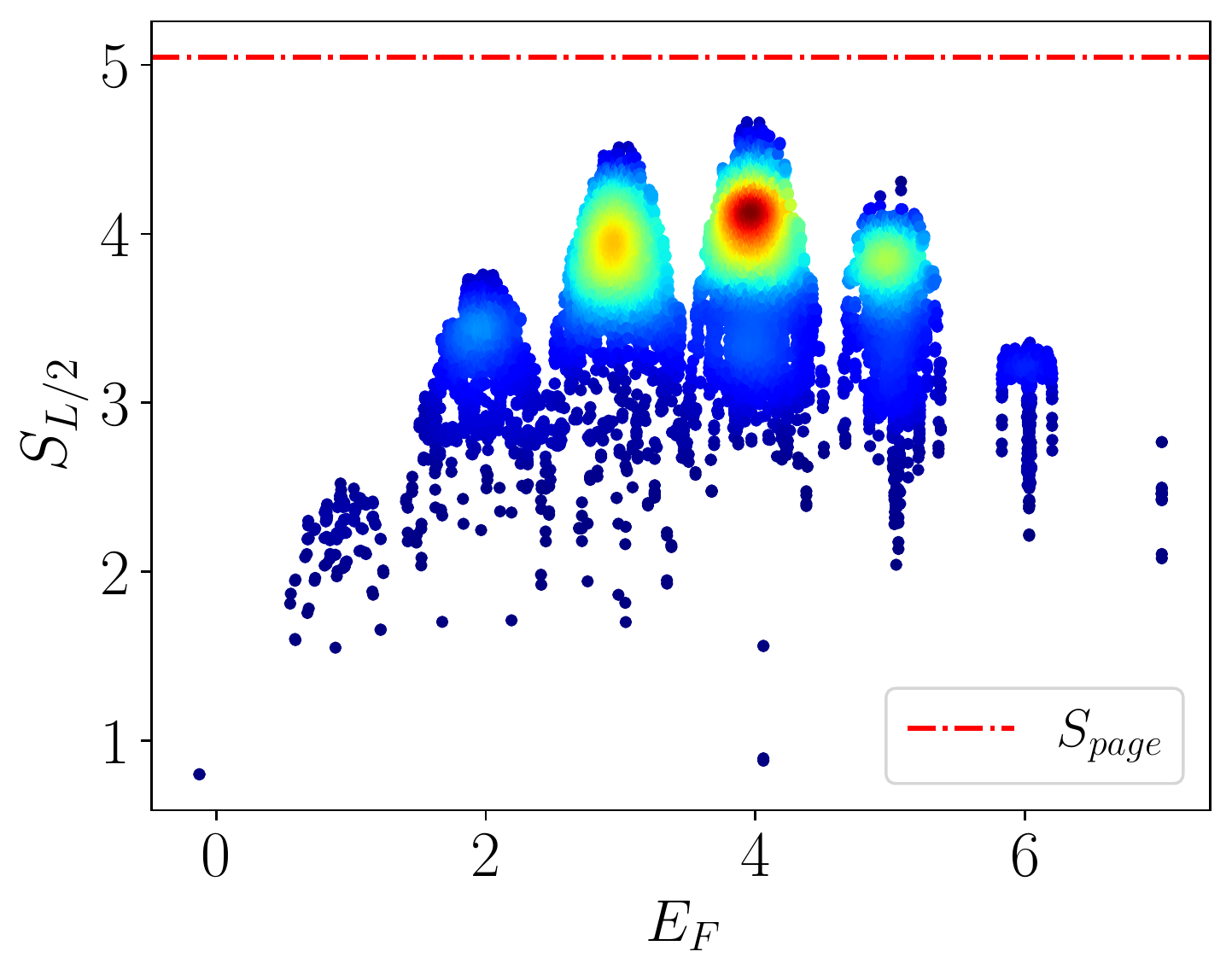}}{\large(d)}
\end{center}
\caption{{\bf Quasienergy and entanglement entropy spectrum of the period-2 model at a DL point and away from a DL point:} (a-b) Plots of the quasienergy spectrum $E_{F}$ 
and half-chain entanglement entropy $S_{L/2}$ as a function of $E_{F}$ exactly at a dynamical localization point with $J=1$, $\mu=\om=20$, and $V=1$, where the 
system exhibits 
many-body flat bands with many low-entanglement states near the middle of the spectrum. 
(c-d) Plots showing the same quantities as in plots (a-b) but away 
from a DL point, with $J=1$,
$\mu=20,~\om=22$ and $V=1$. For this case, the many-body flat bands start to 
disappear as we tune the system away from a DL point.} \label{fig03} \end{figure}

We now consider a system at half-filling with antiperiodic boundary conditions for $J=1$, $\mu=20,~\om=20$, 
$V=1$, and $L=16$, and calculate the spectrum of the Floquet quasienergies and 
half-chain entanglement entropy. The 
dimension of the Hilbert space for $N=L/2$ and $L=16$ is $^{16}C_{8} = 12870$.
We first discuss the Floquet quasienergies and the degeneracies for some of the Floquet eigenstates which can be obtained analytically. We observe that 
there are eight equally spaced quasienergies lying between $0$ and $8V$ with a energy spacing of $V$ for $L=16$, as shown in Fig.~\ref{fig03} (a). Next, 
there are exactly two Floquet 
eigenstates, $\ket{\pm}=1/\sqrt{2}\left(\ket{1'}\pm\ket{2'}\right)$ 
with $E_{F}=0$, where $\ket{1'}$ and $\ket{2'}$ are equal to $\ket{1010101010101010}$ and $\ket{0101010101010101}$ respectively in the
number basis. Furthermore, there are 16 Floquet eigenstates with $E_{F}=8V$. 
We note that the DL induces an emergent integrability which leads to the appearance
of many flat bands and several low-entanglement states (with $S_{L/2}\ll S_{page}$, 
where $S_{page} = (L/2) \ln 2 - 1/2$~\cite{page1,page2}) near the middle of the 
spectrum as can be seen in Figs.~\ref{fig03} (a) and (b) respectively. 
However, this emergent integrability starts to break down as we move away from
a DL point, as we see in Fig.~\ref{fig03} (d), for $J=1,~\mu=20,~\om=22$ and $V=1$. In Figs.~\ref{fig03} (c) and (d), we observe that the flatness of the bands begins to disappear
as we move away from the DL limit.

\begin{figure}[!tbp]
\footnotesize
\begin{center}
\stackunder[5pt]{\includegraphics[width=0.52\hsize]{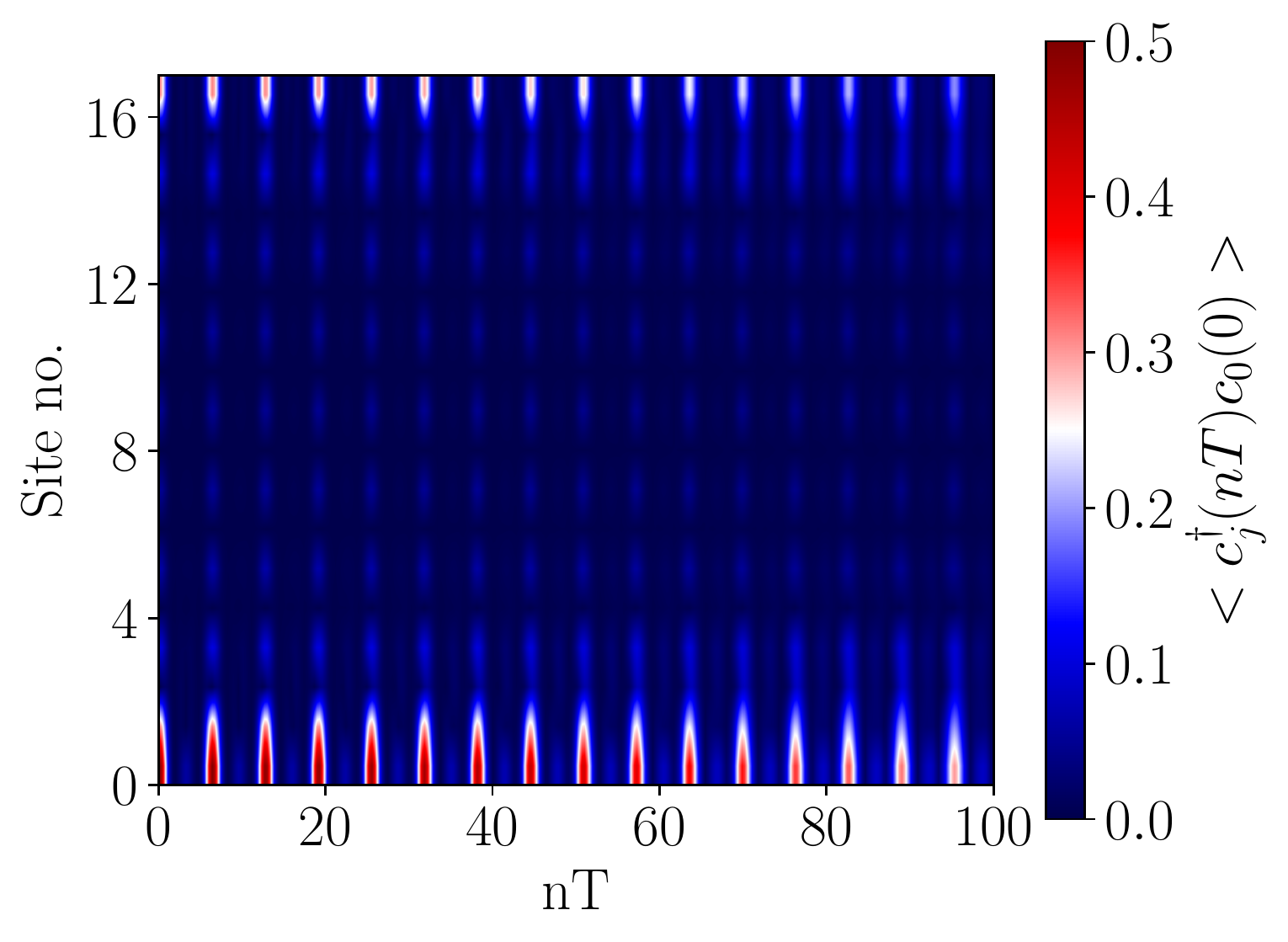}}{\large(a)}
\stackunder[5pt]{\includegraphics[width=0.52\hsize]{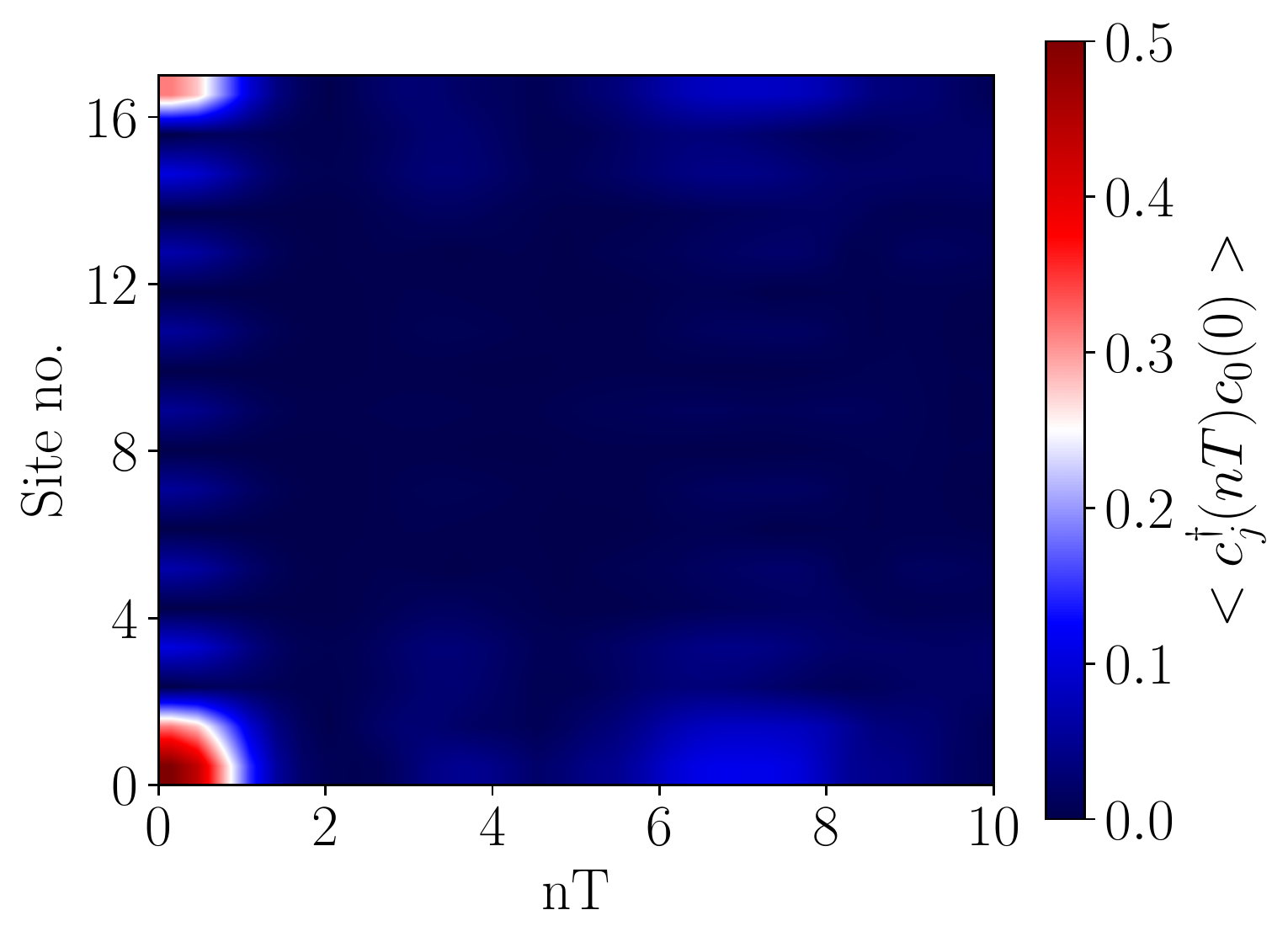}}{\large(b)}\\
\stackunder[5pt]{\includegraphics[width=0.6\hsize]{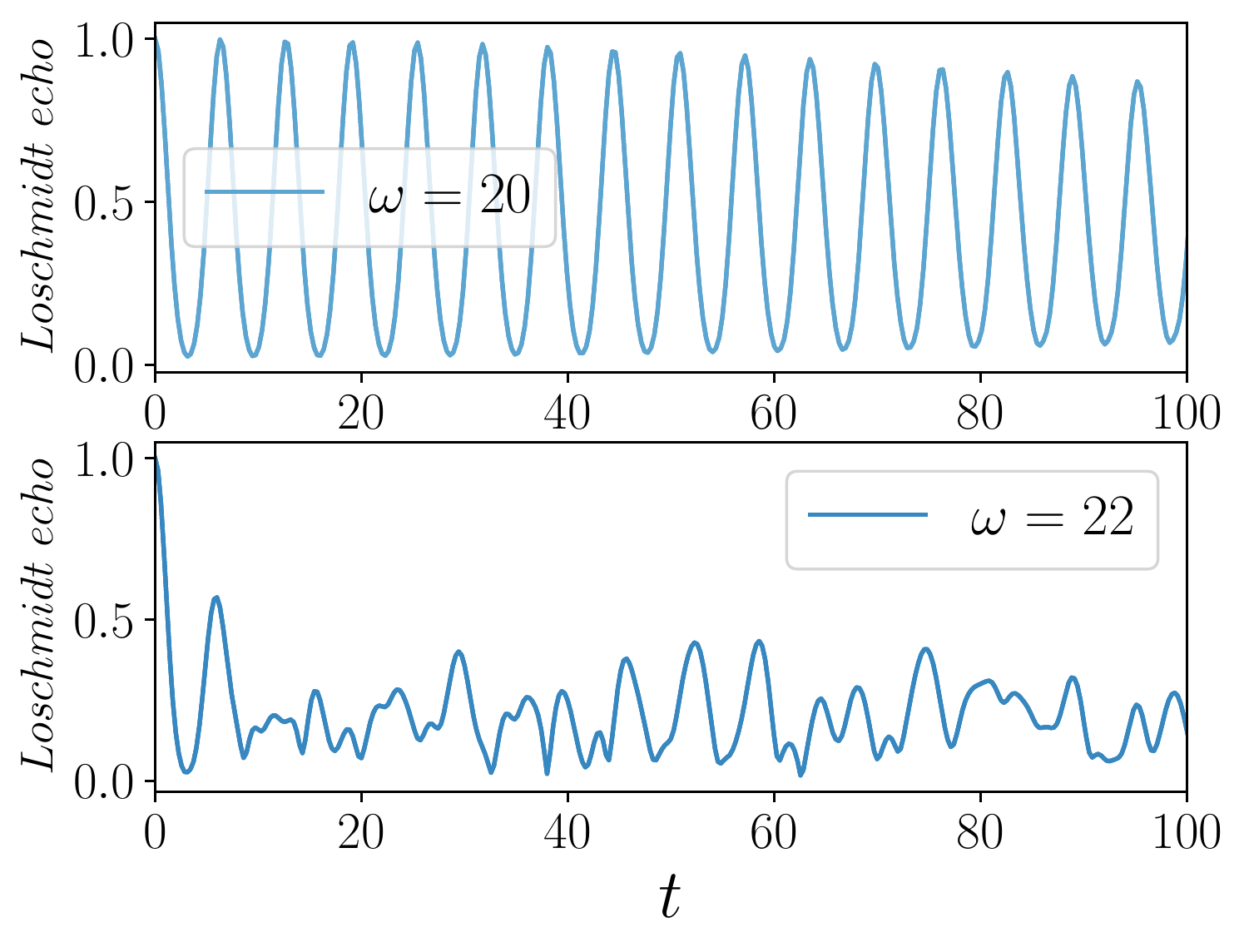}}{\large(c)}
\end{center}
\caption{{\bf Dynamics of the correlation function and Loschmidt echo at a DL point and away from a DL point for the period-2 model:} 
(a-b) Surface plots showing the two-point correlation function as a function of site number and time 
$nT$ with $n$ being the driving cycle number, at a dynamical 
localization point and away from a dynamical localization point. The parameter values chosen for plots (a) and (b) are the same as in Figs.~\ref{fig03}. (c): Plots 
showing the Loschmidt echo versus time for the same parameter values as in Figs.~\ref{fig03}. For all four cases, we choose the initial state to be the ground 
state of the undriven Hamiltonian. As shown in (a) and the upper panel of
(c), the dynamics demonstrates long-time oscillatory behaviors indicating a non-ergodic 
behavior at a DL point. However, both correlation function and Loschmidt echo decay rapidly with time as we move away from a DL point, as can be seen in (b) and the lower panel of (c).} \label{fig04} \end{figure}

In Fig.~\ref{fig04}, we show some dynamical properties of the system, namely, the two-point correlation function and the Loschmidt echo at a DL point and away from a DL point.
The parameter values chosen for this figure are the same as in Fig.~\ref{fig03}. We 
will take the initial state to be the ground state of the undriven 
Hamiltonian. In Fig.~\ref{fig04} (a), we see almost perfect revivals of the correlation function in time, which is expected due to the emergent integrability
at a DL point. Further, the oscillation period of the revivals can be calculated by using the condition $e^{iV \de t}=1$. Hence the time period is given by $\delta 
t=2\pi/V$, where $V=1$ in our case. Away from the DL point, the correlation function decays rapidly in time due to a breakdown of the integrable 
structure. In Fig.~\ref{fig04} (c), we show the Loschmidt echo, $|\langle\psi(t)|\psi(0)\rangle|$, for the same parameter values as in Figs.~\ref{fig04} (a) and \ref{fig04} (b) and with the same initial states. For the first case with $\om=20$, 
we again see perfect revivals with a period of $2\pi$ in Fig.~\ref{fig04} (c), upper panel. For the second case
with $\om = 22$, the amplitude of Loschmidt echo decays rapidly as 
shown in Fig.~\ref{fig04} (c), lower panel. These results indicate that the system
evades thermalization for a long time at a DL point but thermalizes quickly as we go away from a 
DL point~\cite{dynloc1,dynloc2,dynloc3,dynloc4,dynloc5,dynloc6,dynloc7}.

\subsection{Effects of resonances}
\label{sec4}

In this section we will examine the effects of resonances~\cite{rev5,reso} on DL for the period-2 
model. To obtain an analytical insight for this case, we will consider parameter values with $\mu=\om=V \gg J$ and derive an effective Floquet Hamiltonian using 
first-order FPT. We first consider a four-site system so 
that we can easily identify various non-trivial processes, and we will then 
generalize it to larger system sizes. We find that a four-site system only 
offers four distinct non-trivial processes for a particular choice of a periodic potential pattern due to the constraints imposed by DL. We further note that there 
are two possible potential patterns available for such a system consisting of four
sites. Therefore, we need to consider a total of eight distinct non-trivial processes
while formulating the first-order effective Floquet Hamiltonian. These eight processes 
and the corresponding time-dependent effective Hamiltonians are listed below.

\vspace*{.4cm}
\begin{table}[!ht]
\begin{center}
\begin{tabular}{|c|c|c|c|} 
\hline
 Pattern of periodic potential &Process & Effective time-dependent Hamiltonian\\
\hline
+~-~+~-&1100 $\leftrightarrow$ 1010 & $
H(t)=\left(\begin{array}{cc}
V & J\\ J&2\mu(t)\end{array}\right)$\\
\hline
+~-~+~-&0100 $\leftrightarrow$ 0010 & $
H(t)=\left(\begin{array}{cc}
-\mu(t) & J\\ J&\mu(t)\end{array}\right)$\\
\hline
+~-~+~-&0101 $\leftrightarrow$ 0011 & $
H(t)=\left(\begin{array}{cc}
-2\mu(t) & J\\ J&V\end{array}\right)$\\
\hline
+~-~+~-&1101 $\leftrightarrow$ 1011 & $
H(t)=\left(\begin{array}{cc}
V-\mu(t)& J\\ J&V+\mu(t)\end{array}\right)$\\
\hline
-~+~-~+&1100 $\leftrightarrow$ 1010 & $
H(t)=\left(\begin{array}{cc}
V & J\\ J&-2\mu(t)\end{array}\right)$\\
\hline
-~+~-~+&0100 $\leftrightarrow$ 0010 & $
H(t)=\left(\begin{array}{cc}
\mu(t) & J\\ J&-\mu(t)\end{array}\right)$\\
\hline
-~+~-~+&0101 $\leftrightarrow$ 0011 & $
H(t)=\left(\begin{array}{cc}
2\mu(t) & J\\ J&V\end{array}\right)$\\
\hline
-~+~-~+&1101 $\leftrightarrow$ 1011 & $H(t)=\left(\begin{array}{cc}
\mu(t)+V & J\\ J&V-\mu(t)\end{array}\right)$\\
\hline
\end{tabular}
\caption{\label{per2res} Allowed processes and their corresponding effective time-dependent Hamiltonians for a four-site system with all possible patterns of periodic on-
site potential in the case dynamical localization and resonance for a period-2 model.}
\end{center}
\end{table}
\vspace*{.4cm}

As an example, we consider the first process listed above and calculate the 
first-order Floquet Hamiltonian for this case. 
We first note that the Hamiltonian can be recast as
\bea H(t)&=&(\mu(t)+V/2)~ I ~-~ (\mu-V/2) ~\sigma^{z} ~+~ J~ \sigma^{x},\non\\
H_{0}&=&- ~(\mu-V/2)~ \sigma^{z},\non\\
H_{1}&=&J ~\sigma^{x}, \eea
where $I$, $\sigma^x$ and $\sigma^z$ denote the identity and two of
the Pauli matrices,
and $H_{0}$ and $H_{1}$ are the unperturbed Hamiltonian and perturbation, respectively. Assuming $\mu=\om$ and $V \gg J$, the instantaneous
eigenvalues of $H_{0}$ are given by $E_k^{\pm}=\pm(\mu(t)+V/2)$. The eigenfunctions corresponding to $E_{k}^{\pm}$ are given by $\ket{+}=\left(\begin{array}{cc}
0 \\ 1\end{array}\right) ~~~~{\rm and}~~~~
\ket{-}=\left(\begin{array}{cc}
1 \\ 0 \end{array}\right).$ 
These two eigenvalues satisfy the condition given in Eq.~\eqref{per26}, and 
we therefore use degenerate FPT. This gives
\bea \bra{+}H_{F}^{(1)}\ket{+}&=& 0,~~~~~
\bra{-}H_{F}^{(1)}\ket{-}= 0,\non\\
\bra{+}H_{F}^{(1)}\ket{-}&=&J ~I(\mu,V,T),~~~~
\bra{-}H_{F}^{(1)}\ket{+}=J ~I^{*}(\mu,V,T), \non\\
I(\mu,V,\om)&=&\frac{e^{-i(2\mu-V)T/4}\sin ((2\mu-V)T/4)}{(2\mu-V)T/2}~+ ~\frac{e^{-i(2\mu-3V)T/4}\sin ((2\mu+V)T/4)}{(2\mu+V)T/2}.\non\\ \eea
Putting $\mu=V=\omega$, we find that $I(\mu,V,\om)=- 4i/(3\pi)$. Thus, the effective Hamiltonian for this particular process is 
\bea H_{F}^{(1)}&=&-~\frac{4i}{3\pi} ~n_{0}c_{2}^{\dagger}c_{1}(1-n_{3}) ~+~ 
{\rm H.c.},\eea
where we have set $J=1$. Following similar procedures, we can compute the effective
Hamiltonians for all the other processes. These are given below.

\vspace*{.4cm}
\begin{table}[h!]
\begin{center}
\begin{tabular}{|c|c|c|c|} 
\hline
Pattern of periodic potential & Process & First-order Floquet Hamiltonian\\
\hline
+~-~+~-&1100 $\leftrightarrow$ 1010 & $
H_{F}^{(1)}=-\frac{4i}{3\pi}n_{0}c_{2}^{\dagger}c_{1}(1-n_{3})+$ H.c.\\
\hline
+~-~+~-&0100 $\leftrightarrow$ 0010 & $
H_{F}^{(1)}=0$\\
\hline
+~-~+~-&0101 $\leftrightarrow$ 0011 & $
H_{F}^{(1)}=-\frac{4i}{3\pi}(1-n_{0})c_{2}^{\dagger}c_{1}n_{3}+$ H.c.\\
\hline
+~-~+~-&1101 $\leftrightarrow$ 1011 & $
H_{F}^{(1)}=0$\\
\hline
-~+~-~+&1100 $\leftrightarrow$ 1010 & $
H_{F}^{(1)}=\frac{4i}{3\pi}n_{0}c_{2}^{\dagger}c_{1}(1-n_{3})+$ H.c.\\
\hline
-~+~-~+&0100 $\leftrightarrow$ 0010 & $
H_{F}^{(1)}=0$\\
\hline
-~+~-~+&0101 $\leftrightarrow$ 0011 & $
H_{F}^{(1)}=\frac{4i}{3\pi} (1-n_{0}) c_{2}^{\dagger} c_{1}n_{3}+$ H.c.\\
\hline
-~+~-~+&1101 $\leftrightarrow$ 1011 & $
H_{F}^{(1)}=0$\\
\hline
\end{tabular}
\end{center}
\caption{\label{FPTreso}First-order effective FPT Hamiltonians for the allowed processes with all possible patterns of periodic potential in the case of dynamical localization 
and resonance for a period-2 model.}
\end{table}
\vspace*{.4cm}

Taking all these processes into account, the complete effective Hamiltonian for the case where a resonance and DL occur simultaneously is given by
\bea H&=&- ~\frac{4i}{3\pi} ~\sum_{j=1}^{L/2}~ \left[(1-n_{2j})c_{2j+2}^{\dagger}c_{2j+1}n_{2j+3}+n_{2j}c_{2j+2}^{\dagger}c_{2j+1}(1-n_{2j+3})~+~ \rm{H.c.} \right] \non \\
&& +~ \frac{4i}{3\pi} ~\sum_{j=1}^{L/2}\left[n_{2j+1}c_{2j+3}^{\dagger}c_{2j+2}(1-n_{2j+4})+(1-n_{2j+1})c_{2j+3}^{\dagger}c_{2j+2}n_{2j+4} ~+~ {\rm H.c.} \right]. \eea\label{p2reso}
We can perform the unitary transformation $c_{2j}\rightarrow c_{2j}$ and $c_{2j+1}\rightarrow i c_{2j+1}$ to obtain a simpler form of the effective Hamiltonian
\bea H&=&\frac{4}{3\pi} ~\sum_{j=1}^L ~(n_j - n_{j+3})^2 ~\left(c_{j+2}^{\dagger} c_{j+1} ~+~ {\rm H.c.} \right). \label{effp2}\eea
This form implies that hoppings between two nearest-neighbor sites
are forbidden whenever their neighboring sites are both completely empty or 
completely occupied. Hence, these forbidden processes act as kinetic constraints 
in the dynamics~\cite{tomasi,yang,HSF1,HSF2,HSF3,prefrag}, and these constraints 
can, in principle, lead the system towards an anomalous thermalization behavior.
We also find that the Hamiltonian in Eq.~\eqref{effp2} has several zero-energy 
states which consist of single states in the number basis. The number of such states can be found using a 
transfer matrix method as shown in Appendix A. We discover that the
number grows exponentially with system size as $1.466^L$.

This mechanism can be further generalized to a lower
frequency regime by considering the other DL points and resonances given by $\mu=V=
n\om~(n\neq1)$, keeping $\mu,~V>>J$. To obtain an analytical insight in this limit, 
we can derive the first-order FPT Hamiltonian following a similar procedure as 
charted out before. Interestingly, the first-order FPT Hamiltonian 
for $\mu=V=n\om$, where $\mu,~V \gg J$, and $n$ is odd, reads as
\bea H^{(1)}_{F}=\frac{4}{3\pi n}~\sum_{j=1}^L ~(n_j - n_{j+3})^2 ~\left(c_{j+2}^{\dagger} c_{j+1} ~+~ {\rm H.c.} \right), \label{effoddp2}\eea
which is the same Hamiltonian as obtained for the case $\mu=V=\om$, but 
with a hopping strength $4/(3\pi n)$.
On the other hand, the first-order effective Hamiltonian for $\mu=V=n\om$, where 
$\mu,~V \gg J$ and $n$ is {\it even}, turns out to be 
\bea H_{F}^{(1)}=0, \label{effeven2} \eea
which implies that there should be a many-body flat band lying at $E_{F}=0$.

\begin{figure}[!tbp]
\footnotesize
\begin{center}
\stackunder[5pt]{\includegraphics[width=0.5\hsize]{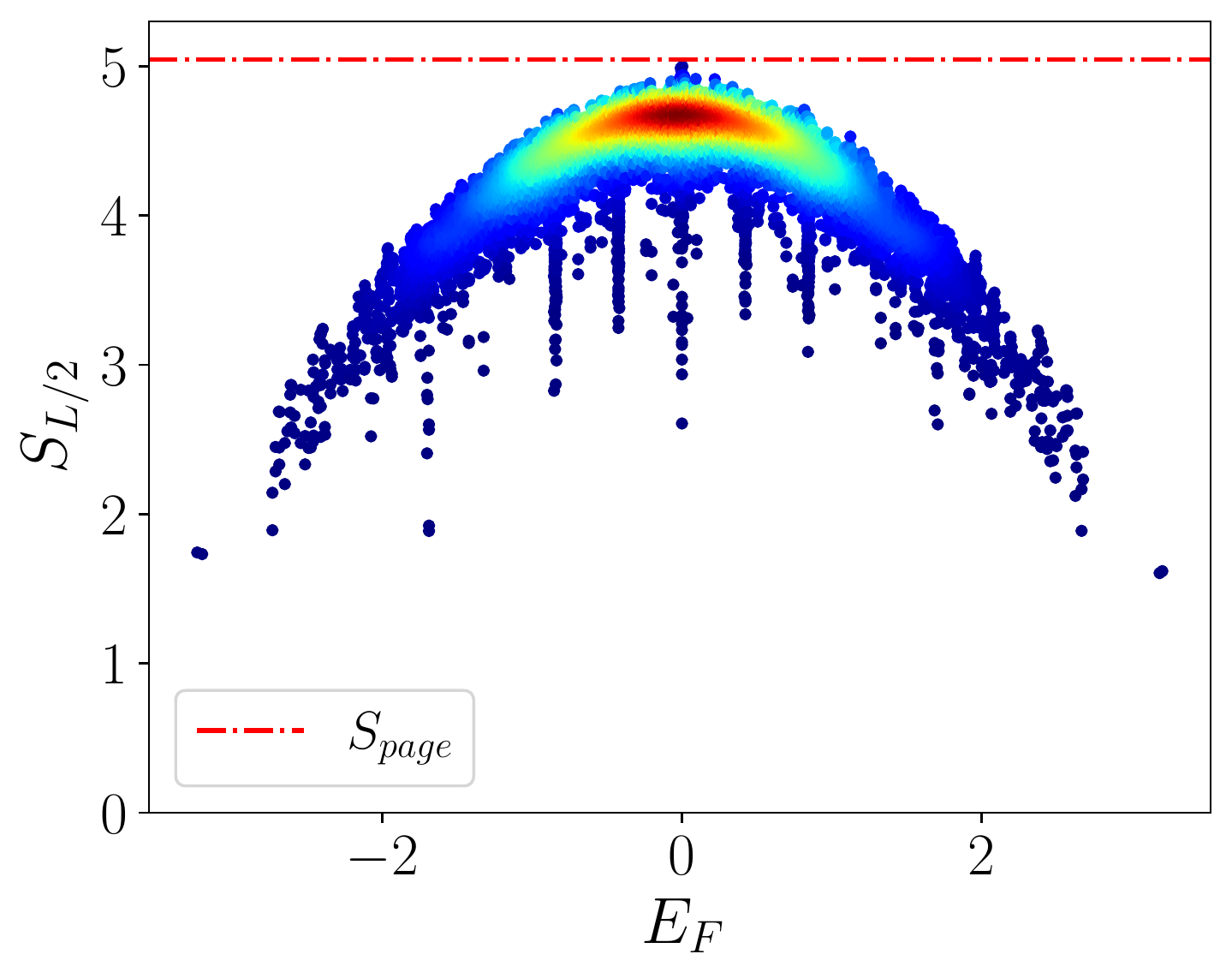}}{\large(a)}
\stackunder[5pt]{\includegraphics[width=0.5\hsize]{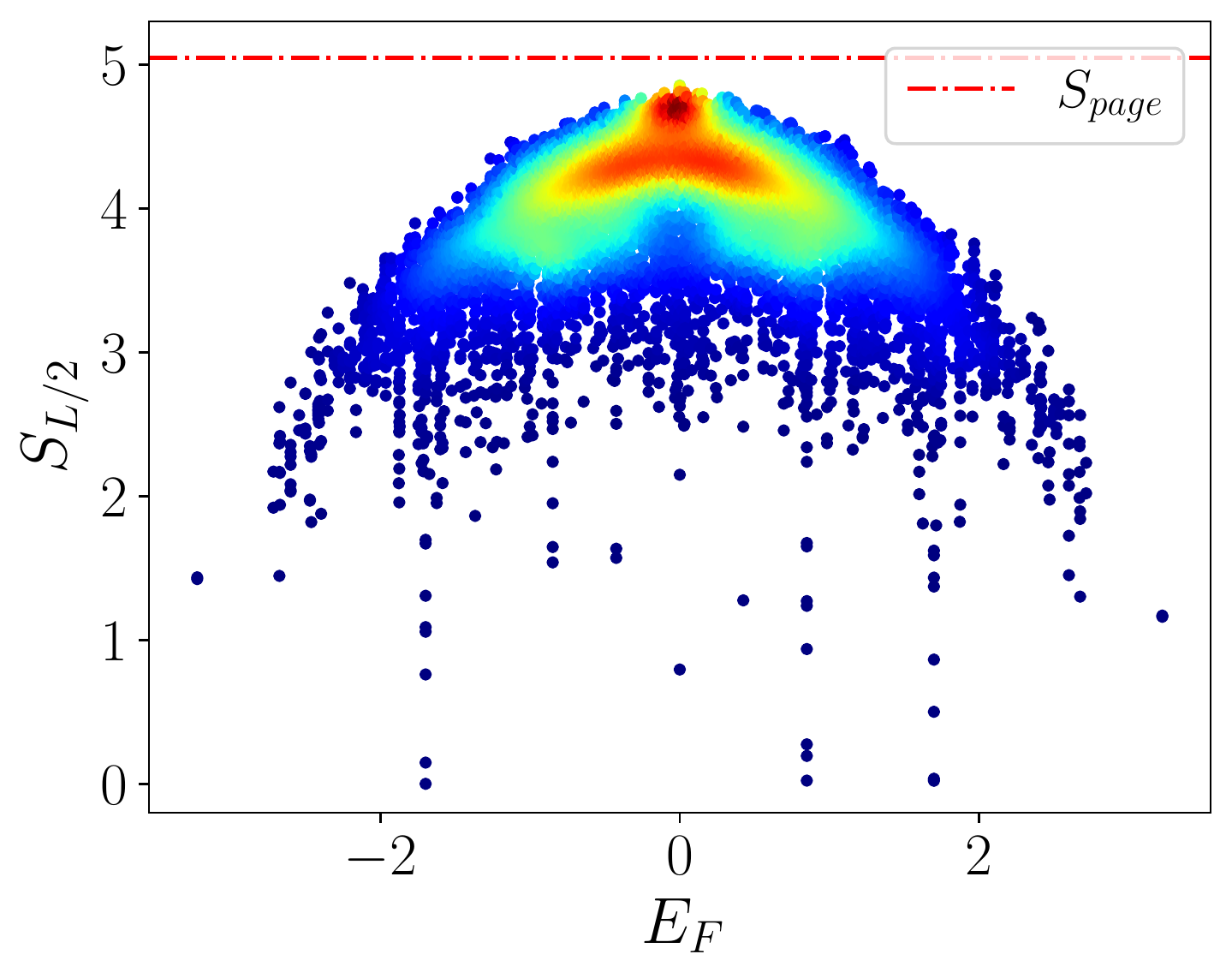}}{\large(b)}
\stackunder[5pt]{\includegraphics[width=0.65\hsize]{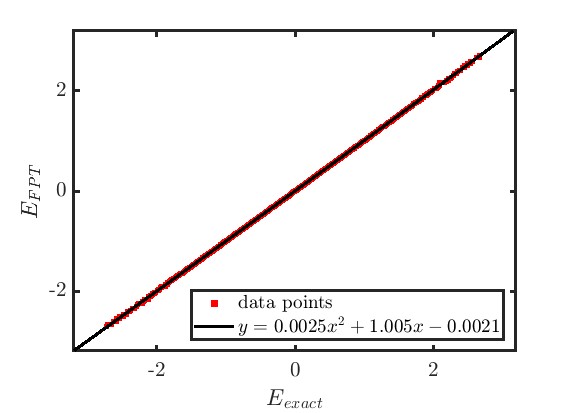}}{\large(c)}
\end{center}
\caption{{\bf Entanglement entropy spectrum of the period-2 model for the case of DL 
and resonance:} Plots showing the entanglement entropy $S_{L/2}$ as a function of 
the Floquet quasienergy $E_{F}$ obtained from (a) the exact numerical calculation and (b) the first-order effective FPT Hamiltonian shown in 
Eq. \eqref{effp2}, for $J=1$, $\mu=\om=V=20$, 
and $L=16$. (c) Plot showing $E_{exact}$ versus $E_{FPT}$ obtained numerically for the same parameter values as in plots (a) and (b). In 
(a) and (b), we see many low-entanglement states near the middle of the spectrum. In 
both plots, the color intensity indicates the density the states, implying 
that the majority of Floquet eigenstates show almost thermal entanglement. 
Plot (c) shows that the quasienergies obtained from the 
first-order FPT agree quite well with the exact numerically computed values.
However, as plots (a) and (b) show, there are a large number 
of Floquet eigenstates for which the entanglement estimated from FPT is much 
smaller than the exact numerically obtained values.} \label{fig161} \end{figure}

\begin{figure}[!tbp]
\begin{center}
\includegraphics[width=0.7\hsize]{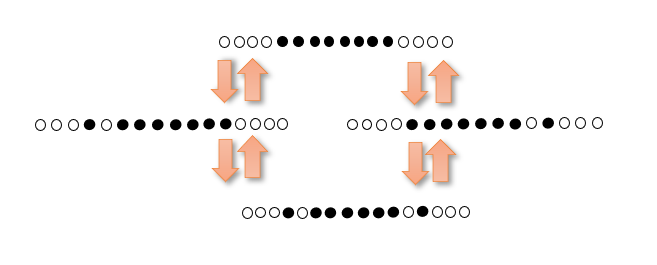}
\end{center}
\caption{{\bf Schematic of a Hilbert space fragment for the period-2 case:} Figure showing a particular Hilbert space sector consisting of four states, 
$\ket{0000111111110000}$, $\ket{0001011111110000}$, $\ket{0000111111101000}$, and 
$\ket{0001011111101000}$. The black and white dots indicate occupied and empty 
sites respectively.} \label{fig30} \end{figure}

In Figs.~\ref{fig161} (a) and (b), we show the variation of the
half-chain entanglement entropy $S_{L/2}$ as a function of $E_{F}$, obtained from exact numerical 
calculations and from the first-order FPT Hamiltonian shown in Eq. \eqref{effp2} for $J=1$, $\mu=\om=20$, and $V=20$. Both these cases point towards many low-entanglement states near the middle of the 
spectrum; these arise due to the kinetic constraints simultaneously imposed by DL and the resonance condition. Before proceeding further, we note that the effective 
Hamiltonian described in Eq.~\eqref{effp2} supports many fragmented Hilbert space sectors~\cite{tomasi,yang,HSF1,HSF2,HSF3,prefrag}, which can be shown as follows. 
First, there are an exponentially large number of fragments each of which 
consists of a single state with zero energy; this is shown in Appendix A.
Next, there are simple fragments consisting of only four states. Consider, for 
example, the fragment containing the states $\ket{0000111111110000}$,
$\ket{0001011111110000}, ~\ket{0000111111101000}$, and $\ket{0001011111101000}$, 
and their translated
partners (we have written all the states in the occupation number basis). The action 
of the effective Hamiltonian on these four states is schematically shown in 
Fig.~\ref{fig30}. Taking into account the action of the 
effective Hamiltonian on these four states, we find an effective $4\times4$ Hamiltonian which represents this particular fragment,
\bea H_{frag}~=~ \frac{4}{3\pi} ~\begin{pmatrix}
0 & 1 & 1 & 0\\ 1&0 & 0 & 1\\1&0 & 0 & 1\\0 & 1 & 1 & 0\end{pmatrix}. \eea
The eigenvalues of this Hamiltonian are given by $E_{1,2}=\pm 0.85$, $E_{3}=0$, and $E_{4}=0$. Hence these four eigenvalues offer two distinct difference in energies, i.e., 
$\Delta E=0.85$ and $\Delta E=1.7$, which will be important later in the discussion of dynamics. 

\begin{figure}[!tbp]
\footnotesize
\begin{center}
\stackunder[5pt]{\includegraphics[width=0.5\hsize]{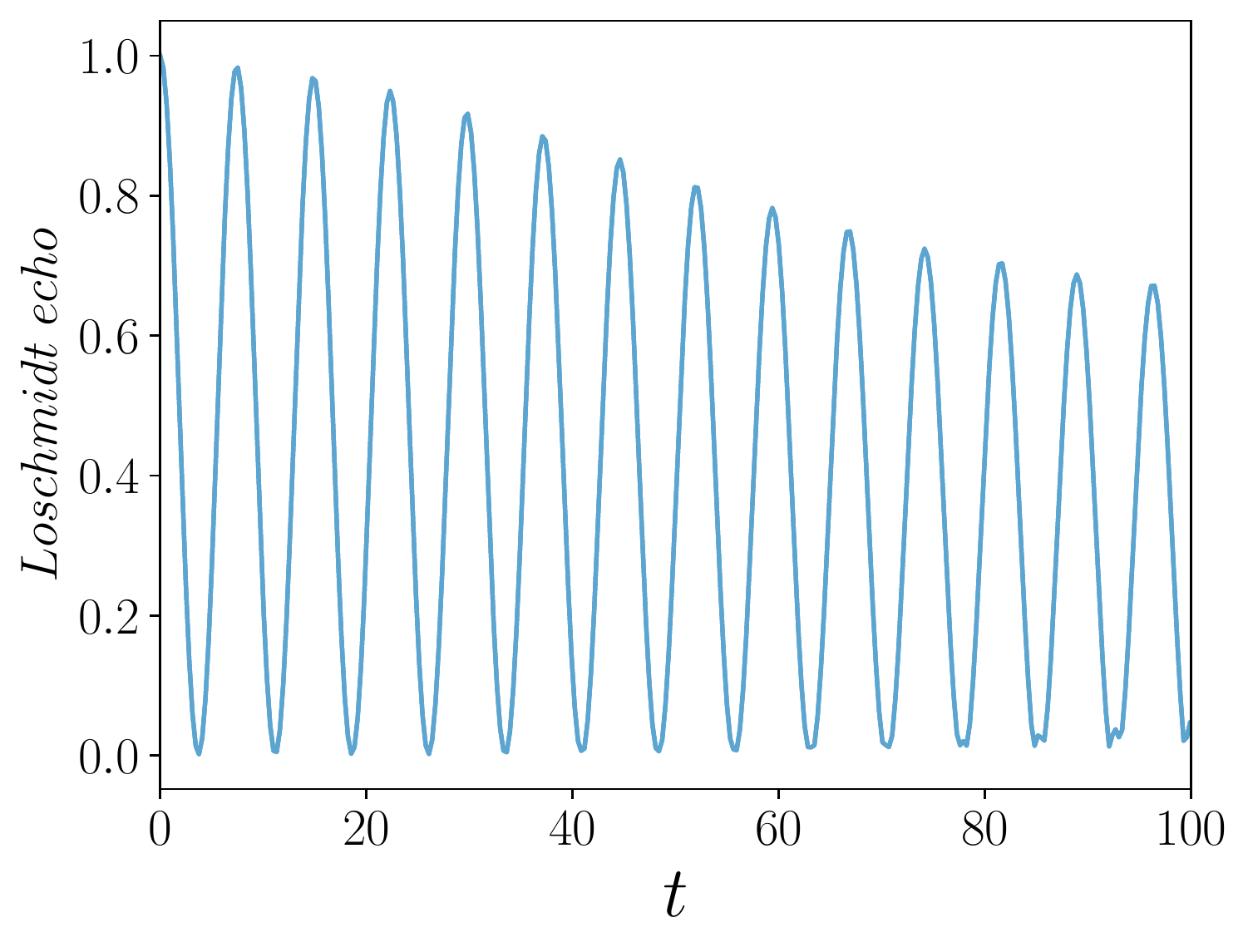}}{\large(a)}
\stackunder[5pt]{\includegraphics[width=0.5\hsize]{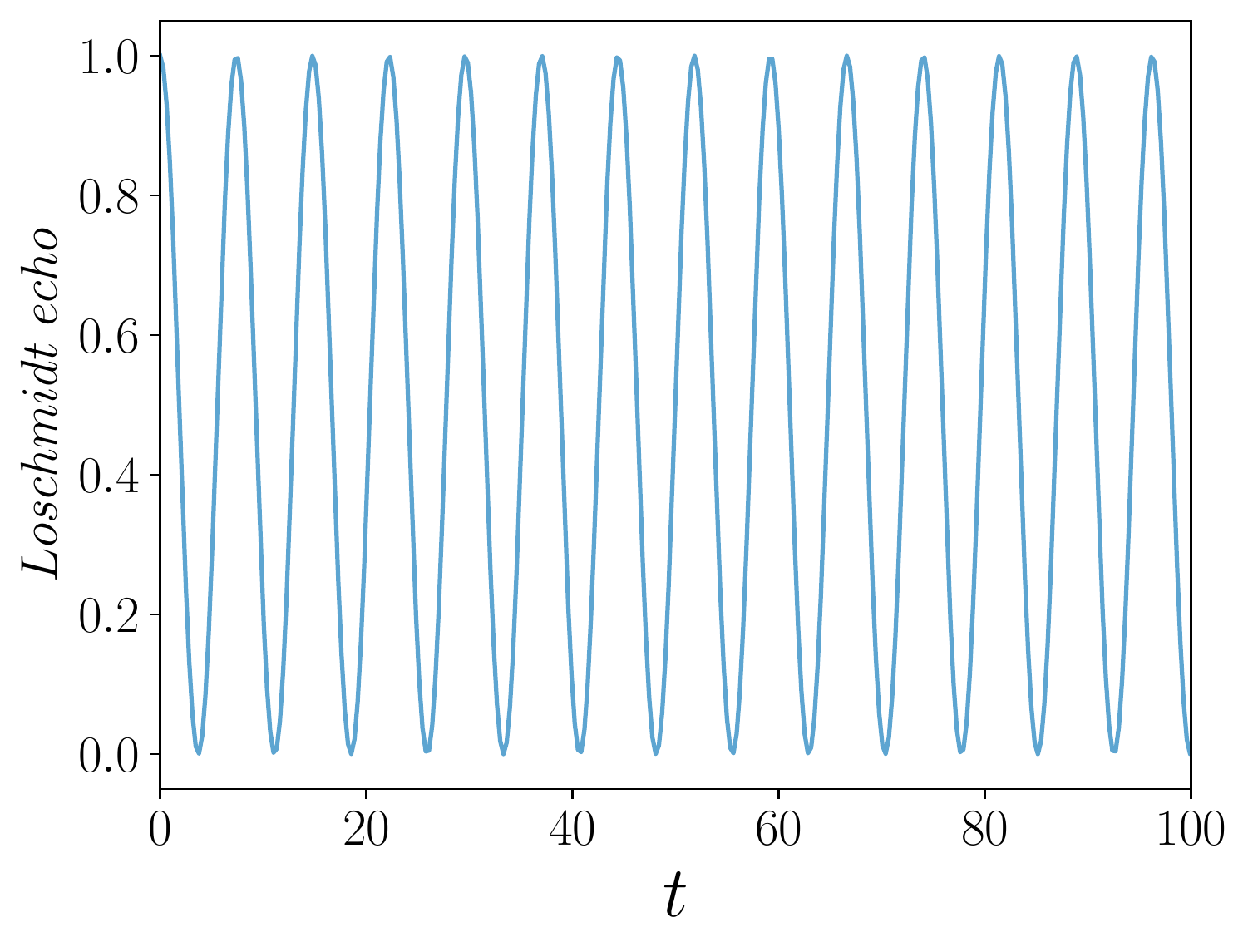}}{\large(b)}\\
\stackunder[5pt]{\includegraphics[width=0.55\hsize]{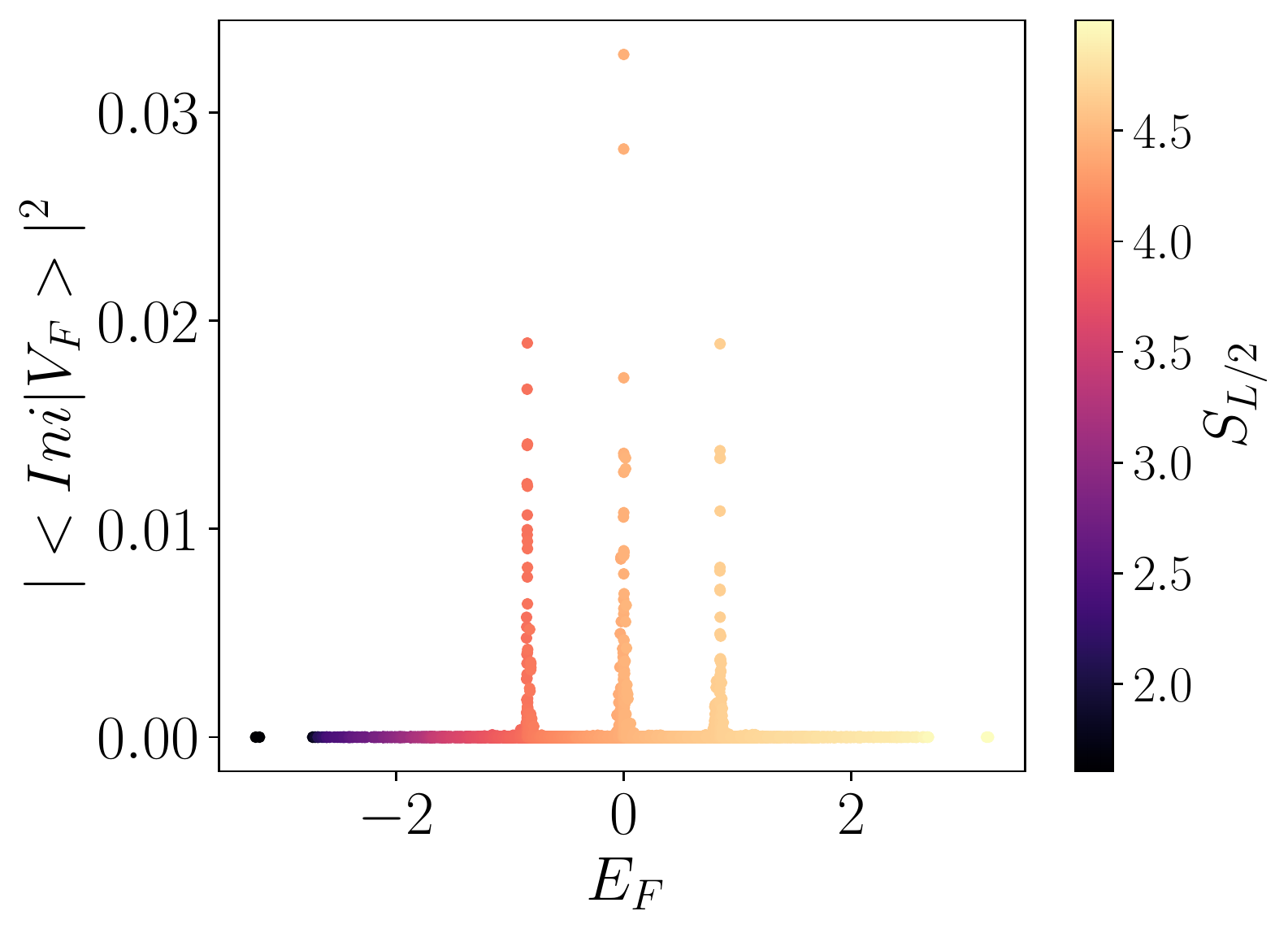}}{\large(c)}\\
\end{center}
\caption{{\bf Dynamics of the Loschmidt echo and overlaps with Floquet eigenstates in a resonant case:} (a-b): Plots of the Loschmidt echo versus time as obtained
from (a) the exact numerical calculation and (b) the first-order effective FPT Hamiltonian, for the same parameter values as in 
Fig.~\ref{fig161}, taking the initial state ($\ket{Ini}$) to be
$\ket{0000111111110000}$. Both plots show show long-time oscillations in time 
showing an anomalous thermalization behavior. (c): Overlaps of the same initial state with the Floquet eigenstates as a function of $E_{F}$ computed from the 
exact numerical calculation for the same parameter values, with a color bar indicating the variation of $S_{L/2}$. The quasienergies of the Floquet eigenstates having the 
highest overlaps with the initial state agree with the analytically predicted values.}
\label{fig162} \end{figure}

\begin{figure}[!tbp]
\footnotesize
\begin{center}
\stackunder[5pt]{\includegraphics[width=0.5\hsize]{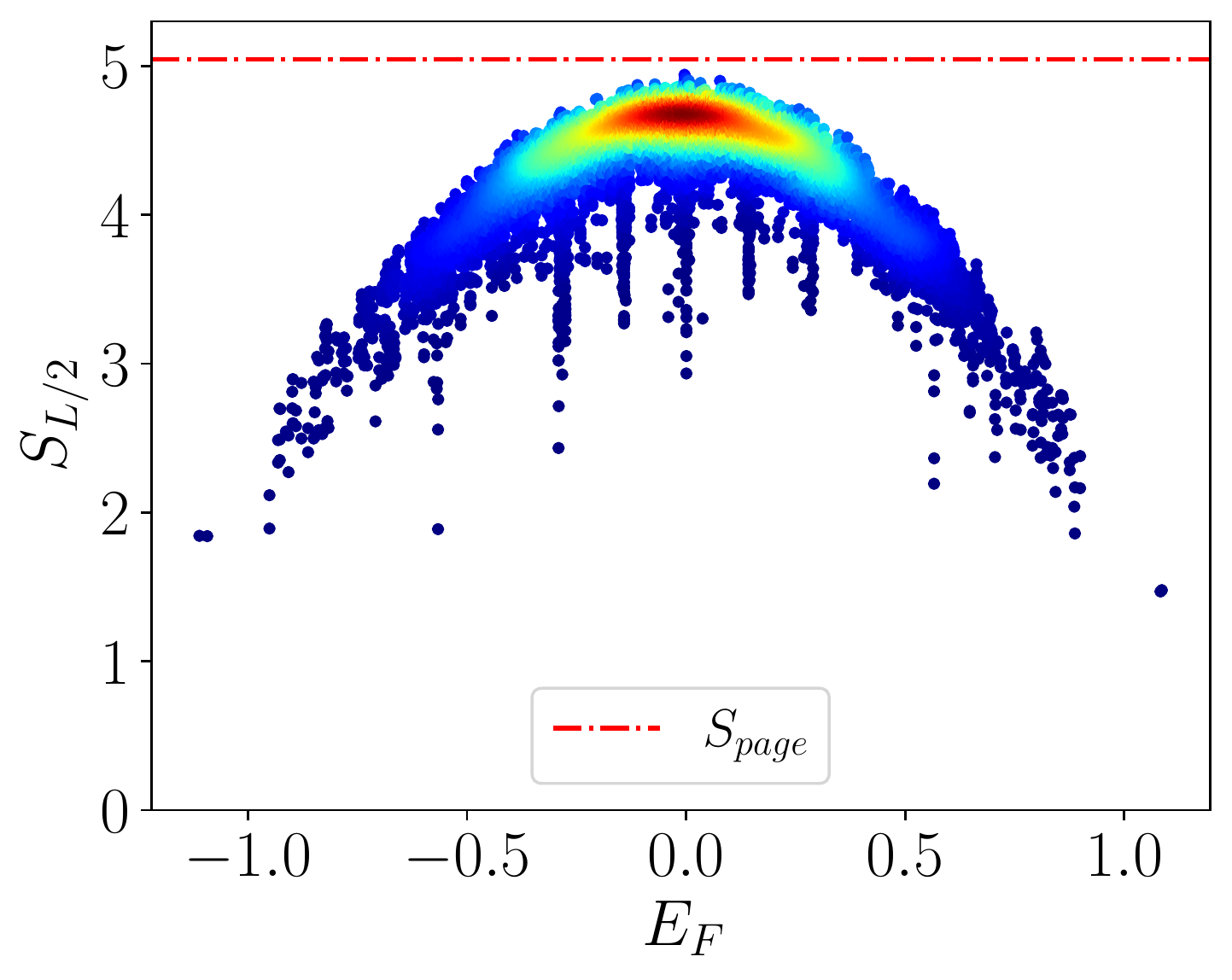}}{\large(a)}
\stackunder[5pt]{\includegraphics[width=0.5\hsize]{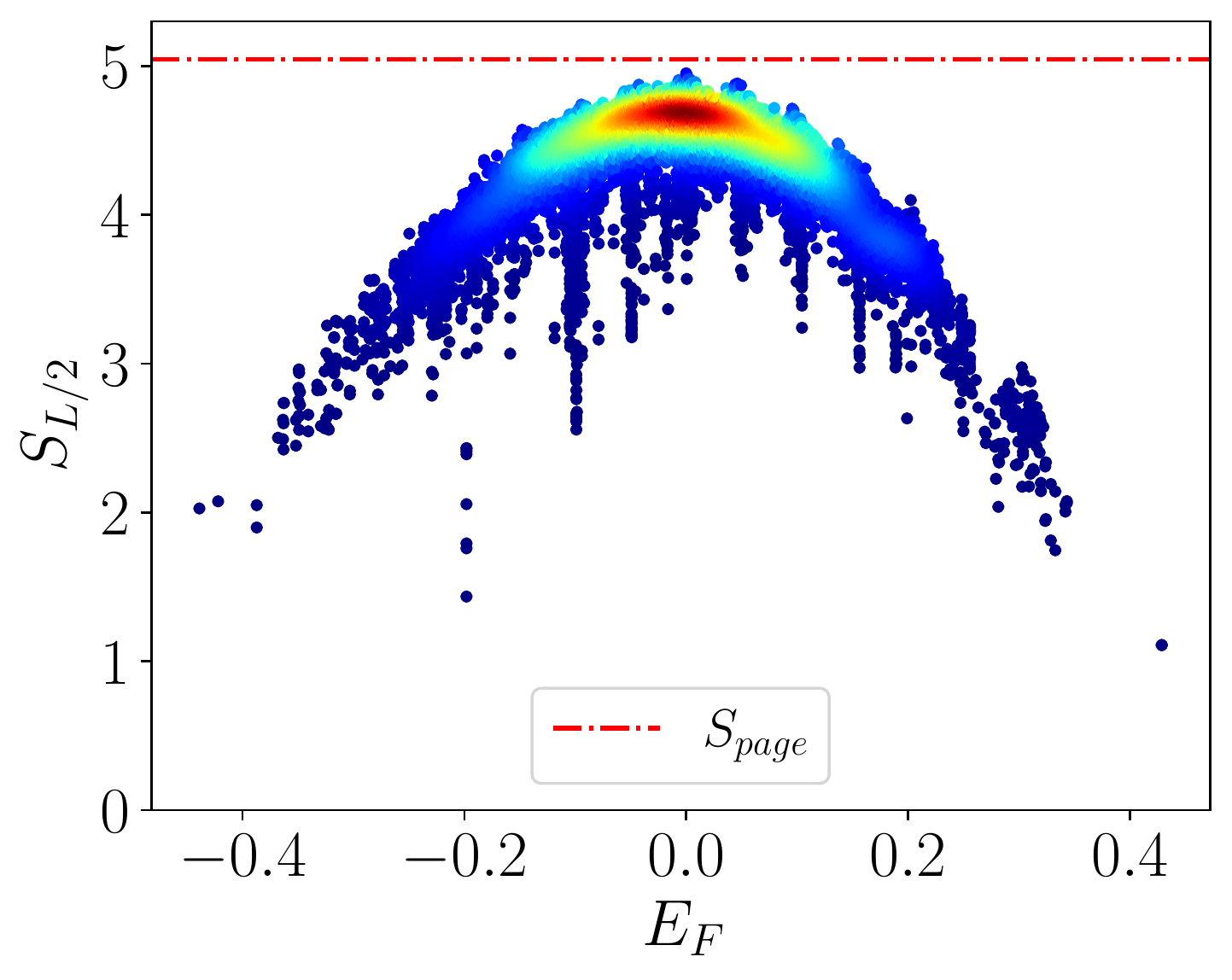}}{\large(b)}
\end{center}
\caption{{\bf Entanglement entropy spectrum of the period-2 model for the case of DL 
and resonances in the intermediate and low driving frequency regime:} (a-b) Plots showing the entanglement entropy $S_{L/2}$ as a function of 
the Floquet quasienergy $E_{F}$ obtained from the exact numerical calculation for $J=1$, $\mu=V=20$, $L=16$, and $\om\approx6.67$ (third DL point) and $\om \approx 2.22$ 
(ninth DL point), respectively. In both cases, we see many low-entanglement states in the middle of the spectrum, which implies that the system breaks ergodicity.} \label{fig1111} \end{figure}

\begin{figure}[!tbp]
\begin{center}
\includegraphics[width=0.85\hsize]{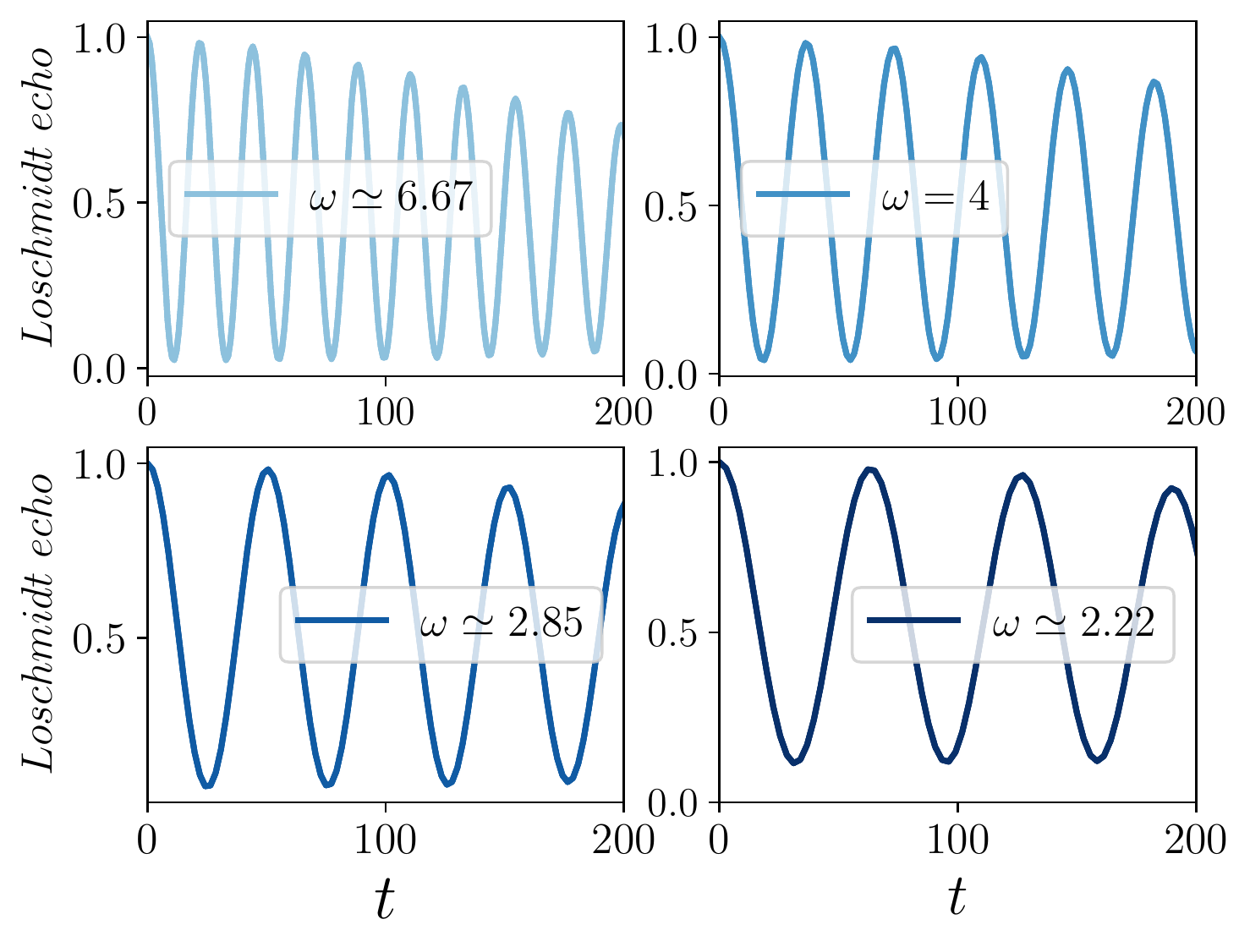}
\end{center}
\caption{{\bf Dynamics of the Loschmidt echo at other dynamical localization points and at resonance:} Plots of the Loschmidt echo versus time 
obtained by an exact numerical calculation for $J=1$, $\mu=V=20$, $L=16$, and $\om\approx6.67,~4,~2.85$, and $2.22$, which correspond to the third, fifth, seventh 
and ninth DL points, respectively. For all four cases, we consider the initial state to be $\ket{0000111111110000}$, which resides in a HSF cluster consisting of four states 
as discussed earlier. The Loschmidt echo for all four cases demonstrates 
long-time revival behavior, indicating that the thermalization is very slow 
at these parameter values.}\label{fig1112} \end{figure}

In Figs.~\ref{fig162} (a) and (b), the variation of the Loschmidt echos with time,
$t=nT$, is shown as found from the exact numerical calculation and the first-order FPT, respectively, for the parameter values, $J=1$, $\mu=\om=V=20$, and $L=16$, taking the 
initial state to be $\ket{0000111111110000}$. We can show analytically that the 
Loschmidt echo for this particular choice of initial state takes the form 
$|a+b\cos(\Delta E t)|$, which implies that it oscillates with a period
$\Delta t=2\pi / \Delta E$. Putting $\Delta E=0.85$, the period of oscillation in the revival pattern turns out to be $\Delta t \simeq 7.4$, which almost 
perfectly captures the numerically obtained value. In
Fig.~\ref{fig162} (c), we show the overlaps of the same initial state with the Floquet eigenstates (obtained from the exact numerical calculation) as a function of 
$E_{F}$, where the color bar indicates the variation of $S_{L/2}$ of the Floquet eigenstates. Interestingly, we observe that the overlap is highest for the Floquet 
eigenstates with $E ~\simeq ~\pm 0.85$ and 0, which almost identically agrees with the analytically predicted values. Similarly in Figs. 
\ref{fig1111} (a-b), we show the entanglement entropy as a function of the quasienergy for $\om \approx 6.67$ (third DL point, i.e., $n = \mu/\om = 3$) and $\om \approx 2.22$, (ninth DL 
point, i.e., $n=9$) with the rest of the parameters being the same as 
in Fig. \ref{fig161}. In both cases, we see many low-entanglement states with the range of quasienergies 
approximately being $1/3$ and $1/9$ times those of the first case shown in Fig. \ref{fig161}; this agrees with the analytically derived first-order FPT 
Hamiltonian shown in Eq. \eqref{effoddp2}. In Fig. \ref{fig1112}, we again examine the dynamics of the Loschmidt echo for $J=1$, $\mu=V=20$, $L=16$, and 
$\om \approx 6.67,~4,~2.85$, and $2.22$, which correspond to the third, 
fifth, seventh, and ninth DL points. [Note that the last value of $\om$ is not
much larger than the hopping amplitude $J$, and therefore does not lie
in the high-frequency regime. Nevertheless we see that the Loschmidt echo decays
very slowly. This shows that DL and resonances lead to very slow thermalization 
even when the
driving frequency is not very large]. For all four cases, we consider the same 
initial state as before, and we see that the Loschmidt echo demonstrates long-time revivals, indicating that the system shows very slow thermalization. 
Furthermore, the period of oscillations in the revival pattern for all four cases can be explained by considering the first-order FPT Hamiltonian and the effective 
Hamiltonian for the HSF cluster consisting of the four states described in 
Fig.~\ref{fig30}. As shown in Eq. \eqref{effoddp2}, the DL points for $\mu=V=n\om$ 
($n=3,~5,~7,~\cdots$) renormalizes the effective hopping strength of the effective Hamiltonian obtained for the case with $\mu=V=\om$ by a factor of $1/n$. 
We can then argue that the period of oscillations in the revival pattern for 
these four cases will be $\Delta t_{n}=n~ 2\pi / \Delta E$, where 
$n=3,~5,~7,...$ and $\Delta E \approx 0.85$. This implies that the period of 
oscillations corresponding to the third, fifth, seventh and ninth DL points will be 
$\Delta t_{n} \approx 22.2,~37,~51.8$, and $66.6$, which almost perfectly agrees with the exact numerical calculation as seen in Fig. \ref{fig1112}.

\section{Thermodynamic stability of Hilbert space fragmentation in period-2 model}

\begin{figure}[!tbp]
\footnotesize
\begin{center}
\stackunder[5pt]{\includegraphics[width=0.45\hsize]{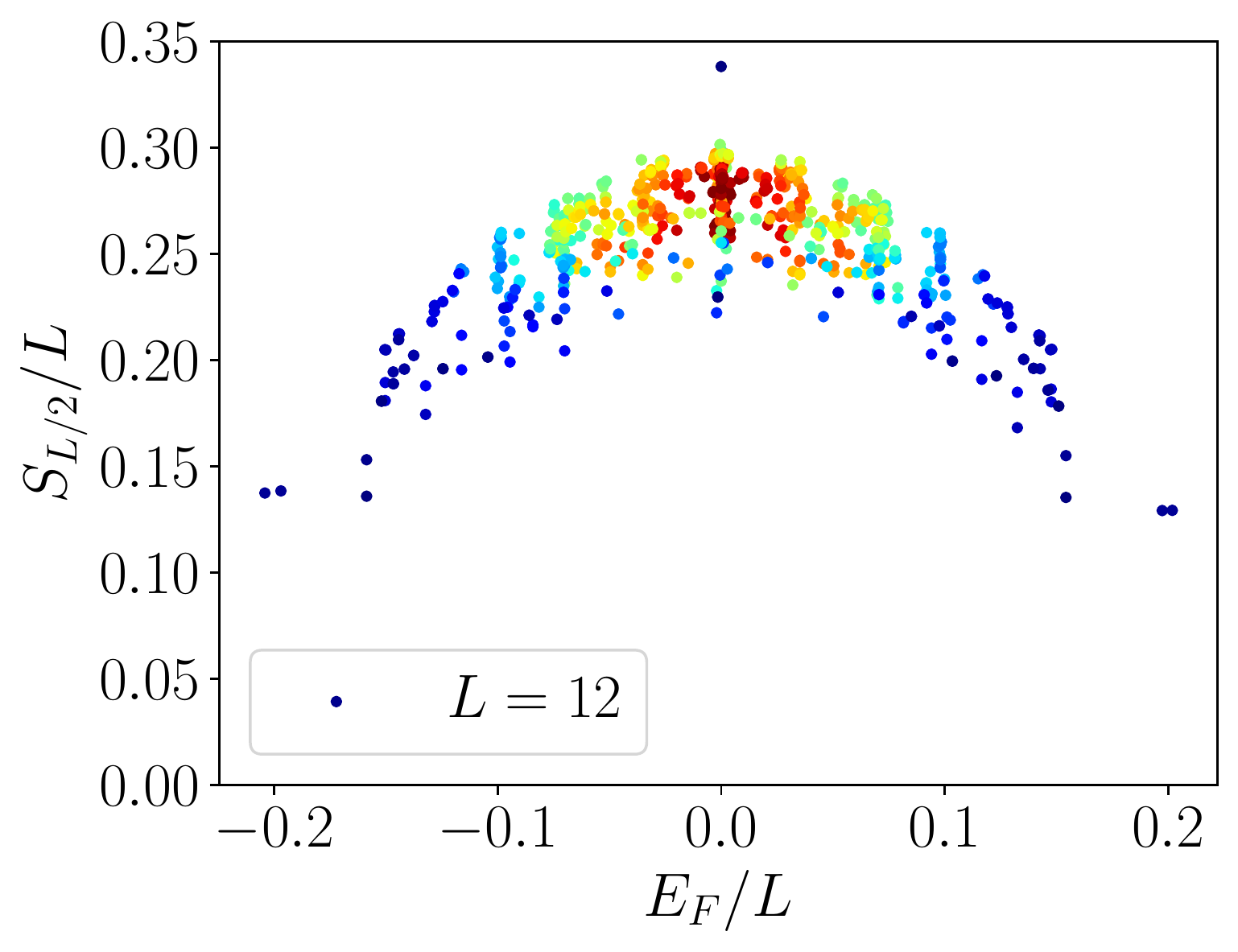}}{\large(a)}
\stackunder[5pt]{\includegraphics[width=0.45\hsize]{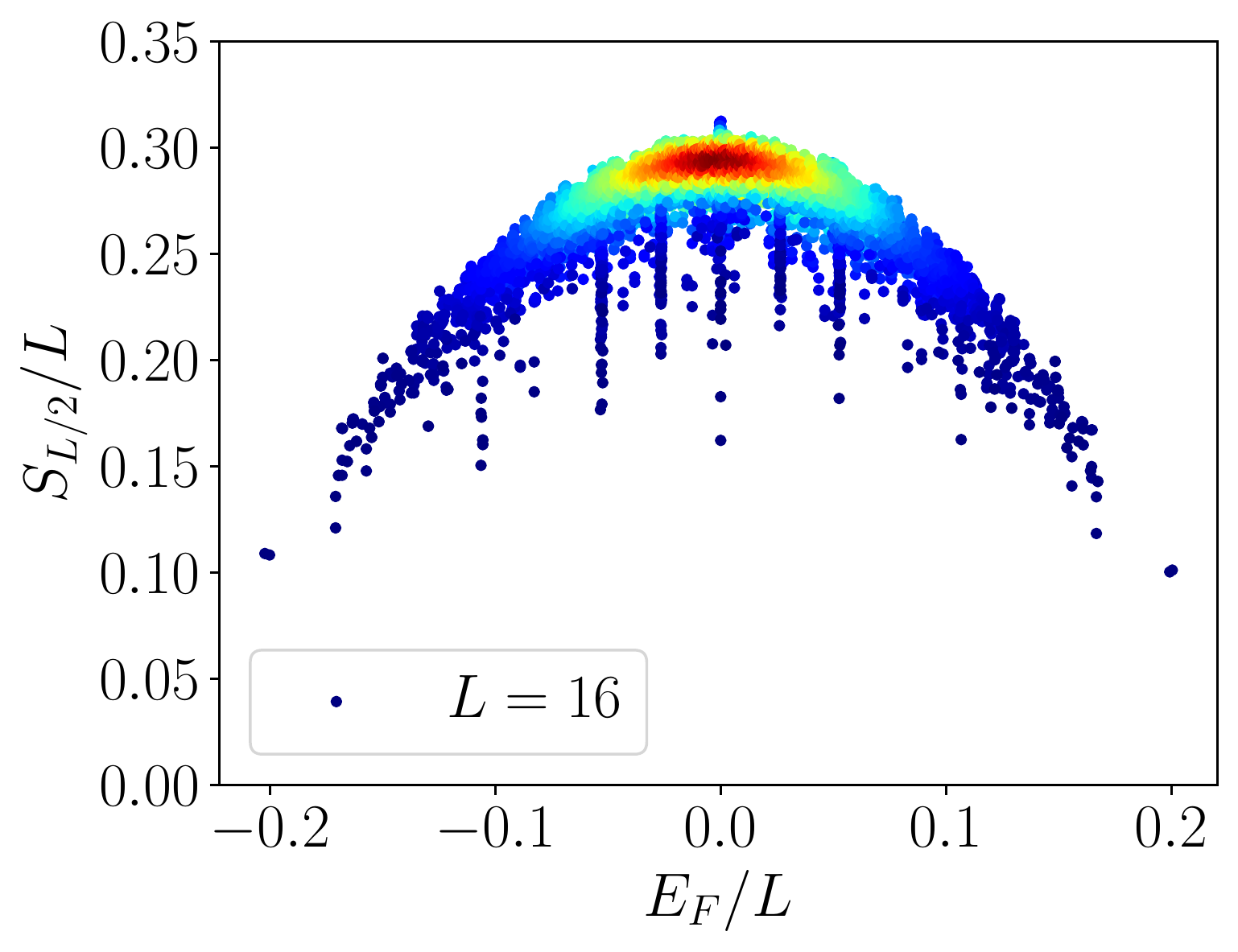}}{\large(b)}\\
\stackunder[5pt]{\includegraphics[width=0.45\hsize]{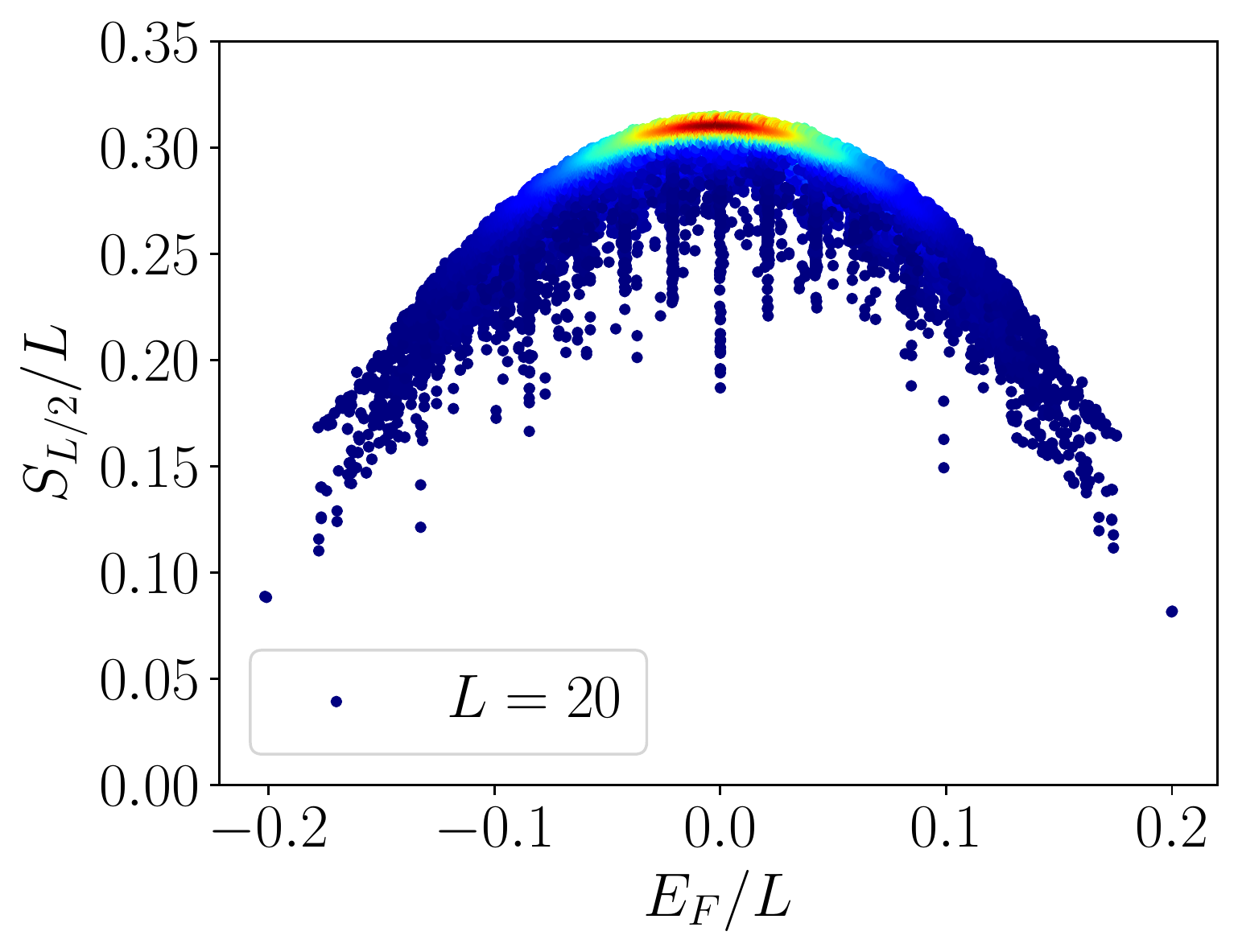}}{\large(c)}
\end{center}
\caption{{\bf Scaled entanglement spectrum as a function of scaled quasienergy 
at another DL point and at resonance for the period-2 model:} (a-c) 
Plots showing the scaled entanglement entropy $S_{L/2}$ obtained from exact 
numerical calculations as a function of the scaled quasienergy at a DL point with 
$J=1$, $\mu=\om=V=20$ for $L=12,~16$ and $20$, respectively. In all three cases, 
the range of the scaled quasienergy appears to be the same. However, the spectrum 
of the scaled entanglement entropy broadens with increasing system 
size, as can be seen in plots (a-c).} \label{fig22} \end{figure}

\begin{figure}[!tbp]
\footnotesize
\begin{center}
\stackunder[5pt]{\includegraphics[width=0.445\hsize]{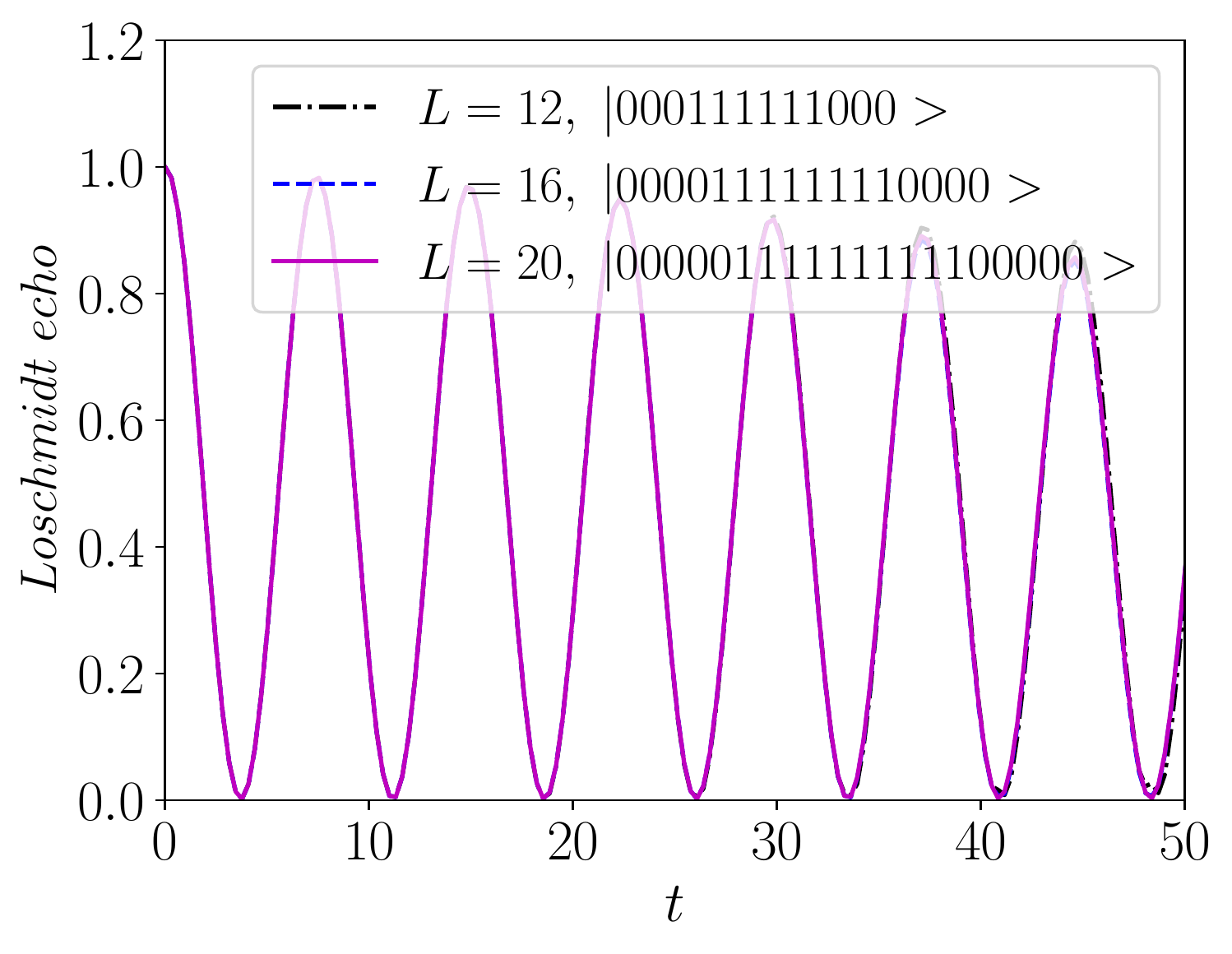}}{\large(a)}
\stackunder[5pt]{\includegraphics[width=0.445\hsize]{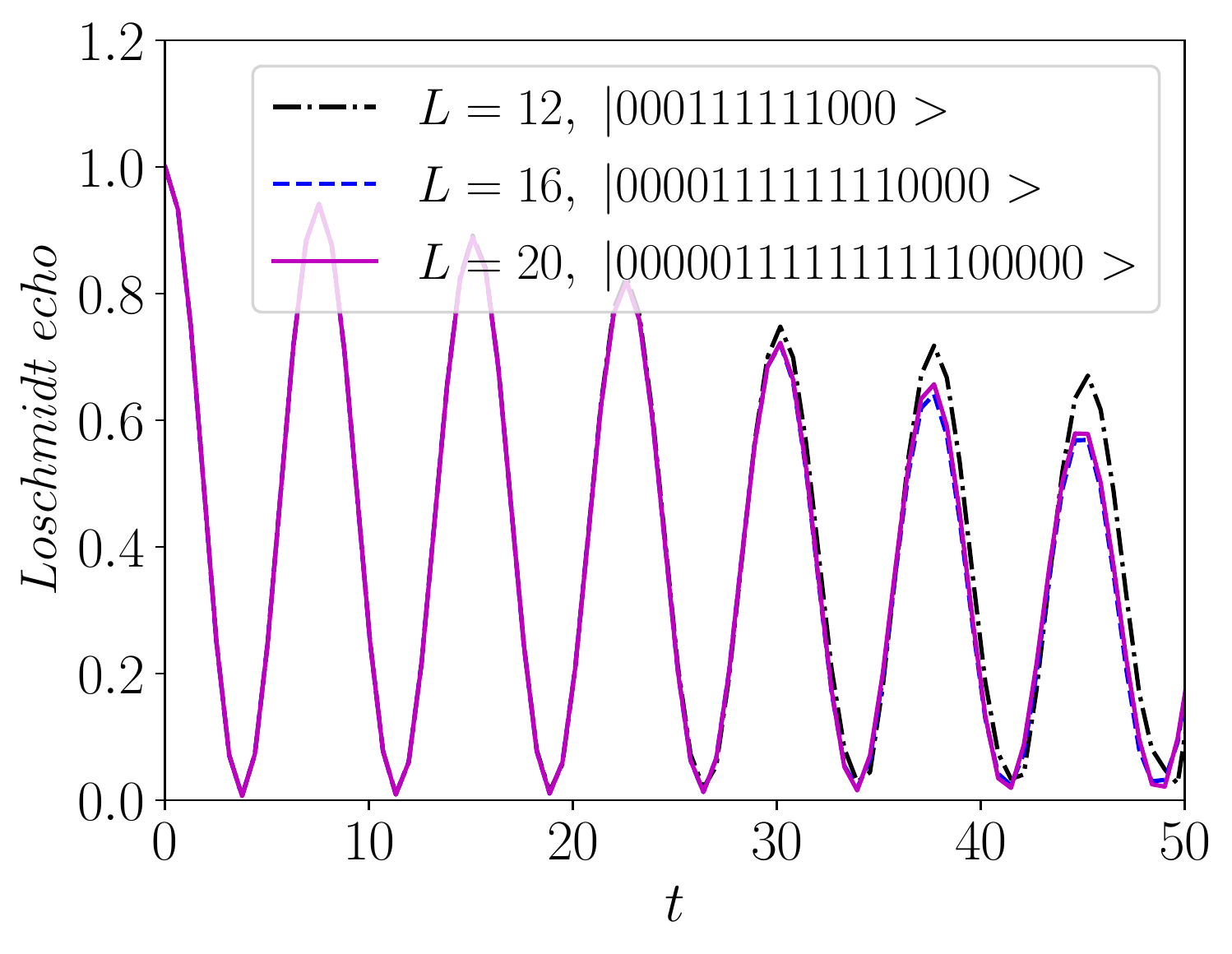}}{\large(b)}\\
\stackunder[5pt]{\includegraphics[width=0.47\hsize]{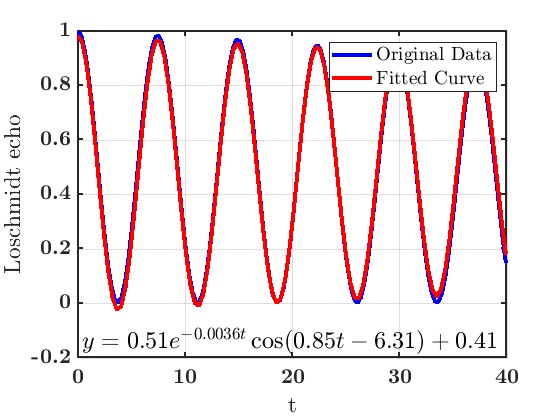}}{\large(c)}
\stackunder[5pt]{\includegraphics[width=0.47\hsize]{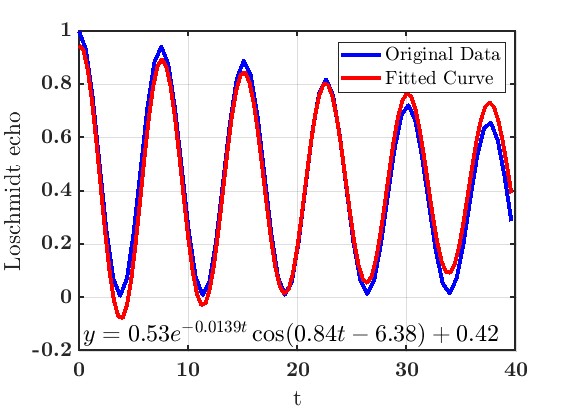}}{\large(d)}
\end{center}
\caption{{\bf Variation of Loschmidt echo with time for different system sizes 
at a DL point and at resonance:} (a-b) Plots showing the variation 
of Loschmidt echo with time for three different system sizes, $L=12,~16$ and $20$,
and $\mu=V=\om=20$ and $\mu=V=\om=10$, respectively. (c-d) The fitting of the 
envelop of the Loschmidt echo for $L=20$ for the same parameter values as in plots
(a) and (b). In plots (a-b), we consider three different initial states for the 
three system sizes, all of them being kinetically constrained to lie within a single 
Hilbert space cluster due to HSF. Consequently, these states exhibit long-time 
persistent oscillations. We see that the Loschmidt echo in (a) falls off very slowly
compared to (b).
Plots (c-d) show the functional forms of the 
envelops of the Loschmidt echo as extracted from a fitting analysis. 
In both cases, the period of oscillation of the Loschmidt echo is almost the same, 
with $\Delta t \simeq 2\pi$. However, the decay rate significantly increases as 
$\mu,~V$ and $\om$ decreases, as is clear from the fitting form of the envelop.}
\label{fig23} \end{figure}

\begin{figure}[!tbp]
\footnotesize
\begin{center}
\stackunder[5pt]{\includegraphics[width=0.45\hsize]{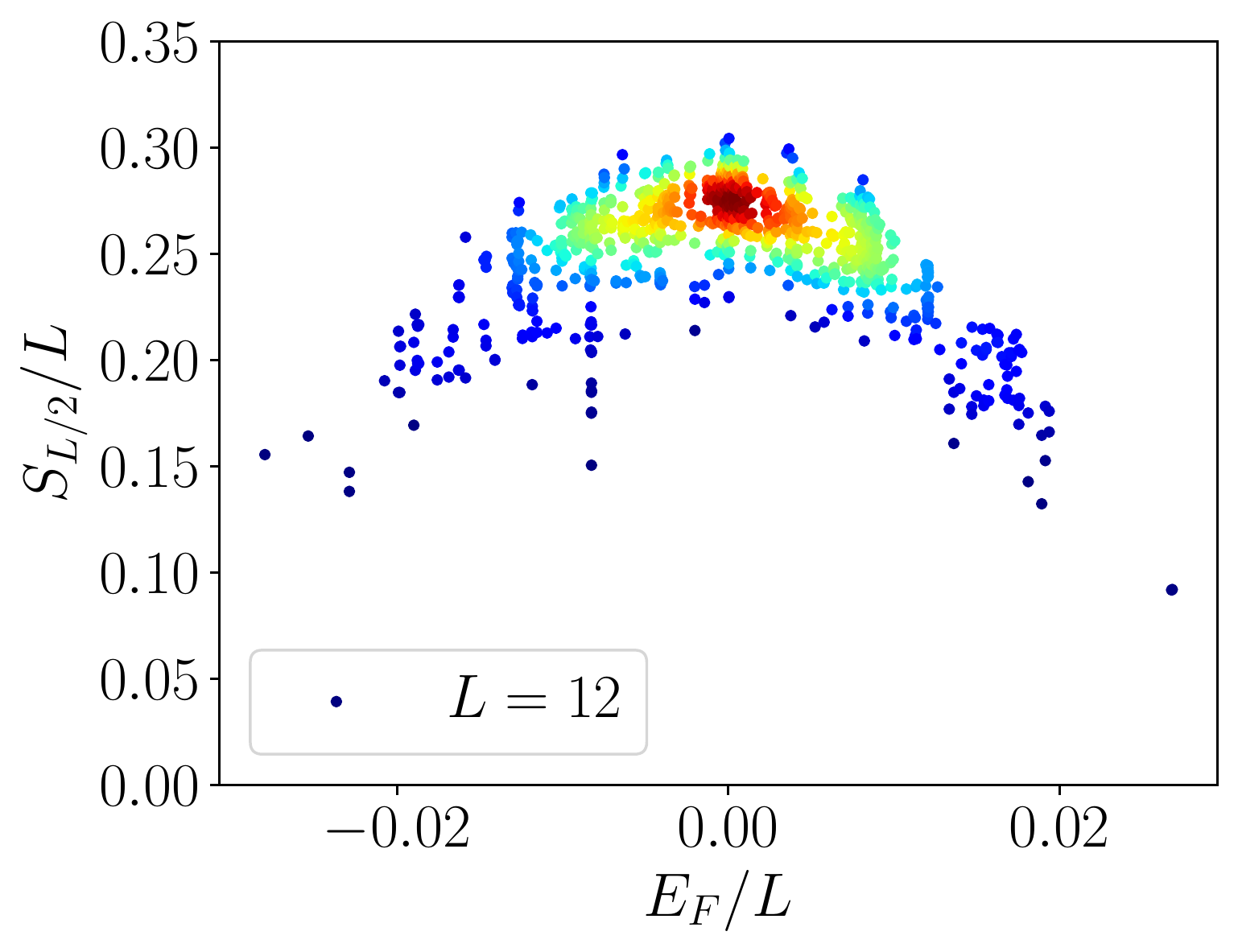}}{\large(a)}
\stackunder[5pt]{\includegraphics[width=0.45\hsize]{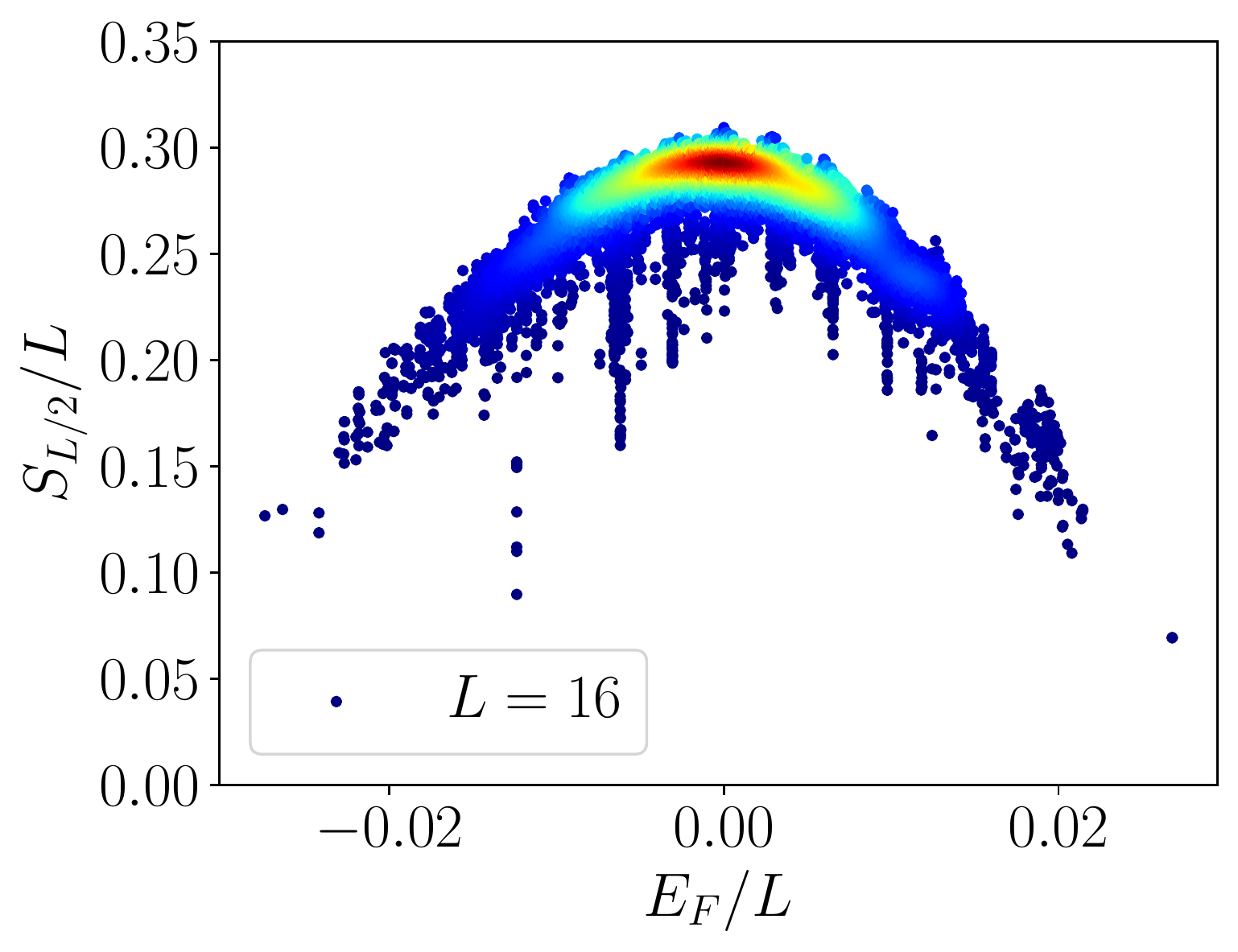}}{\large(b)}\\
\stackunder[5pt]{\includegraphics[width=0.45\hsize]{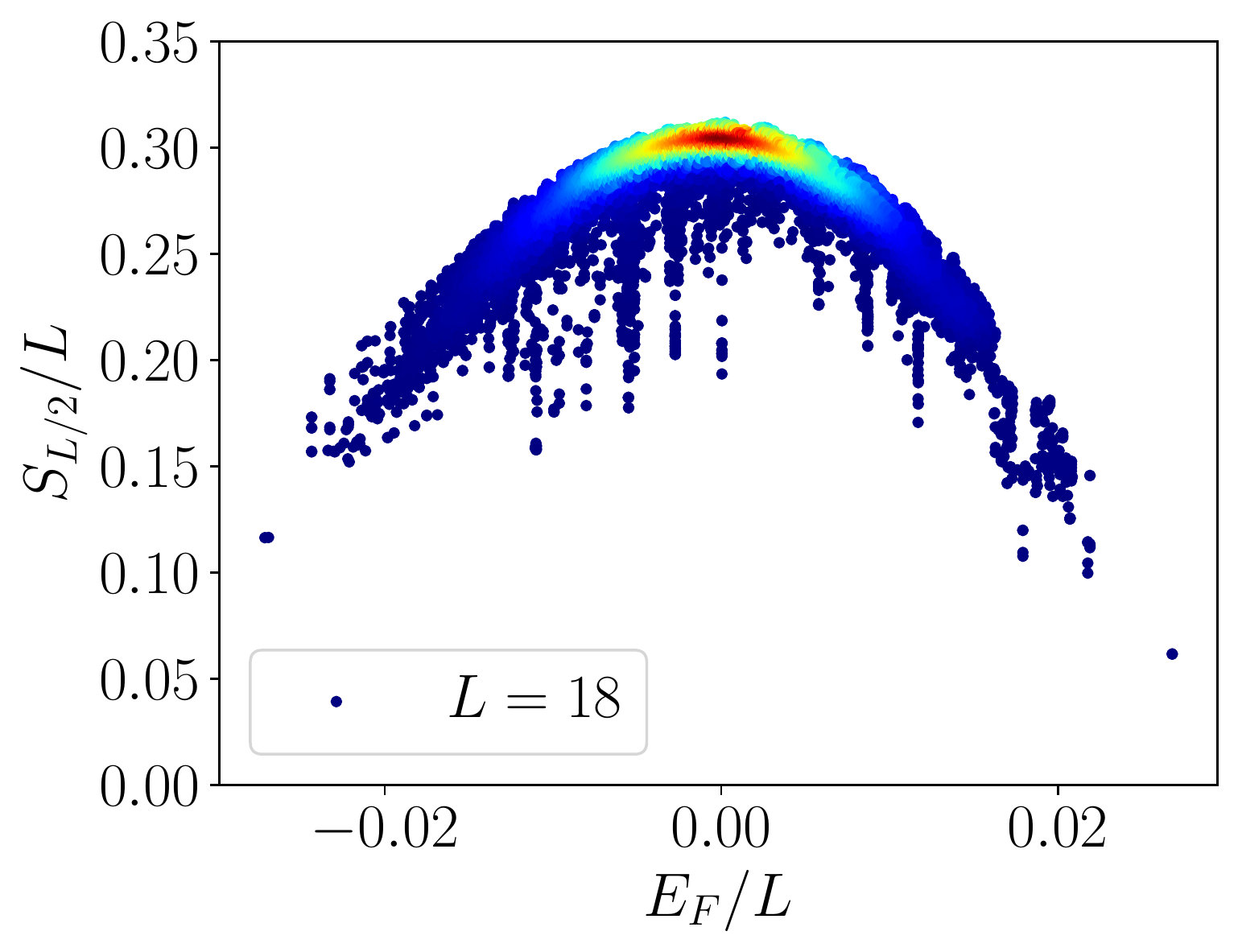}}{\large(c)}
\end{center}
\caption{{\bf Scaled entanglement spectrum as a function of scaled quasienergy 
at another DL point and at resonance for the period-2 model:} (a-c) 
Plots showing the scaled entanglement entropy $S_{L/2}$ obtained from exact 
numerical calculations as a function of the scaled quasienergy at a DL point with 
$J=1$, and $\mu=V=20$ and $\om\approx2.22$ for $L=12,~16$ and $18$, respectively. In all 
three cases, the range of the scaled quasienergy appears to be the same with many low-entanglement states just as we
see in Fig.~\ref{fig22}.} \label{fig24} \end{figure}

\begin{figure}[!tbp]
\footnotesize
\begin{center}
\stackunder[5pt]{\includegraphics[width=0.45\hsize]{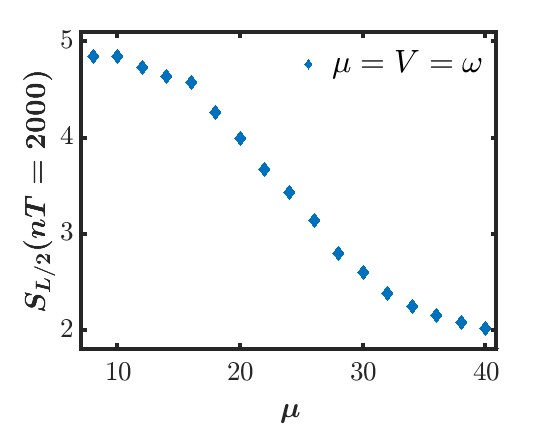}}{\large(a)}
\stackunder[5pt]{\includegraphics[width=0.48\hsize]{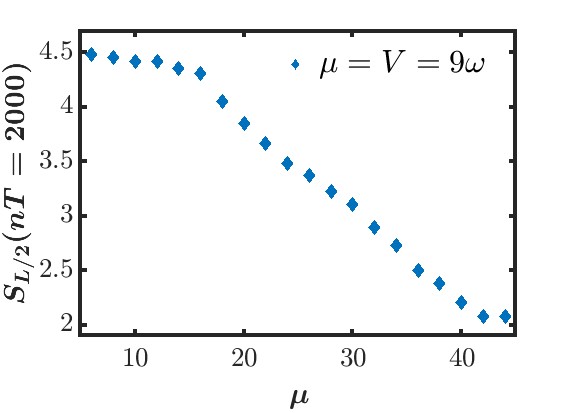}}{\large(b)}\\
\end{center}
\caption{{\bf Slow relaxation behavior in the high and low frequency regimes observed from the dynamics of entanglement entropy at different two dynamical 
localization and resonance points:} (a-b) The half-chain entanglement entropy $S_{L/2}$ versus
$\mu$ at the stroboscopic number $n=t/T=2000$ for $\mu=V=\om$ and $\mu=V=9\om$, respectively. For this analysis, we have taken $L=16$ and the initial state to be 
$\ket{0000111111110000}$. In both cases, we see that $S_{L/2}(nT)$ gradually decreases with increasing values of $\mu = V$, showing a slow relaxation behavior 
in the large driving amplitude regime. We 
further see a smooth crossover from ergodic to 
non-ergodic behavior occurring in both cases with increasing values of $\mu$ and $V$. In Fig. (a), we see the crossover occurring for $18 \lesssim \mu=V=\om
\lesssim 30$, which shows a slow relaxation behavior in the high-frequency regime. In Fig. (b), we observe a similar crossover occurring for $18 
\lesssim \mu=V=9\om \lesssim 40$, where $\om$ lies in the range of $[2,4.44]$. This regime shows a slow relaxation behavior when the ratio of the driving frequency to 
$J$ lies in the regime of low to intermediate values.} \label{fig25} \end{figure}

In this section, we will discuss the stability of HSF 
in the thermodynamic limit ($L \to\infty$) in the context of the period-2 model
which is at a DL point along with a resonance. In Figs.~\ref{fig22} (a-c),
the scaled entanglement entropy $S_{L/2} /L$ is shown as a function of the scaled 
quasienergy $E_F /L$ for $L=12,~16$ and $20$, respectively, for $\mu=V=\om=20$.
For this analysis, we employed the exact diagonalization method for individual 
momentum sectors by using the translation symmetry in order to access larger system 
sizes. For all three values of $L$, the range of the scaled quasienergy turns out to 
be the same since the many-body bandwidth increases linearly with $L$. 
Interestingly, we find that the scaled entanglement spectrum broadens with 
increasing system size, as can be seen from Figs.~\ref{fig22} (a-c). The broadening 
of the entanglement spectrum indicates an increasing number of Hilbert 
space fragments and inert configurations of states for larger system sizes. In 
Figs.~\ref{fig23} (a-b), the dynamics of the Loschmidt echo is shown for 
$L=12,~16$ and $20$ for the parameter values $\mu=V=\om=20$ and $\mu=V=\om=10$, 
respectively. For both sets of parameters, we consider three different choices of 
initial states, $\ket{000111111000}$, $\ket{0000111111110000}$, and 
$\ket{00000111111111100000}$ for the three different system sizes, with the 
dynamics of all three states being kinetically constrained to lie within a single
Hilbert space fragment consisting of four states, as 
shown in Fig.~\ref{fig30} for $L=16$. As a result, all three states execute 
long-time oscillations in the dynamics, as seen in Figs.~\ref{fig23} (a-b). 
Furthermore, the envelop of the Loschmidt echo in the first case falls off very 
slowly compared to the second case. This is due to the larger values of $\mu,~V$ 
and $\om$ in the first case; hence the first-order FPT is a better approximation 
to the exact Floquet Hamiltonian because the higher order corrections are smaller.
In Figs.~\ref{fig23} (c-d), we numerically fit the envelop of the Loschmidt echo 
with time for $L=20$ for the same parameter values same as in Figs.~\ref{fig23}
(a-b). In both cases, we see that the period of oscillations of the Loschmidt 
echo is almost the same, with $\Delta t \simeq 2\pi$. However, 
the decay rate of the envelop for the first case is seen to be $1/\tau_1
\simeq 0.0036$, whereas the same quantity for the latter case is almost four times 
larger, $1/ \tau_2 \simeq 0.0139$. The faster decay rate in the 
second case can be explained by the following argument. Due to the symmetry 
discussed in Eq.~\eqref{hamsym}, the correction to the first-order FPT effective 
Hamiltonian will be of third order, which should scale as $J^3 /\mu^2$; this is
derived in \hyperlink{Appendix C}{Appendix C} at the DL points for $V=0$. Therefore,
the decay rate $1/\tau$, whose dominant contribution is expected to come from the 
third-order correction, should scale such that $\tau_{2}/ \tau_{1} = (\mu_{2}
/\mu_{1})^2$. Putting $\mu_1=20$ and $\mu_2=10$, we expect $1 /\tau_2 =4 /\tau_1$,
which agrees quite well with the numerically fitted decay rate.
In Figs. \ref{fig24} (a-c), we examine the thermodynamic stability of 
HSF in a regime with lower frequency by setting $J=1$, $\mu=V=20$, $\om\approx2.22$ 
(which corresponds to the ninth DL point), and system sizes $L=12,~16$ and $18$. The
rescaled quasienergies for all three cases are again observed to be the same due to
the reason mentioned in the earlier case. The low-entanglement states are also
found to be quite stable with increasing system size, which 
indicates the stability of the effective model even in the lower 
frequency regime. In Fig. \ref{fig25} (a-b), we show the variation of the half-chain entanglement entropy $S_{L/2}$ with
$\mu$ at the stroboscopic number $n=t/T=2000$ for $\mu=V=\om$ and $\mu=V=9\om$, respectively. We have taken $L=16$ and the initial state to be 
$\ket{0000111111110000}$ for this analysis. In both cases, $S_{L/2}(nT)$ gradually decreases with increasing values of $\mu = V$, which shows a slow relaxation 
behavior in the large driving amplitude regime. It further reveals that there is a smooth crossover from ergodic to 
non-ergodic behavior occurring in both cases with increasing values of $\mu$ and $V$. In Fig. \ref{fig25} (a), we observe a crossover occurring for $18 \lesssim \mu=V=\om
\lesssim 30$, which implies a slow relaxation behavior in the high-frequency regime. On the other hand, in Fig. (b), we see a similar behavior occurring
for $18 \lesssim \mu=V=9\om \lesssim 40$, where $\om$ lies in the range of $[2,4.44]$, which necessarily lies in the low to intermediate driving frequency 
regime. Therefore, we can conclude that DL and resonance-induced HSF in this 
model is valid for a broad range of driving frequencies even in the thermodynamic 
limit provided that the amplitude of the driving and the strength of 
the interaction are strong enough to stabilize this ergodicity-breaking 
mechanism. It is well-known in the literature that a generic many-body interacting 
Floquet system is most susceptible to heating in the slow 
driving regime. However, our model shows that the HSF mechanism emerging from the 
interplay between DL, interaction and resonances provides significant protection 
against heating even in the case of slow driving; this is a quite ubiquitous
feature of this class of models.

\section{Period-4 model}
\label{sec5}

We will now discuss a model with the second type of periodic potential, namely,
the period-4 model. The general form of the Hamiltonian for this class is given by
\bea H&=& ~\sum_{j} ~ [ J\,(c_{j}^{\dagger}c_{j+1} ~+~ c_{j+1}^\dagger c_j) ~+~ \mu (t) \cos(\pi j/2+\phi) ~c_j^\dagger c_j ~+~ V n_{j}n_{j+1}], 
\label{ham44} \eea
where $J$ denotes the nearest-neighbor hopping, $\mu$ is the amplitude of the periodic potential with $m=4$, $V$ defines the nearest-neighbor density-density interaction, and 
$\phi$ refers to a generalized phase. This model possesses a mirror symmetry~\cite{mirror} for certain special values of $\phi$. Since we are interested in 
mirror-symmetric configurations for our analysis, it turns out that we
can only have two possible realizations, namely, $\phi=0$ and $\phi=7\pi/4$. 
We will denote these as type-1 and type-2 cases respectively.

\begin{figure}[!tbp]
\begin{center}
\includegraphics[width=0.75\hsize]{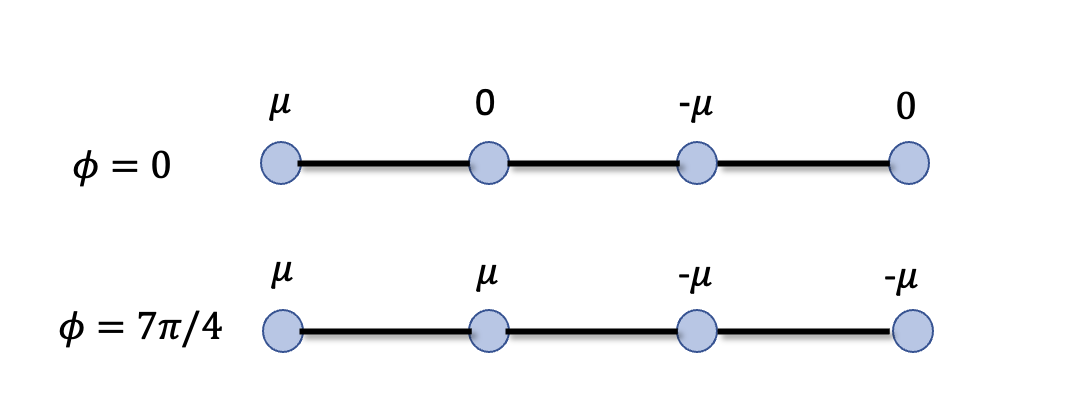}
\end{center}
\caption{{\bf Schematic of the mirror-symmetric periodic potential pattern for the period-4 model:} Schematic diagrams showing the potential patterns for the
two mirror-symmetric configurations of the period-4 model corresponding to 
$\phi = 0$ and $7 \pi/4$.} \label{fig40} \end{figure}

\subsection{Type-1 mirror-symmetric case}

Taking $\mu (t)$ to be proportional to $\mu$ in Eq.~\eqref{ham44}, we
find that the on-site potential pattern for $\phi = 0$ is given by $(\mu,0,-\mu, 0)$ on four consecutive sites numbered $(4n,4n+1,4n+2,4n+3)$; 
this is shown in Fig.~\ref{fig40} (a). Assuming $\mu \gg J$, and using the results presented in Appendix B, we find that the first-order FPT Hamiltonian is given by
\bea H_{F1}^{(1)} &=& J ~\sum_{j=1}^{L/4} ~[M c_{4j}^{\dagger}c_{4j+1}+ 
M c_{4j+1}^{\dagger}c_{4j+2}+M^{*}c_{4j+2}^{\dagger}c_{4j+3}+
M^{*}c_{4j+3}^{\dagger} c_{4j+4}+ {\rm H.c.}], \non \\
M &=& e^{i\mu T/4} \left(\frac{\sin(\mu T/4)}{\mu T/4}\right). \eea
Since the interaction term commutes with the unperturbed Hamiltonian, $H_{0}$, that part of the Hamiltonian will just be given by 
\bea H_{F2}^{(1)} ~=~ V~ \sum_{j=1}^L ~n_{j}n_{j+1} \eea
to first order in $V$.

\begin{figure}[!tbp]
\begin{center}
\footnotesize
\stackunder[5pt]{\includegraphics[width=0.48\hsize]{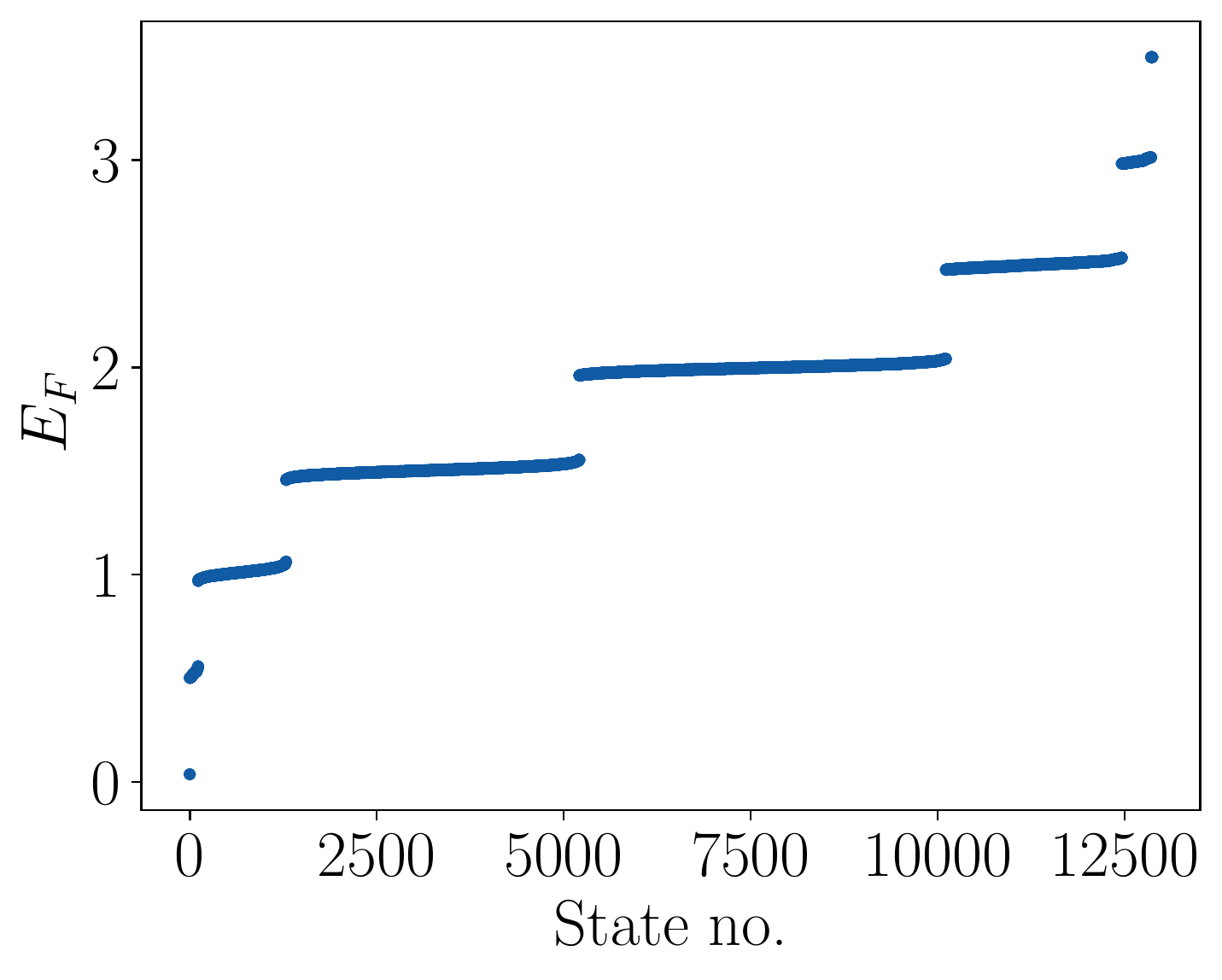}}{\large(a)}
\stackunder[5pt]{\includegraphics[width=0.48\hsize]{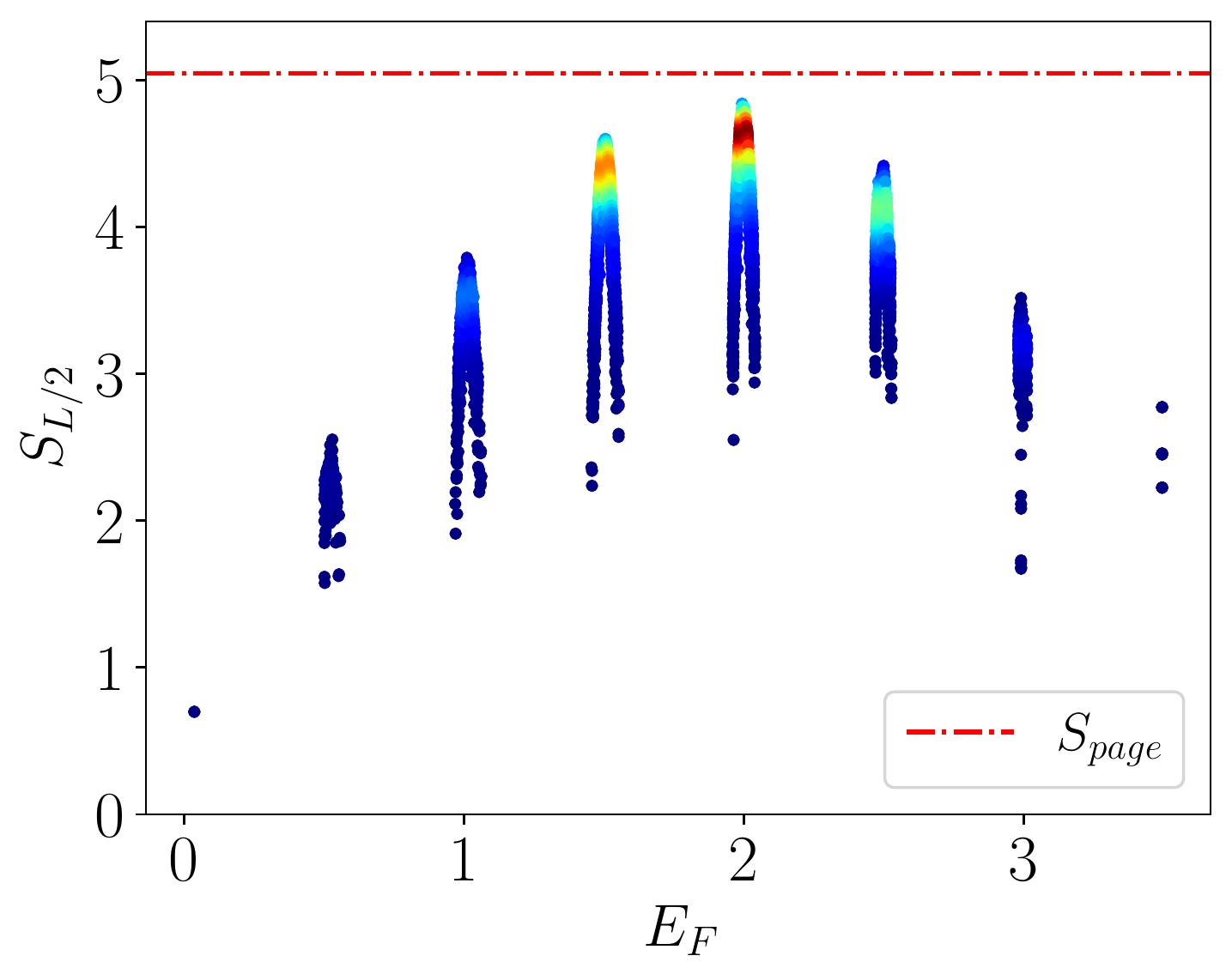}}{\large(b)}\\
\caption{{\bf Quasienergy and entanglement entropy spectrum of the period-4 model with $\boldsymbol{\phi=0}$:} Plots showing the spectrum 
of (a) $E_{F}$ and (b) $S_{L/2}$ as a function of $E_{F}$ at a DL point 
with $\mu=2\om=20$ and $V=0.5$. Both the spectra look identical to those of 
the period-2 case at a DL point.} \label{fig12}\end{center} \end{figure}

The symmetry property discussed in Eq.~\eqref{hamsym} again holds for this model, 
and therefore, similar to the period-2 case,
this system will also exhibit DL when $M=0$, i.e., when $\mu=2n\omega$ where $n=1,2,3,\cdots$. Thus, this mirror-symmetric configuration~\cite{mirror} of the period-4 
model with $\phi=0$ and the period-2 model are identical to each other exactly at a DL point.
In Figs.~\ref{fig12} (a) and \ref{fig12} (b), we show the Floquet quasienergy spectrum 
$E_F$ and the variation of the entanglement entropy $S_{L/2}$ with $E_{F}$ 
for $J=1$, $\mu=20,~\om=20$, and $V=0.5$. The spectrum looks almost identical to the spectrum of the period-2 model at DL with many 
low-entanglement states near the middle of the spectrum.

\subsection{Type-2 mirror-symmetric case}

The period-4 model with $\phi=7\pi/4$ is another mirror-symmetric configuration 
with many interesting properties, which we will now discuss in detail. Taking $\mu (t)$ to be proportional to 
$\mu \sqrt{2}$ in Eq.~\eqref{ham44}, the on-site potential pattern for $\phi = 7 \pi/4$
is given by $(\mu, \mu, -\mu, -\mu)$ on four consecutive sites numbered
$(4n,4n+1,4n+2,4n+3)$.
Assuming $\mu \gg J$, we obtain the following first-order FPT Hamiltonian, 
\bea H_{F1}^{(1)} &=& J \sum_{j=1}^{L/4} (c_{4j}^{\dagger}c_{4j+1}
+ M_{1}c_{4j+1}^{\dagger}c_{4j+2} + c_{4j+2}^{\dagger}c_{4j+3} + M_{1}^{*} c_{4j+3}^{\dagger}c_{4j+4} + {\rm H.c.}) ~+ ~V \sum_{j=1}^L n_{j}n_{j+1}, \non \\
M_{1} &=& e^{i\mu T/2}\left(\frac{\sin(\mu T/2)}{\mu T/2}\right). \eea
Remarkably, we see that the non-interacting part of the Hamiltonian
exactly describes the SSH model~\cite{SSH},
with nearest-neighbor hoppings which have alternating strengths
given by $J$ and $J |M_1|$. (It is clear that $|M_1|$ is always smaller than 1.
The phase of $M_1$ can be removed by 
doing a unitary transformation of the form $c_j \to c_j e^{i \al_j}$
with appropriately chosen $\al_j$'s). We thus see that the periodicity 
of the model has effectively reduced from 4 to 2.

\begin{figure}[!tbp]
\begin{center}
\includegraphics[width=0.8\hsize]{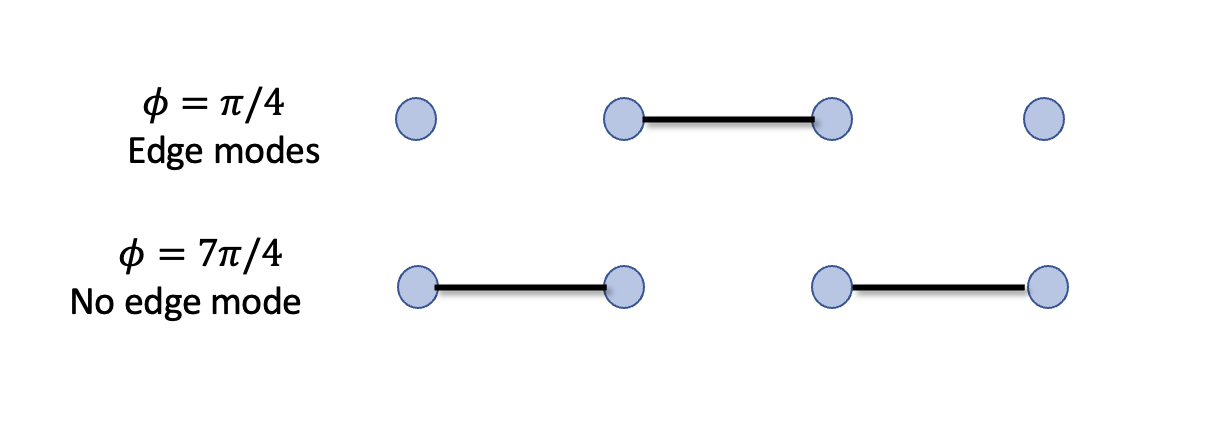}
\end{center}
\caption{{\bf Schematic of the topologically protected edge modes for the period-4 model at a DL point:} Schematic picture showing topologically protected 
zero-energy edge modes for the $\phi=\pi/4$ at a DL point and no edge modes for $\phi=7\pi/4$.} \label{fig50} \end{figure}

\begin{figure}[!tbp]
\footnotesize
\begin{center}
\stackunder[5pt]{\includegraphics[width=0.5\hsize]{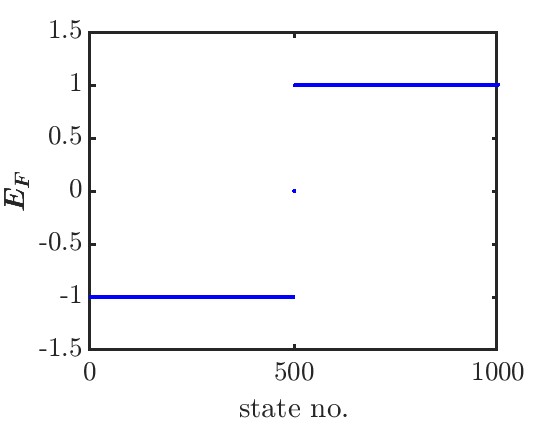}}{\large(a)}
\stackunder[5pt]{\includegraphics[width=0.5\hsize]{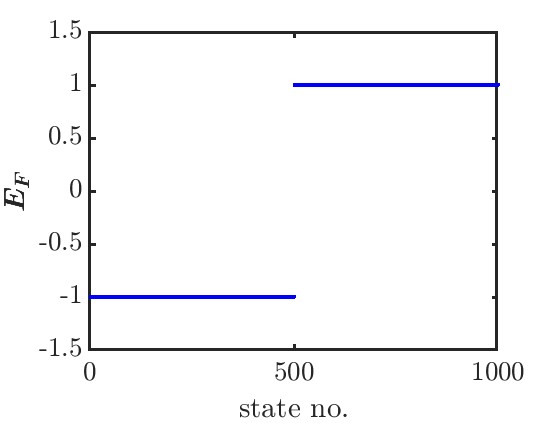}}{\large(b)}\\
\stackunder[5pt]{\includegraphics[width=0.5\hsize]{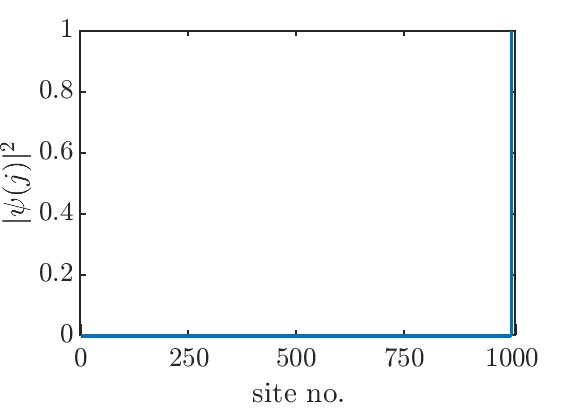}}{\large(c)}
\stackunder[5pt]{\includegraphics[width=0.5\hsize]{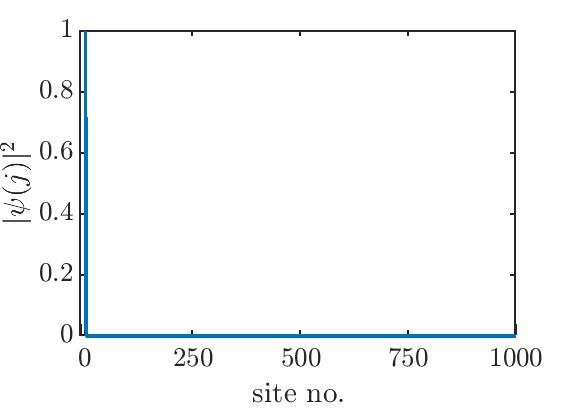}}{\large(d)}
\end{center}
\caption{{\bf Quasienergies and wave function probabilities of the edge modes:} (a-b): Plots of $E_{F}$ in increasing order versus state number for a system with 
open boundary conditions with $J=1,~\mu=\om=20$. (a) A system with 
$\phi=\pi/4$ supports two zero-energy edge modes, while (b) a system with 
$\phi=7\pi/4$ does not. (c-d): Plots showing the probability versus site number
for the edge modes showing that the modes are localized at the ends 
of the system.} \label{fig09} \end{figure}

The expression for $M_1$ implies that this model will exhibit DL for 
$\mu=n\omega$, where $n=1,2,\cdots$. Exactly at these points, the 
effective first-order Hamiltonian is given by
\bea H_{F1}^{(1)}&=&J ~\sum_{j=1}^{L/4} (c_{4j}^{\dagger}c_{4j+1} +c_{4j+2}^{\dagger}c_{4j+3}+ {\rm H.c.}) ~+~ V ~\sum_j n_{j}n_{j+1}. \eea
The non-interacting part of this Hamiltonian is an extreme limit of the
SSH model with alternating nearest-neighbor hoppings $\ga_{1}=1$, and $\ga_{2}=0$. 
Our model therefore inherits the property of the SSH model
that a system with open boundary conditions has 
topologically protected zero-energy edge modes 
provided that the hopping strength on the leftmost or rightmost bond 
is weaker than the strength of the bond next to it.
For $\phi=7\pi/4$, we find that the leftmost bond (between sites
numbered 0 and 1) has a hopping strength which is larger than the
strength of the next bond, and therefore, the system has
no edge modes. However, the system has edge modes for
$\phi=\pi/4$, i.e., when the bonds are shifted by one unit cell and the 
stronger and weaker bonds
get interchanged (see the schematic pictures in Fig.~\ref{fig50}).
In Figs.~\ref{fig09} (a) and \ref{fig09} (b), we see two zero-energy 
edge modes and no edge modes for $\phi=\pi/4$ and $\phi=7\pi/4$, 
respectively, for $J=1$, $\mu=\om=20$ and $L=2000$ with open boundary conditions. 
As shown in Figs.~\ref{fig09} (c) and \ref{fig09} (d), the two modes are localized 
at the two edges of the 
system, which can be seen from a plot of the probability $|\psi (j)|^2$ versus the site number $j$. Note that the parameter values $J=1$ and 
$\mu=\om=20$ imply that the system is at a DL point. 
Another interesting point to observe is that a static model with a 
nearest-neighbor hopping $J$ and a period-4 time-independent potential 
with strength $\mu$ does not have any such zero-energy end modes.

Since the interaction part again commutes with unperturbed Hamiltonian, $H_{0}$, the effective Hamiltonian to first order in $V$ again reads as $H_{F2}^{(1)}=V\sum 
n_{j}n_{j+1}$. The interplay between interaction and DL in this case gives rise to various intriguing phenomena. We will study this in the next few sections using 
an effective spin model based on the first-order Floquet Hamiltonian. 

\subsubsection{Effective spin model based on first-order effective Hamiltonian}

\begin{figure}[!tbp]
\begin{center}
\includegraphics[width=0.8\hsize]{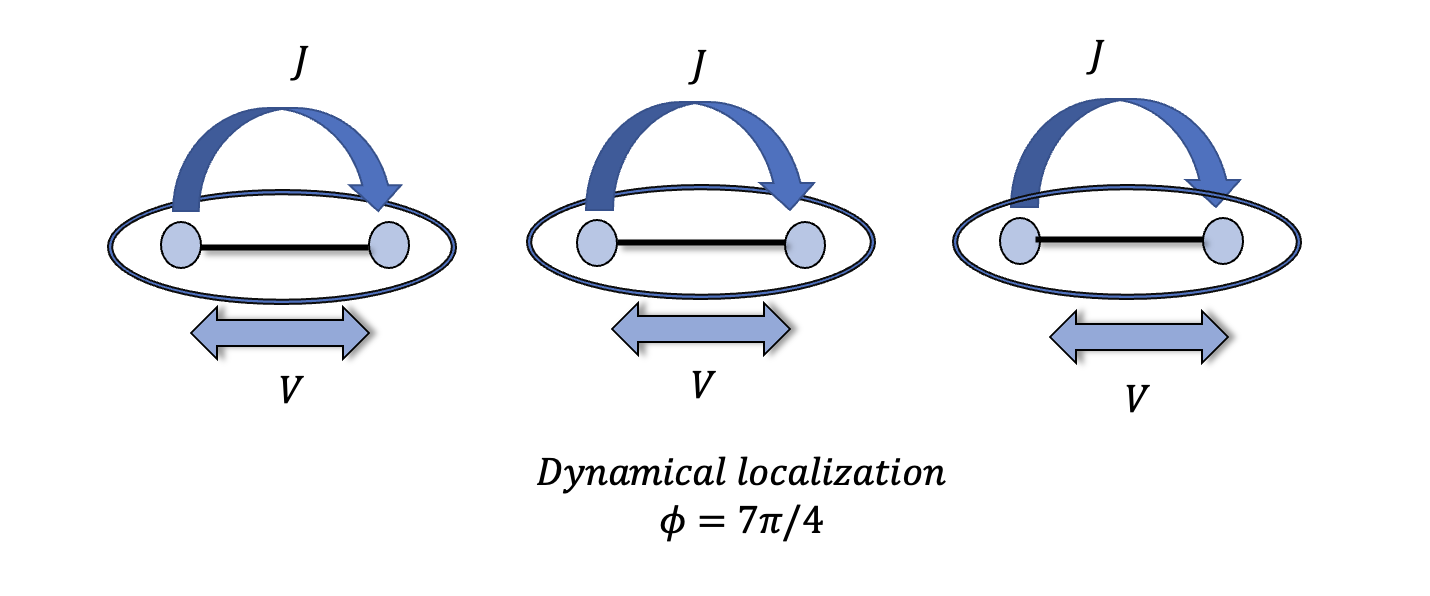}
\end{center}
\caption{{\bf Schematic of period-4 model with interaction at a DL point:} The schematic picture of the period-4 model at a DL point, with a nearest-neighbor
hopping which alternates between $J$ and zero, and an interaction $V$.} 
\label{fig70} \end{figure}

To derive the effective spin model, it is 
convenient to recast the effective Hamiltonian in terms of unit cells
\bea H ~=~ J~ \sum_{j=1}^{L/2} ~\left[(a_{j}^{\dagger}b_{j}+ {\rm H.c.})+V(n_{j,a}n_{j,b}+n_{j,a}n_{j-1,b})\right], \eea
where $a_{j}$ and $b_{j}$ denote the annihilation operators on the even and odd 
numbered sites of the $j$-th unit cell. (Henceforth we will refer to the $j$-th unit
cell as the $j$-th site for convenience). The schematic of the periodic-4 model at a DL point is shown in 
Fig.~\ref{fig70}. This above form suggests that the particle number $n_j$ at the
$j$-th site (unit cell) commutes with $H_{F}^{(1)}$. Hence $H_{F}^{(1)}$ has $L/2$
conserved quantities. (Note that these will be only approximately conserved quantities. The exact effective Hamiltonian will have higher order terms 
which do not commute with these quantities). For a 
system consisting of two sites with $n_{j}^{\rm{max}}=2$, we can have nine possible effective Hamiltonians which are shown in the table below.

\vspace*{.4cm} 
\begin{table}[h!]
\begin{center}
\begin{tabular}{|c|c|c|c|} 
\hline
$n_{1}$ & $n_{2}$ & Effective Hamiltonian \\
\hline
0 & 0 & $E=0$ \\ 
\hline
 0 & 2 & $E=V$ \\
 \hline
2 &0 & $E=V$ \\
\hline
2&2&$E=3V$\\
\hline
0&1&$H=J(a_{2}^{\dagger}b_{2}+ {\rm H.c.})$\\
\hline
1&0&$H=J(a_{1}^{\dagger}b_{1}+ {\rm H.c.})$\\
\hline
1&2&$H=J(a_{1}^{\dagger}b_{1}+ {\rm H.c.})+Vb_{1}^{\dagger}b_{1}+V$\\
\hline
2&1&$H=J(a_{2}^{\dagger}b_{2}+ {\rm H.c.})+Va_{2}^{\dagger}a_{2}+V$\\
\hline
1&1&$H=J(a_{2}^{\dagger}b_{2}+a_{1}^{\dagger}b_{1}+ {\rm H.c.})+Vn_{b,1}n_{a,2}$\\
\hline
\end{tabular}
\caption{\label{effHa_twosite} Allowed configurations and the corresponding 
effective Hamiltonians for two unit cells at a DL point with $\mu \gg V$ for a 
period-4 model.}
\end{center}
\end{table}
\vspace*{.4cm}

In the table above, $n_{1}$ and $n_{2}$ are the occupation numbers of the first and second unit cells, respectively. We can see that eight out of the nine possibilities 
shown above can be mapped to a non-interacting problem. The only instance when the
effects of the interaction is non-trivial is when both the unit cells are singly 
occupied. For this case, an effective spin degrees of freedom can be defined as
\beq \ket{\uparrow} ~=~ \ket{10}, ~~~~\ket{\downarrow} ~=~ \ket{01}, \eeq
where $\ket{10}$ defines a unit cell with the left and the right sites being occupied and empty, respectively, and $\ket{01}$ means the other way around. With this
definition, the correlated two-site problem takes the following form 
\bea H ~=~ J ~(\sigma_{1}^{x}+\sigma_{2}^{x}) ~+~ \frac{V}{4}~ (1-\sigma_{1}^{z})~(1+\sigma_{2}^{z}), \eea
where $\sigma^{z}$ and $\sigma^{x}$ are Pauli matrices. This two-site problem can now be generalized to larger system sizes. To do so, we first consider the case 
where
all the sites are singly occupied. The effective spin Hamiltonian for this case is given by
\bea H ~=~ \sum_{j=1}^{L/2} ~[J \sigma_{j}^{x} ~+~ \frac{V}{4}~(1-\sigma_{j}^{z}\sigma_{j+1}^{z})], \eea
which is essentially the transverse field Ising model with the interaction term $-V/4$ and the transverse field $J$. The other four cases for a system with $L/2-2$ unit cells being 
singly occupied and with the two boundary unit cells being either empty or doubly occupied have effective spin Hamiltonians as follows.

\vspace*{.4cm}
\begin{table}[h!]
\begin{center}
\begin{tabular}{|c|c|c|c|} 
\hline
$n_{L}$ & $n_{R}$ & Effective spin Hamiltonian \\
\hline
0 & 0 & $H=J\sum_{j=2}^{L/2-1}\sigma_{j}^{x}-(V/4)\sum_{j=2}^{L/2-2}\sigma_{j}^{z}\sigma_{j+1}^{z}+(V/4)(-\sigma_{2}^{z}+\sigma_{L/2-1}^{z}) +V(L - 4)/8$\\ 
\hline
 0 & 2 & $H=J\sum_{j=2}^{L/2-1}\sigma_{j}^{x}-(V/4)\sum_{j=2}^{L/2-2}\sigma_{j}^{z}\sigma_{j+1}^{z}+(V/4)(-\sigma_{2}^{z}-\sigma_{L/2-1}^{z}) +VL/8$\\ 
 \hline
2 &0 &$H=J\sum_{j=2}^{L/2-1}\sigma_{j}^{x}-(V/4)\sum_{j=2}^{L/2-2} \sigma_{j}^{z}\sigma_{j+1}^{z}+(V/4)(\sigma_{2}^{z}+\sigma_{L/2-1}^{z})
+VL/8$ \\ 
\hline
2&2&$H=J\sum_{j=2}^{L/2-1}\sigma_{j}^{x}-(V/4)\sum_{j=2}^{L/2-2}
\sigma_{j}^{z}\sigma_{j+1}^{z}+(V/4)(\sigma_{2}^{z}-\sigma_{L/2-1}^{z}) +V(L +4)/8$ \\
\hline
\end{tabular}
\end{center}
\caption{The four possible effective spin Hamiltonians emerging from
a system with $L/2 - 2$ unit cells being singly occupied and two boundary 
unit cells, each of them either completely occupied or completely empty, at a DL point with $\mu \gg V$ for a period-4 model.}
\end{table}
\vspace*{.4cm}

In Table 5, $n_{L}$ and $n_{R}$ denote the occupation numbers of the leftmost and rightmost unit cells labeled $j=1$ and $L/2$, respectively. Therefore, we see from Table 5 that the effective spin Hamiltonian has
the form of the transverse field Ising model with additional longitudinal magnetic field terms of strength $\pm V/4$ at the boundary sites, depending on the adjoining sites 
having $n=0$ or 2. Before proceeding further, we perform 
the transformation $\sigma_j^{x} \to \sigma_j^{z}$, $\sigma_j^z \to - \sigma_j^x$, 
and $\sigma_j^y$ remains unchanged. The Hamiltonian then takes the form
\beq H ~=~ J ~\sum_{j=2}^{L/2-1} ~\sigma_{j}^{z} ~-~ \frac{V}{4} \sum_{j=2}^{L/2-2}\sigma_{j}^{x}\sigma_{j+1}^{x}
~+~ \frac{V}{4} ~(\pm ~\sigma_{2}^{x}~ \mp~ \sigma_{L/2-1}^{x}) ~+~ 
\frac{V(L-4)}{8} ~+~ (0,\frac{V}{2},\frac{V}{2},V), \label{ham52} \eeq
where the last term depends on the four possible boundary conditions.
It may appear that the longitudinal field terms at the boundary sites 2 and $L/2-1$
would make it difficult to find the energy spectrum analytically for this
model. To overcome this problem, we add two more sites, labeled 1 and $L/2$, with 
Pauli operators $\sigma_1^x$ and $\sigma_{L/2}^x$, which couple to $\sigma_2^x$ 
and $\sigma_{L/2-1}^x$ respectively~\cite{bouIsing}. The Hamiltonian then becomes 
\beq H ~=~ J ~\sum_{j=2}^{L/2-1} ~\sigma_{j}^{z} ~-~ \frac{V}{4} \sum_{j=1}^{L/2-1}\sigma_{j}^{x}\sigma_{j+1}^{x}
~+~ \frac{V(L-4)}{8}, \label{ham53} \eeq
where we have ignored some constants. Note that $\sigma_1^x$ and $\sigma_{L/2}^x$
commute with the Hamiltonian. Hence, there are four decoupled sectors of states
corresponding to $\sigma_1^x = \pm 1$ and $\sigma_{L/2}^x = \pm 1$. These
four sectors precisely cover the four possible combinations of $\pm$ signs
in Eq.~\eqref{ham52}. 

The Hamiltonian in Eq.~\eqref{ham53} can now be solved analytically by writing
it in terms of Majorana fermion operators using the Jordan-Wigner transformation,
\bea \sigma_{j}^{x} &=& \left( \prod_{i=1}^{j-1} \sigma_i^z \right) ~\alpha_{j}, 
\non \\
\sigma_{j}^{y} &=& \left( \prod_{i=1}^{j-1} \sigma_i^z \right) ~\beta_{j}, \eea 
where $\alpha_{j},~\beta_{j}$ are Majorana operators. 
In terms of these operators, the Hamiltonian takes the form
\bea H &=& -~ iJ ~\sum_{j=2}^{L/2-1} ~\alpha_{j}\beta_{j} ~-~ \frac{iV}{4} ~
\sum_{j=1}^{L/2-1} ~\alpha_{j+1}\beta_{j}. \eea
Since this Hamiltonian is quadratic in terms of Majorana operators, it describes
a non-interacting system and its spectrum can be found exactly~\cite{bouIsing}.

\begin{figure}[!tbp]
\begin{center}
\footnotesize
\stackunder[5pt]{\includegraphics[width=0.47\hsize]{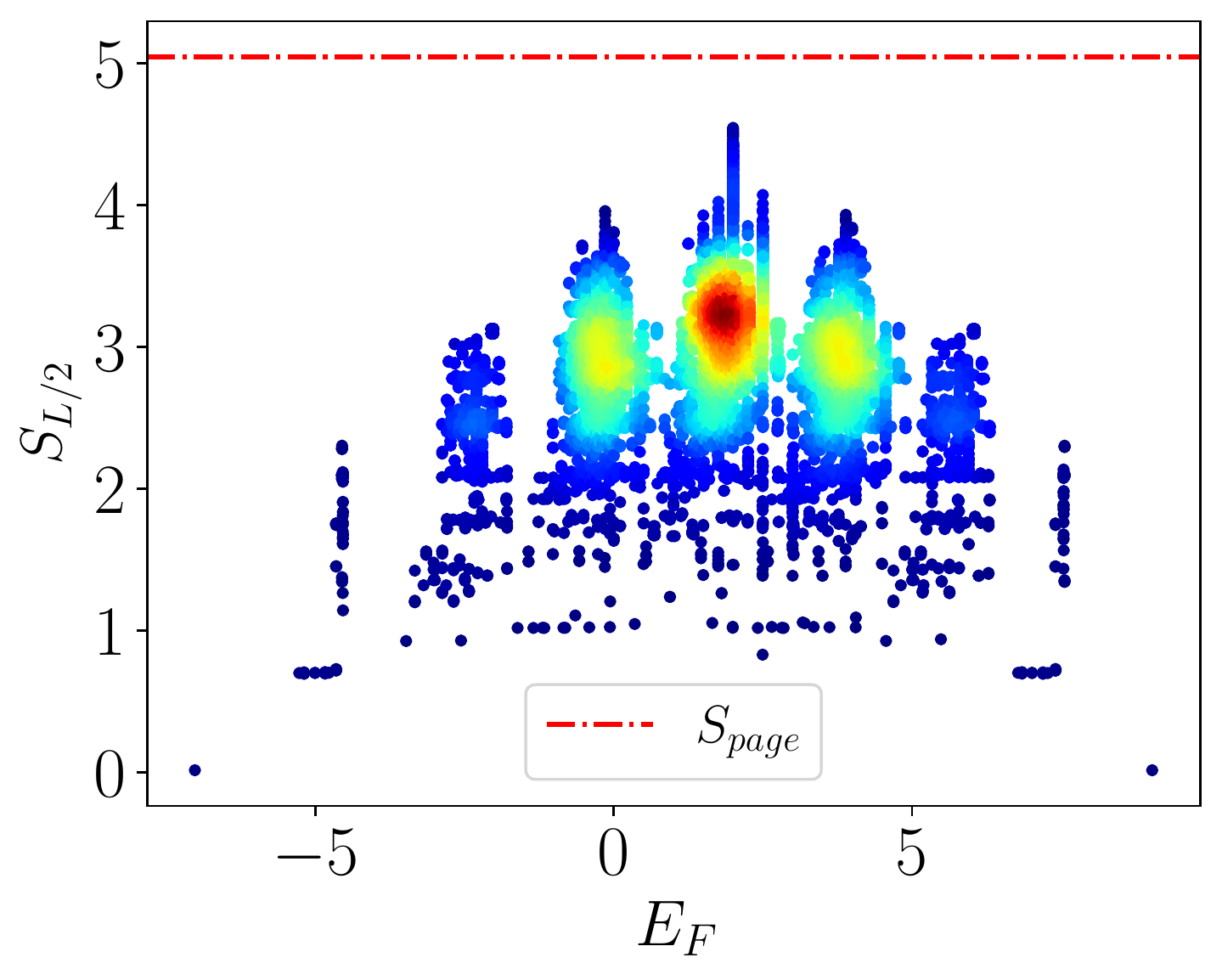}}{\large (a)}
\stackunder[5pt]{\includegraphics[width=0.47\hsize]{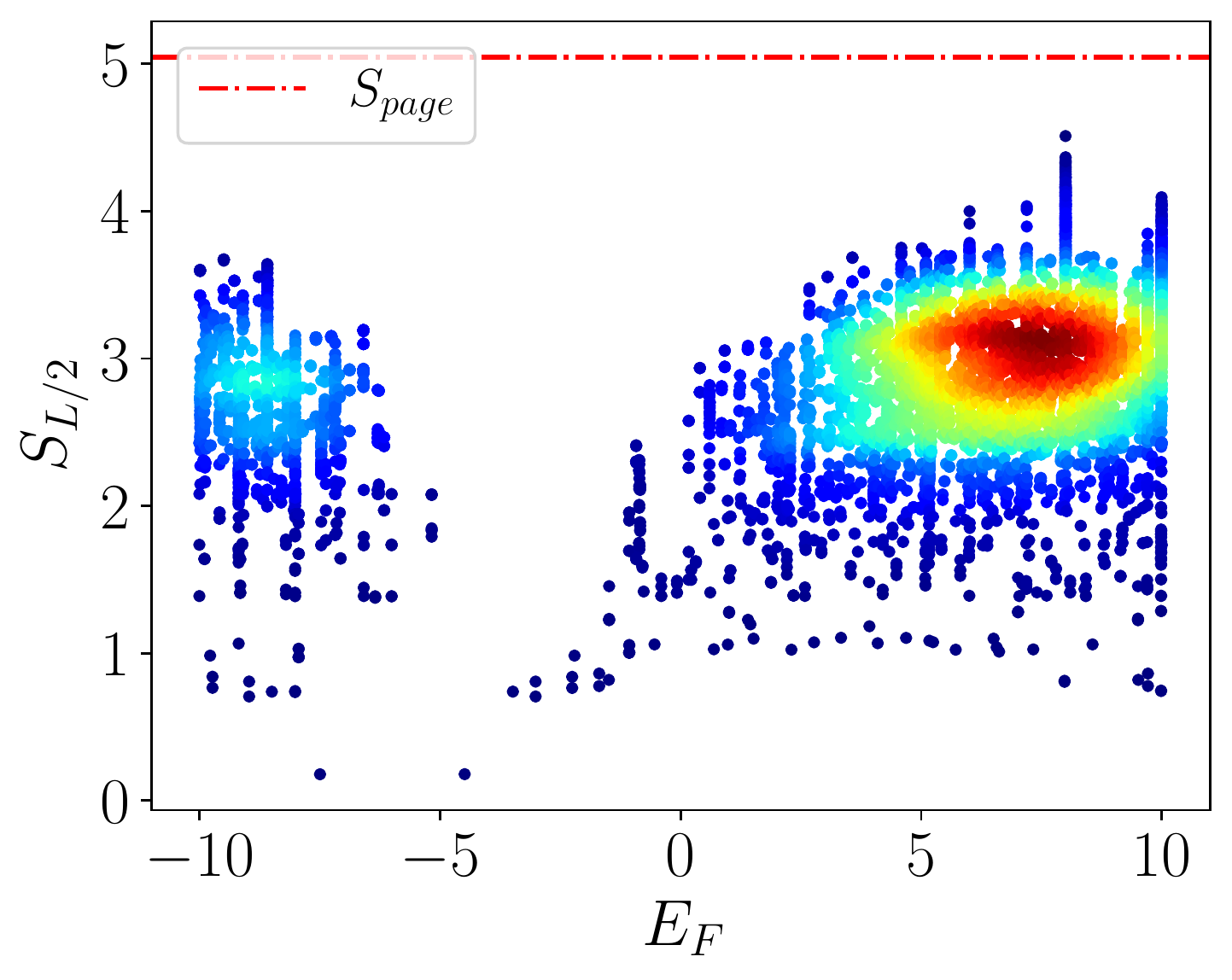}}{\large(b)}\\
\stackunder[5pt]{\includegraphics[width=0.5\hsize]{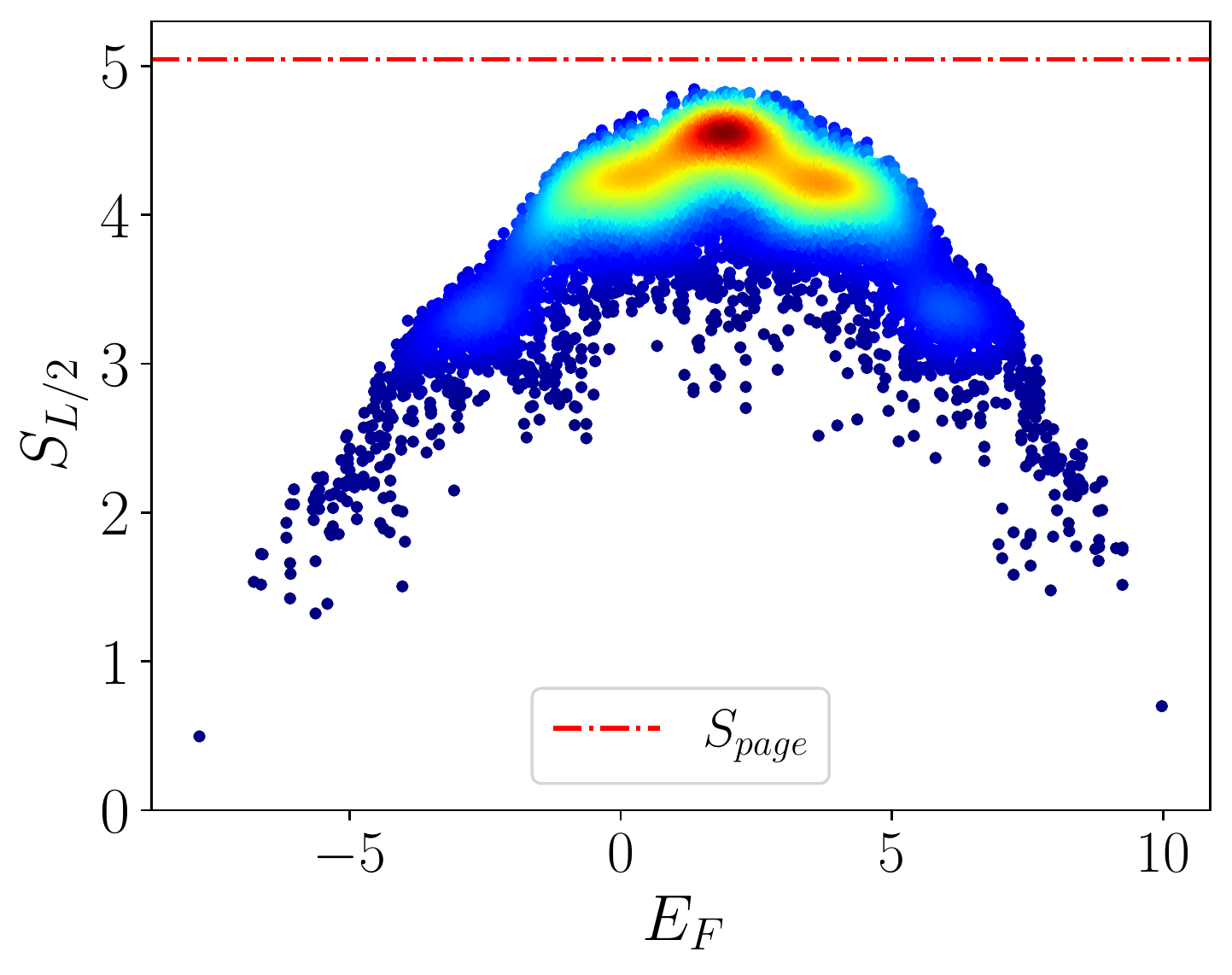}}{\large(c)}
\caption{{\bf Entanglement entropy spectrum for the period-4 case at a DL point and away from a DL point:} Plots showing the half-chain entanglement entropy $S_{L/2}$ versus the 
quasienergy $E_{F}$. For all three cases, we take $J=1$ and $\om=20$. For the first two cases, we consider a DL point with $\mu=20$, and (a) $V=0.5$ 
and (b) $V=2$, respectively. (a): The entanglement entropy consists of many finger-like structures with multiple low-entanglement states near the middle 
of the spectrum.
(b): No finger-like structure is present; however, the system still exhibits 
multiple low-entanglement states near the middle of the spectrum.
(c): We consider a point away from DL, with $\mu=10$, and $V=0.5$. As 
opposed to the behavior at a DL point, we see the system exhibits low-entanglement states (except at the end points of the quasienergy spectrum
where the entanglement is always low). In all the plots, the color 
intensity indicates the density of states. In plot (c) we see that the majority of the Floquet eigenstates show thermal entanglement.}
\label{fig05} \end{center}\end{figure}

To examine the effects of DL on the 
thermalization of the system, we consider the variation of the half-chain entanglement entropy $S_{L/2}$ with the quasienergy $E_{F}$, which gives a 
static measure 
of ergodicity. As shown in Figs.~\ref{fig05} (a) and \ref{fig05} (b), we consider the system at a DL point with $J=1$, $\mu=20$, and $\om=20$, and 
take $V$ to be 
0.5 and 2, respectively. In the first case, we observe many finger-like structures\cite{scarz2} in the entanglement spectrum, which are due to the 
presence 
of an extensive numbers of approximate conserved quantities arising due to the DL. Furthermore, the DL offers many frozen states with extremely 
low-entanglement values, 
i.e., $S_{L/2}=\ln 2 \simeq 0.693$ or $2 \ln 2 \simeq 1.386$ near the middle of the spectrum. Some of the frozen states can be found easily from the 
effective Hamiltonian, such as $\ket{22220000},~\ket{20220200}$,
$\ket{22202000}$ and their translated partners. Nevertheless, as shown in Fig.~\ref{fig05} (b), these finger-like structures are absent 
for $V=2$ due to the 
disappearance of these approximate conserved quantities with increasing interaction strength. The low-entanglement states near the middle of the 
spectrum are still 
present, which again indicates that this system would thermalize very slowly.
We can, therefore, conclude that this slow thermalization occurs due to two possible mechanisms: \\
\noi (i) the existence of extensive numbers of conserved quantities arising 
due to the DL, which grows exponentially with the system size as $3^{L/2}$ (which grows less rapidly than the Hilbert space dimension which goes as 
$2^{L}$)~\cite{emercon}. We emphasize again that these quantities are conserved to a good approximation only for $\mu \gg J,~V$. \\ 
\noi (ii) the presence of many frozen state configurations, which do not participate in the dynamics at a DL point. \\
In Fig.~\ref{fig05} (c), we consider a system away from a DL point with $\mu=10,~\om=20$ and $V=0.5$, and we see that the low-entanglement states have disappeared, signaling that the system should thermalize quickly.

\begin{center}
\begin{figure}[!tbp]
\footnotesize
\stackunder[5pt]{\includegraphics[width=0.51\hsize]{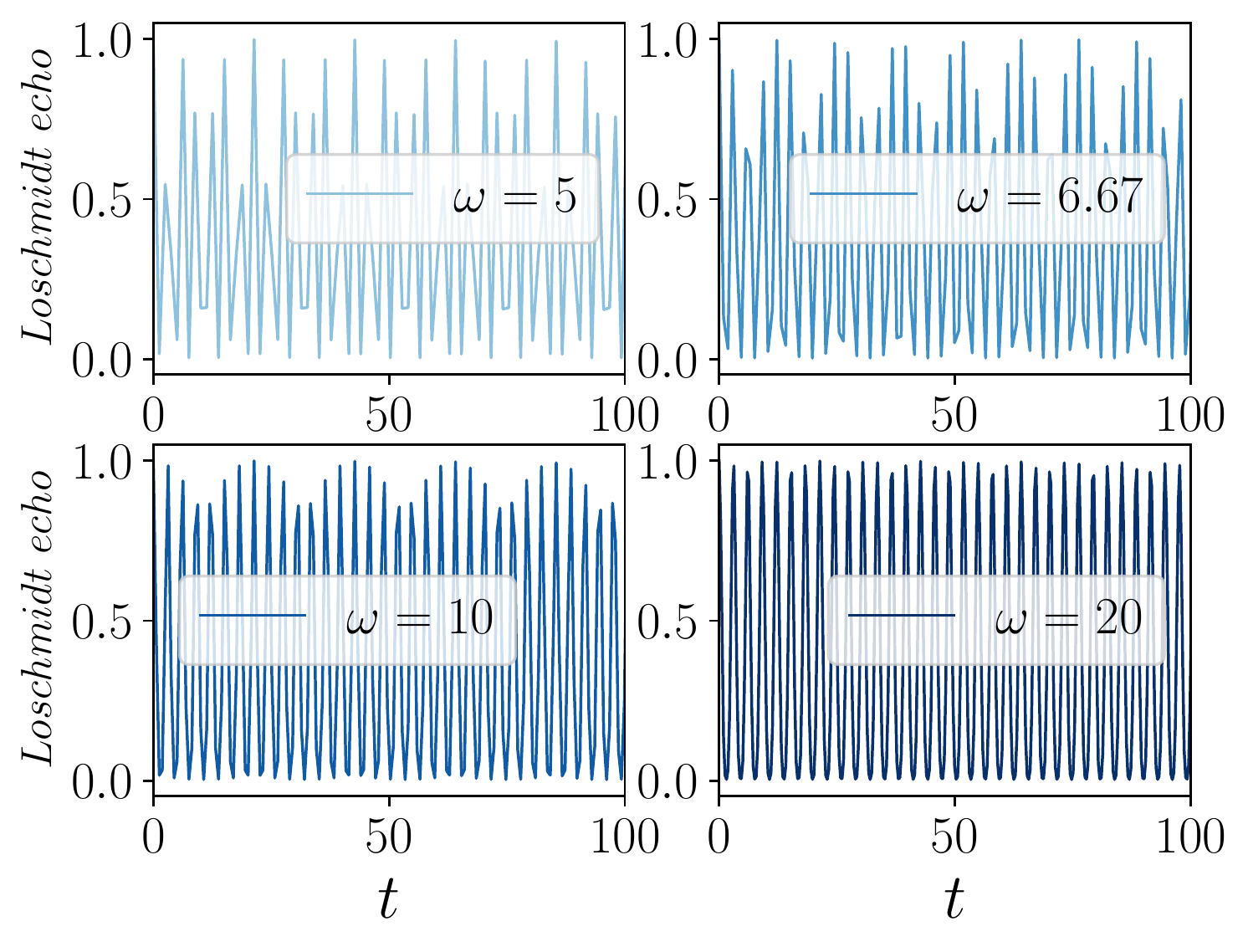}}{\large(a)}
\stackunder[5pt]{\includegraphics[width=0.51\hsize]{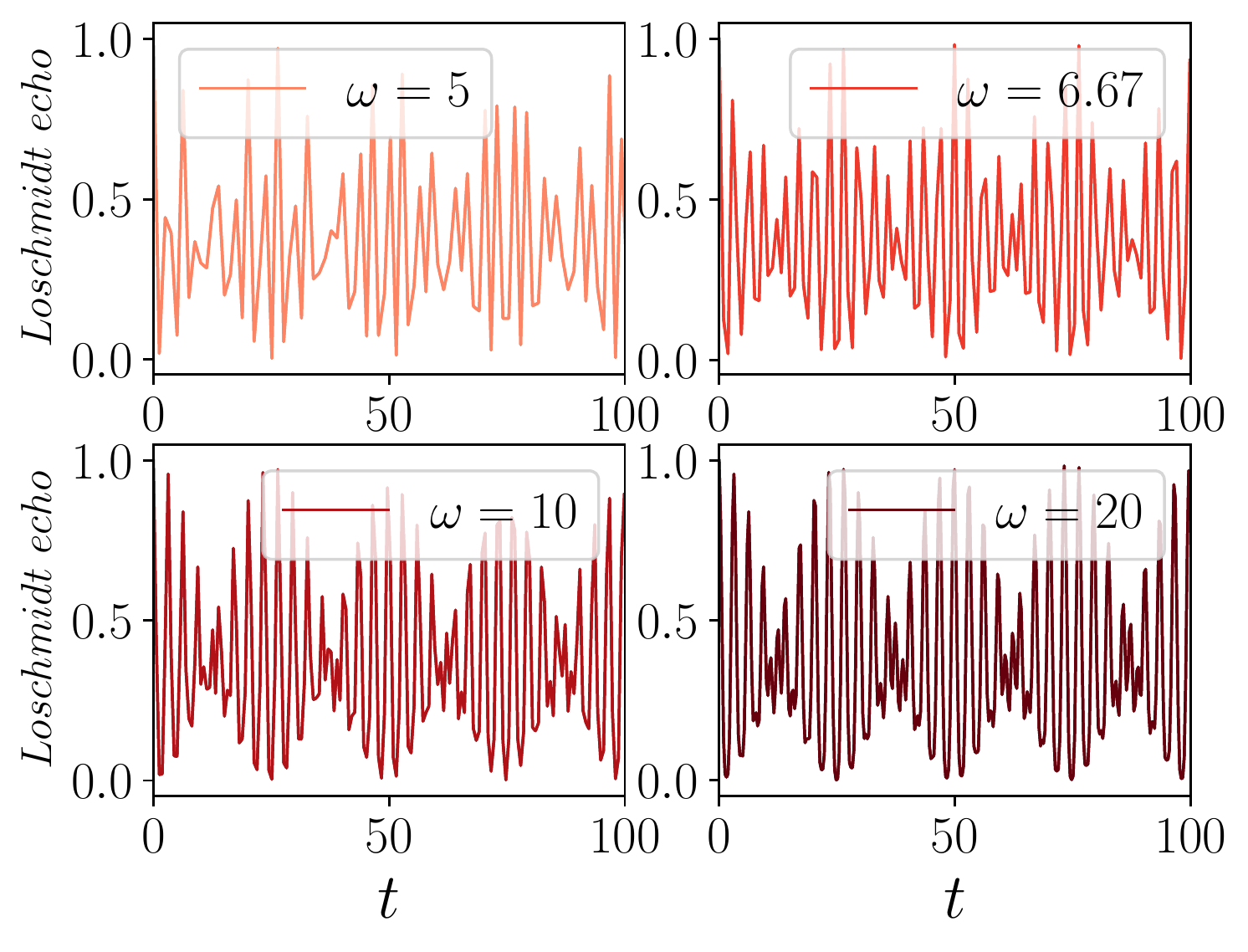}}{\large(b)}\\
\stackunder[5pt]{\includegraphics[width=0.51\hsize]{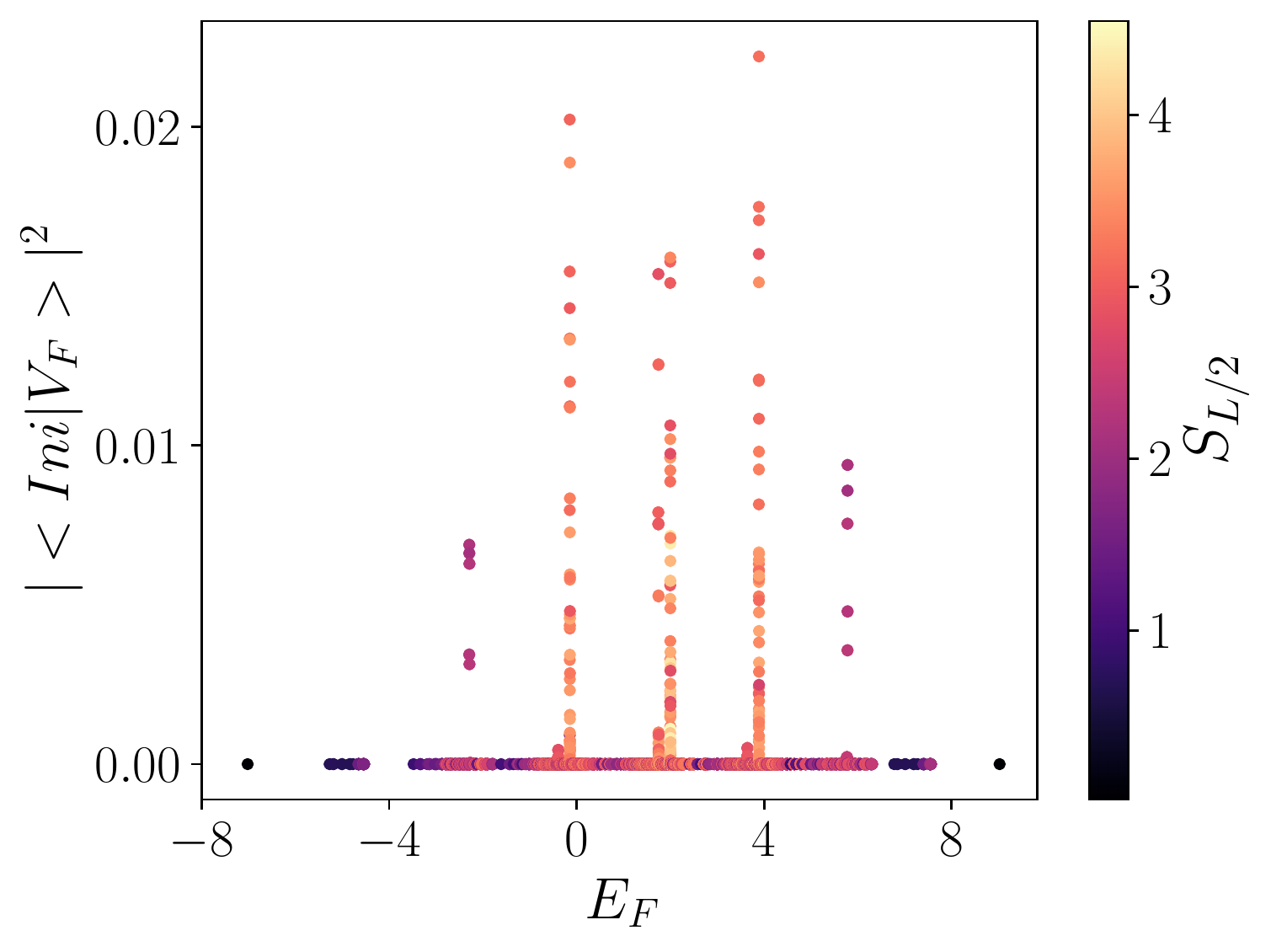}}{\large(c)}
\stackunder[5pt]{\includegraphics[width=0.51\hsize]{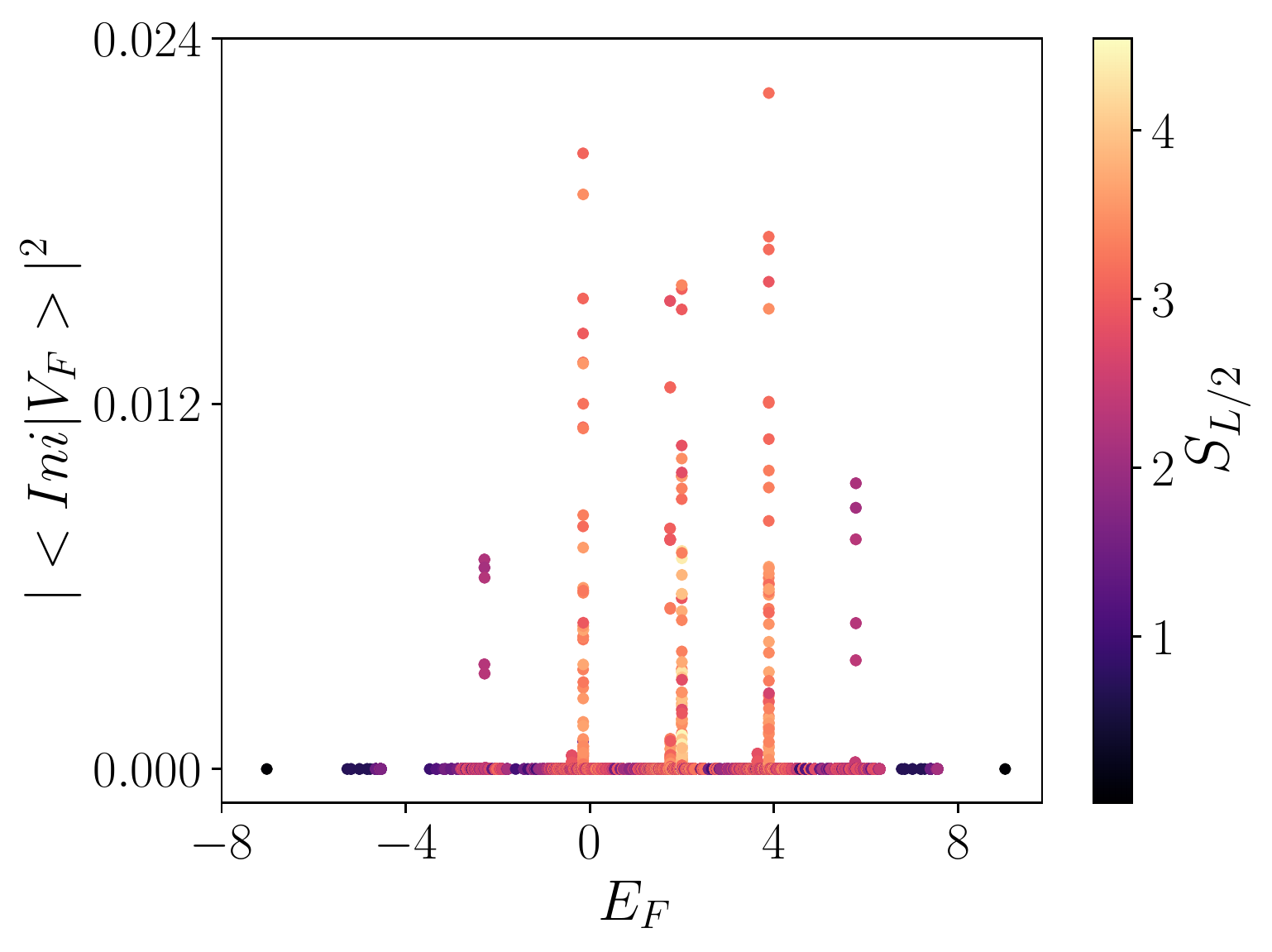}}{\large(d)}
\caption{{\bf Dynamics of the Loschmidt echo starting from an initial state and the overlap with Floquet eigenstates at a DL point for the period-4 model:} (a-b):
Variation of Loschmidt echo with time for two different initial states ($\ket{Ini}$), $\ket{210120101}$, and 
$\ket{21102110}$, respectively, for four DL points with $\mu=20$, $V=0.5$, and $\om=5,~6.67, ~10$ and $20$ (satisfying $\om = \mu /n$). In all these cases, the initial 
states show perfect revivals due to the presence of approximate conserved charges. (c-d): The overlaps of these two initial states with the Floquet 
eigenstates as a function of $E_{F}$, with the color bar indicating the variation of $S_{L/2}$. The initial states have significant amounts of overlap with a large number 
of Floquet eigenstates, and some of these eigenstates have extremely low entanglement.} \label{fig061} \end{figure}
\end{center}

To see a dynamical signature of slow thermalization, we study the time 
evolution of the Loschmidt echo precisely at a DL point with two non-trivial initial states, i.e., $\ket{1}$=$\ket{21012101}$ and 
$\ket{2}$=$\ket{21102110}$.
As shown in Fig.~\ref{fig061} (a), the Loschmidt echo exhibits long-time revivals, 
showing that the system shows very slow thermalization. To gain an
analytical understanding, we first derive an effective Hamiltonian for the state $\ket{210}$, $H_{eff}=\left(\begin{array}{cc}
V & J\\ J&0\end{array}\right)$. This Hamiltonian has the 
energy eigenvalues
$E_{\pm}=\frac{V}{2}\pm \sqrt{J^{2}+\frac{V^{2}}{4}}$. Taking this into account, 
we see that the initial state $\ket{1}$ would have the highest overlap with 
$4\times2^{4}=64$ such Floquet eigenstates with $E_{F}=4V$, $4V+4\sqrt{J^{2}+\frac{V^{2}}{4}}$, $4V-4\sqrt{J^{2}+\frac{V^{2}}{4}}$, $4V+\sqrt{J^{2}+
\frac{V^{2}}
{4}}$, $4V+2\sqrt{J^{2}+\frac{V^{2}}{4}}$, $4V-\sqrt{J^{2}+\frac{V^{2}}{4}}$, and $4V-2\sqrt{J^{2}+\frac{V^{2}}{4}}$. From the numerically obtained data for a system 
with $\mu=\om=20$, and $V=0.5$, we find that two Floquet eigenstates with $E_{F}=4V
= 2$ and $4V+2\sqrt{J^{2}+\frac{V^{2}}{4}} = 4.06$ have the highest overlaps with this particular initial state, which agrees quite well with the 
analytically 
predicted values. Within the approximation of these two highest overlapping states, the Loschmidt echo will have a time-dependence of the form be $|1+e^{i(E_{1}-
E_{2})t}|=2|\cos ((E_{1}-E_{2})t/2)|$ which oscillates with a period given by $\Delta t=2\pi/(E_{1}-E_{2})$. For the parameter values given above,
we find 
$\Delta t\sim 3$, which agrees with what we see in Fig.~\ref{fig061} (a). In a similar manner, we can find the period of oscillation for the other three 
DL points with $\om=5, ~6.67$ and $10$. For Fig.~\ref{fig061} (b), we choose the initial state 
$\ket{2}$ which has the highest overlaps with $64$ such Floquet eigenstates. As shown in Fig.~\ref{fig061} (b), we see long-time 
oscillating 
behaviors in the Loschmidt echo at the four DL points, which again indicates that the system will not thermalize for a long time. In Figs.~\ref{fig061} 
(c) and (d), we plot the overlaps of these two initial states, $\ket{1}$ and $\ket{2}$, with all the Floquet eigenstates at a DL point with 
$\mu=\om=20$, where the color bar shows the variation of the entanglement entropy of the Floquet eigenstates. In both cases, we observe that these two 
initial states have overlaps with multiple Floquet eigenstates, some of them having extremely low entanglement entropy which possibly causes the long-
time persistent oscillations in the Loschmidt echo.

\begin{figure}[!tbp]
\footnotesize
\begin{center}
\stackunder[5pt]{\includegraphics[width=0.47\hsize]{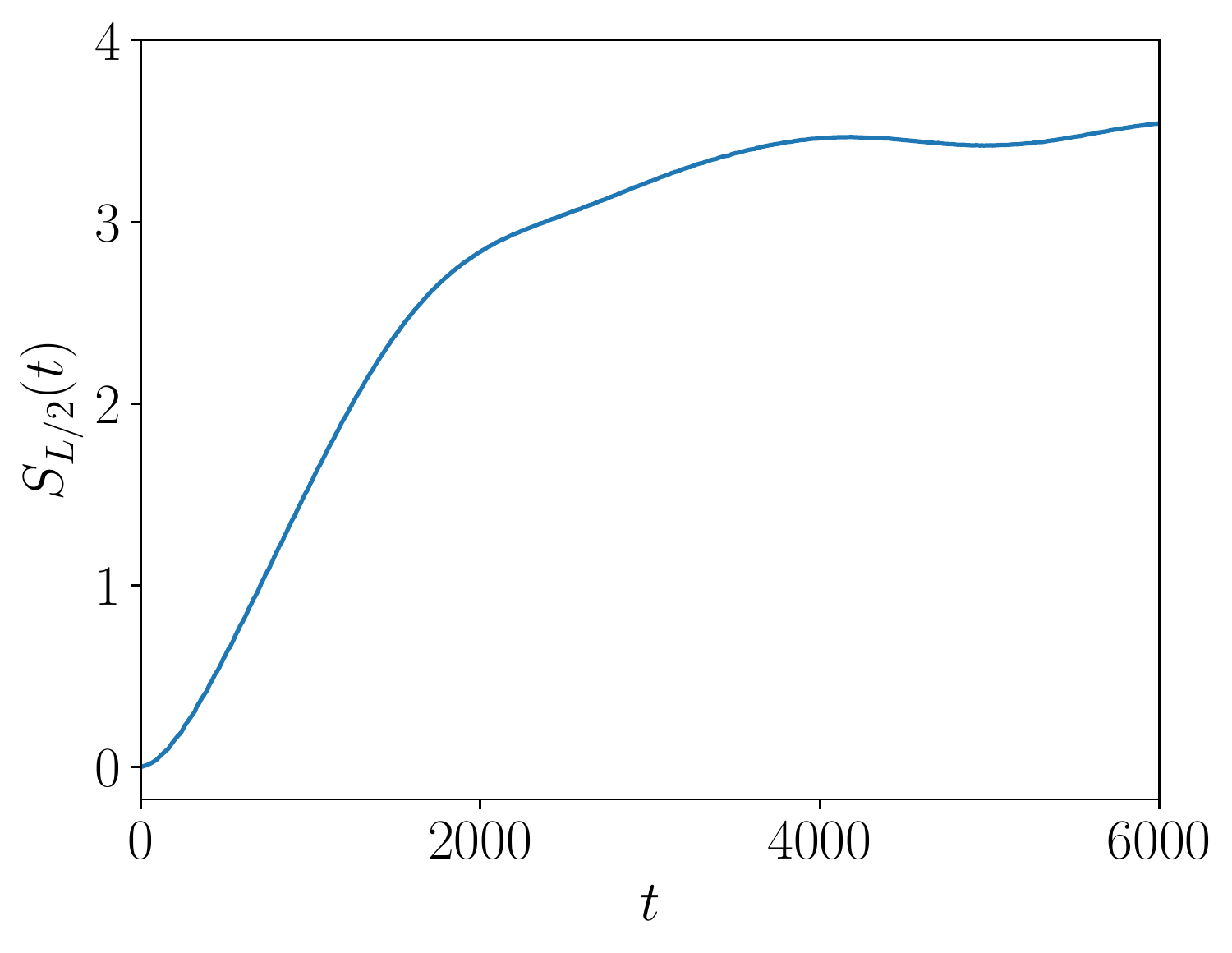}}{\large(a)}
\stackunder[5pt]{\includegraphics[width=0.45\hsize]{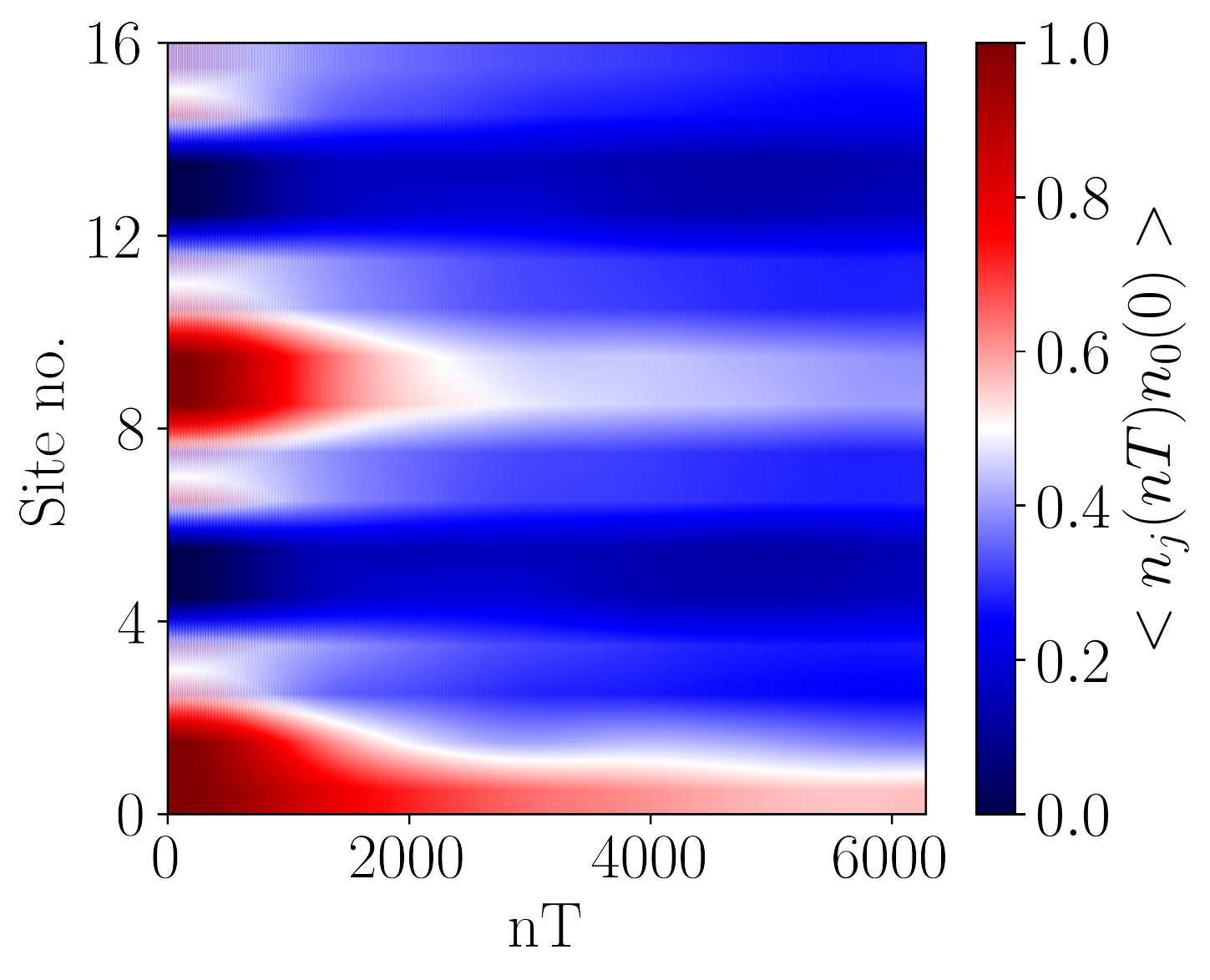}}{\large(b)}\\
\stackunder[5pt]{\includegraphics[width=0.47\hsize]{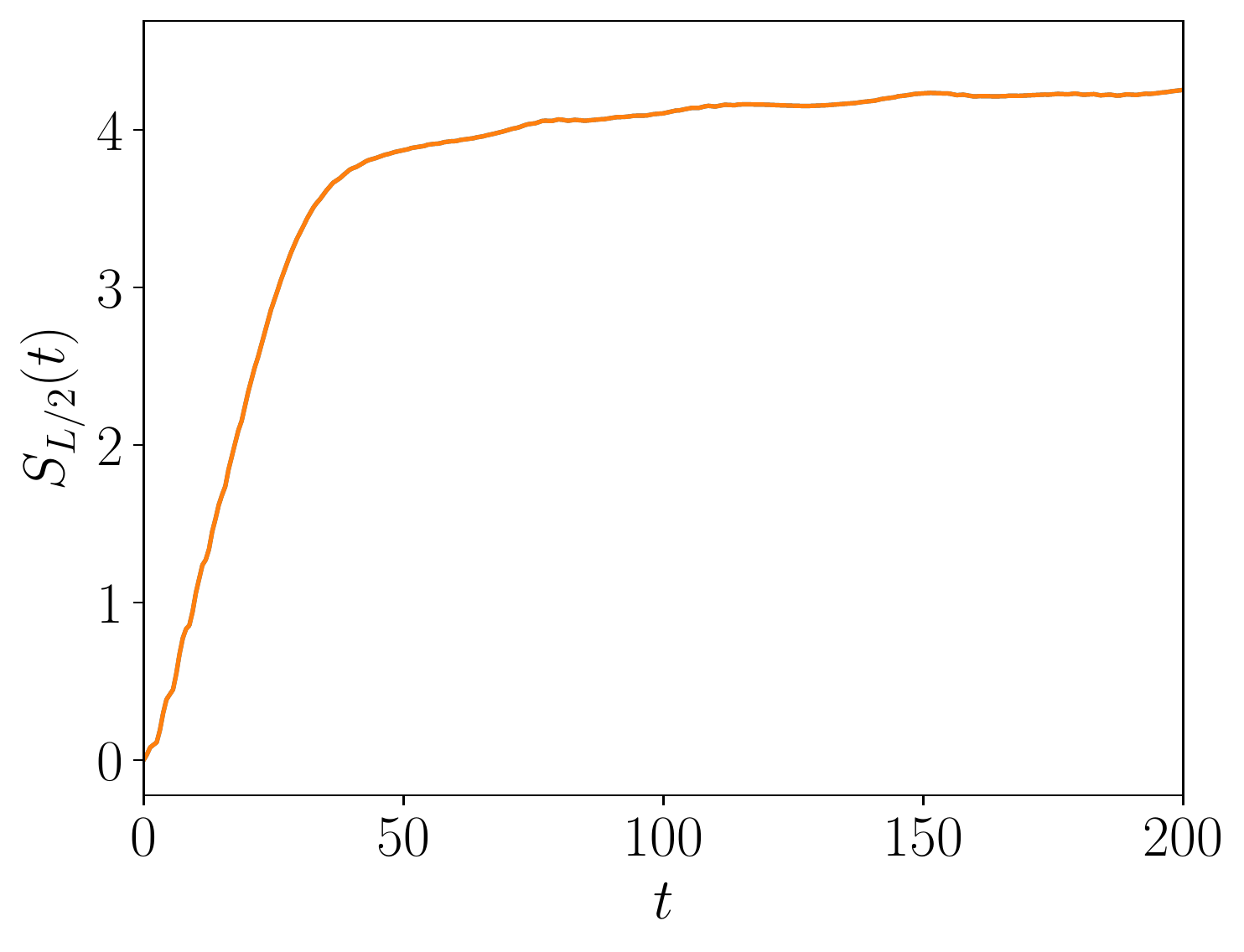}}{\large(c)}
\stackunder[5pt]{\includegraphics[width=0.45\hsize]{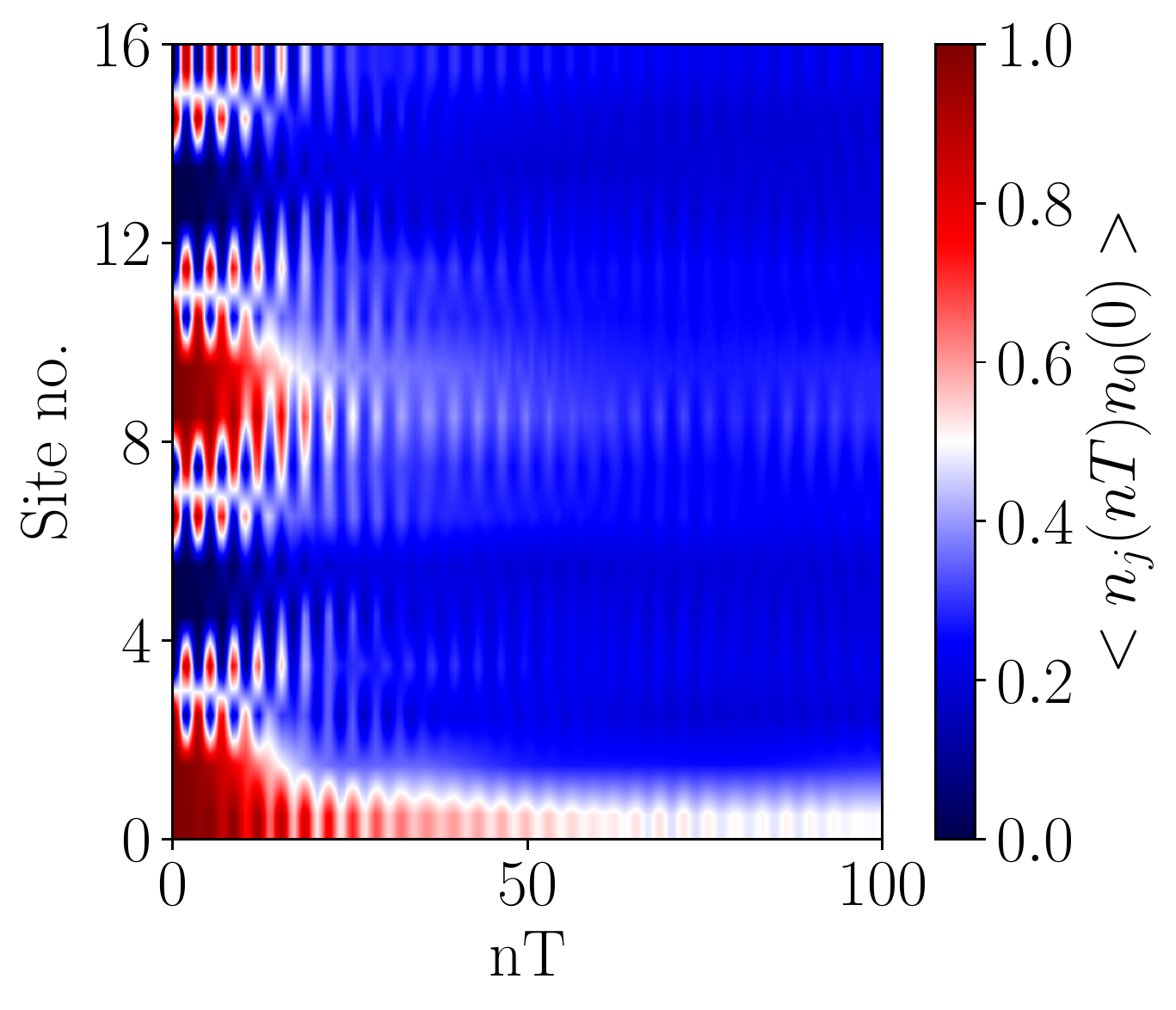}}{\large(d)}
\end{center}
\caption{{\bf Dynamics of entanglement entropy and correlation function at a 
DL point and away from a DL point for the period-4 model:} (a-b): Plots showing the dynamics of $S_{L/2}$ and the correlation function $\la 
n_{j}(nT)n_{0}(0) \ra$ for the initial state $\ket{21012101}$ at a DL point 
with $J=1$, $\mu=20$, $\om=10$, and $V=0.5$. (c-d): Same plots away from a DL point with $J=1$, $\mu=20$, $\om=11$, and $V=0.5$. (a): At a DL point,
we see that the 
entanglement entropy increases extremely slowly before reaching a saturation value which is less than the thermal value. (b): The correlation 
function shows a behavior similar to $S_{L/2}$ with a long-time revival pattern. (c): Away from a DL point, $S_{L/2}$ reaches a saturation 
value soon after an initial growth in time. (d): The correlation function
demonstrates a similar behavior, suggesting thermalizing behavior 
away from a DL point. Note that the time scales in 
(c-d) are much shorter than in (a-b)} \label{fig062} \end{figure}

To further confirm the above findings, we investigate the dynamics of $S_{L/2}$ and the 
unequal-time two-point density-density correlation function with the initial 
state taken to be $\ket{21012101}$. As shown in 
Figs.~\ref{fig062} (a), we see that the entanglement entropy slowly increases before reaching a saturation value for $\mu=2\om=20$ and $V=0.5$, as expected. Further, the 
saturation value ($\sim 3.5$) is much less than $S_{page} \sim 
5.1$ signaling a deviation from a volume law behavior. In Fig.~\ref{fig062} (b), we examine the two-point correlation function with the same initial 
state and see a behavior similar to $S_{L/2}$. Thus, the dynamics also confirms
the behaviors predicted from the static signatures. In Figs.~\ref{fig062} (c) and \ref{fig062} (d), we repeat the same analysis for the system away from a DL point, namely, for $\mu=20$, $\om=11$, and $V=0.5$. As opposed to 
Fig.~\ref{fig062} (a), we see in Fig.~\ref{fig062} (c) that the entanglement entropy saturates quite quickly. In Fig.~\ref{fig062} (d), the 
two-point correlation function exhibits a similar behavior with the saturation value of 
$\la n_{j}\ra^{2}=1/4$ at half-filling, which suggests that thermalization has occurred.

\begin{figure}[!tbp]
\footnotesize
\begin{center}
\stackunder[5pt]{\includegraphics[width=0.45\hsize]{fig5a.pdf}}{\large(a)}
\stackunder[5pt]{\includegraphics[width=0.45\hsize]{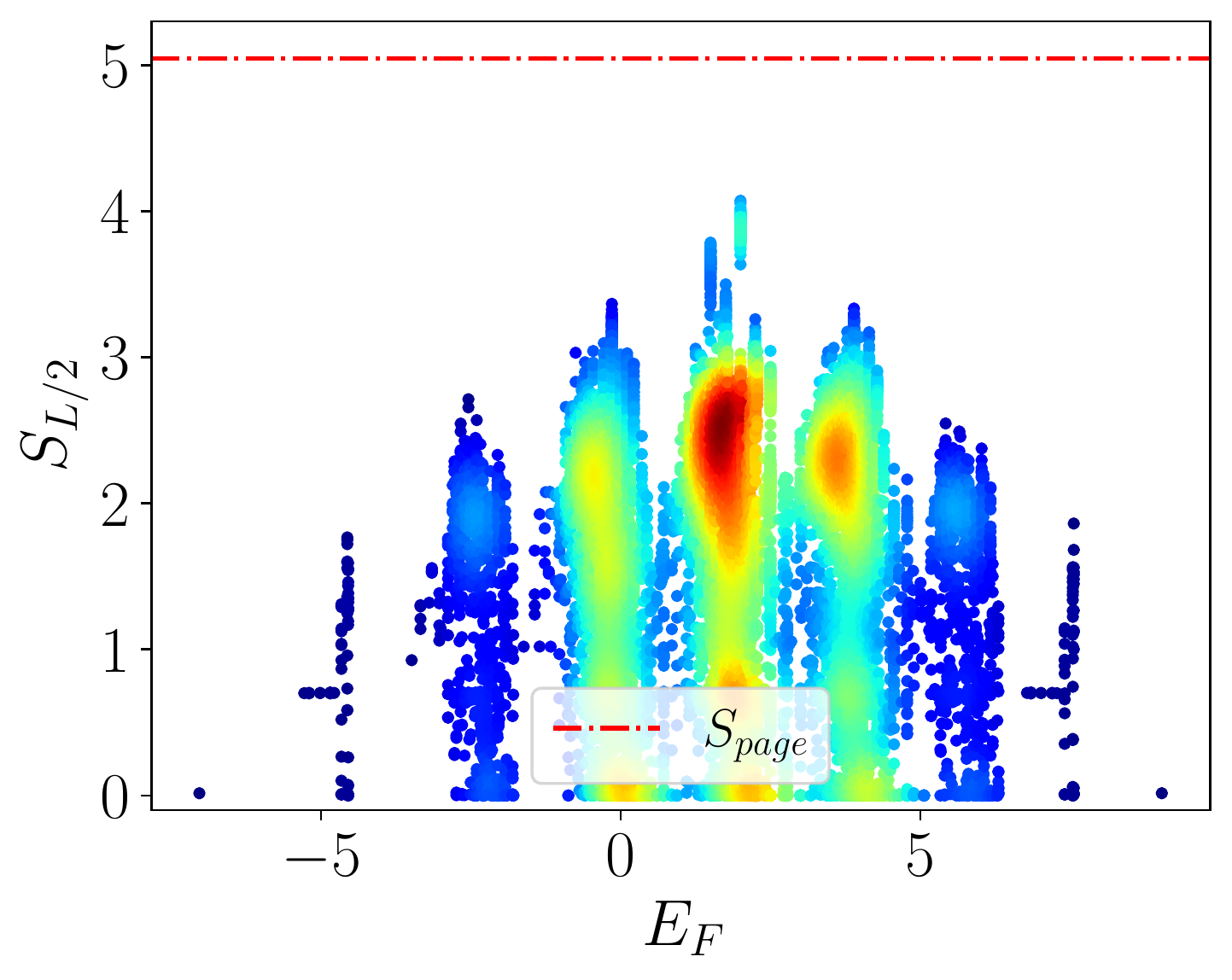}}{\large(b)}\\
\stackunder[5pt]{\includegraphics[width=0.45\hsize]{fig5c.pdf}}{\large(c)}
\stackunder[5pt]{\includegraphics[width=0.45\hsize]{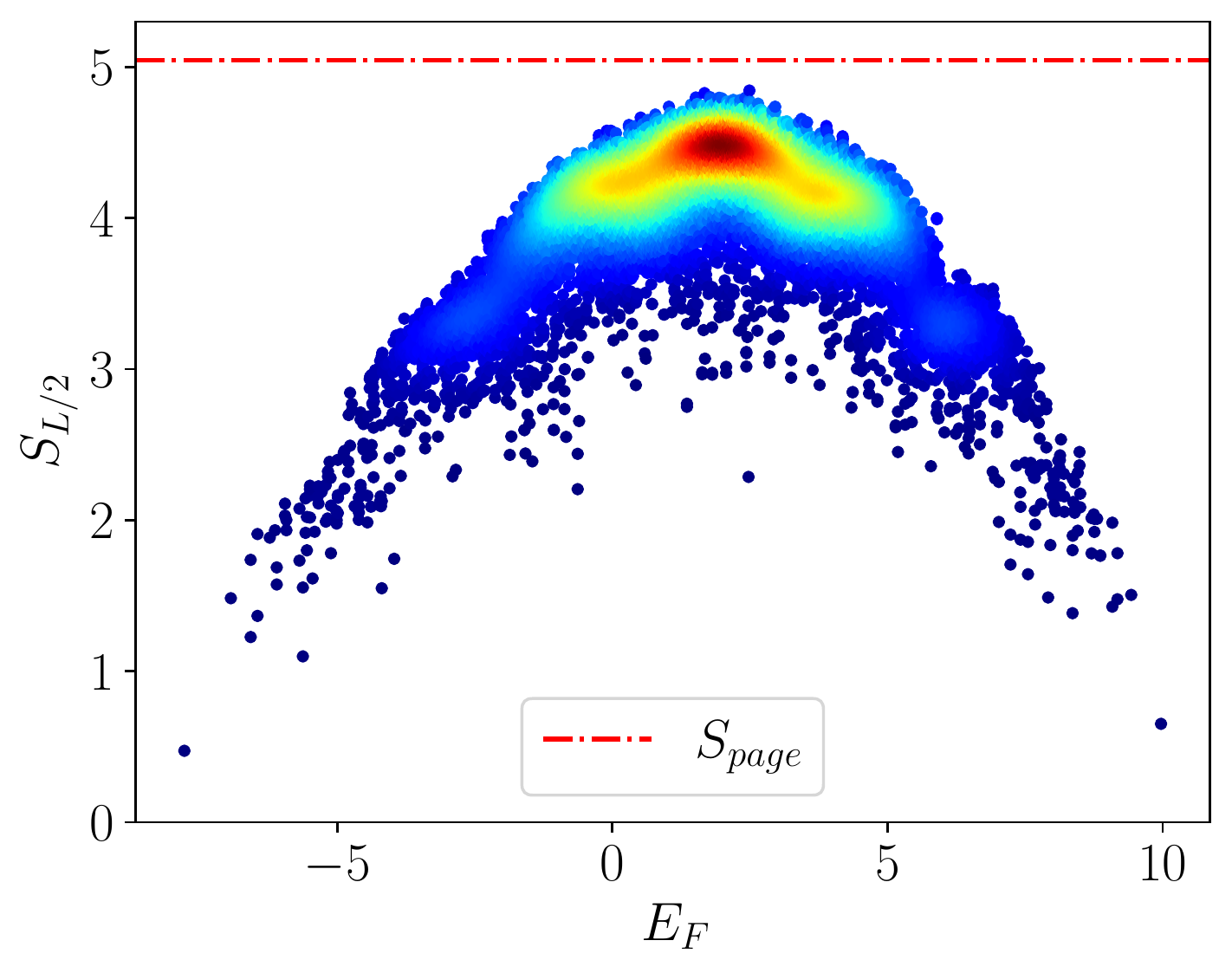}}{\large(d)}%
\end{center}
\caption{{\bf Comparison of the exact numerical results with the 
first-order FPT:} Entanglement entropy $S_{L/2}$ obtained from (a) exact numerical computations and (b) first-order FPT, for a system with $J=1$, $\mu=\om=20$, and $V=0.5$ 
(a DL point). The quasienergies agree quite well; however, $S_{L/2}$ turns out to be generally smaller from the first-order FPT in comparison to the exact numerically obtained 
values. (c-d) show the same plots as in (a-b) but for a system with $J=1$, $\mu=10$, $\om=20$, and $V=0.5$ (away from a DL point). For these parameter values, both the 
quasienergies and the entanglement entropy obtained from the exact numerical computation agree almost perfectly with the first-order FPT results.} \label{fig80} \end{figure}

Finally, we compare the results obtained from the exact numerics and the 
first-order FPT calculation for $\mu \gg J,~V$. In Figs.~\ref{fig80} (a) and \ref{fig80} 
(b), we 
see that the quasienergies obtained from exact numerics and from first-order FPT for $J=1$, $\mu=\om=20$ (lying at a DL point), and $V=0.5$ agree very well with each 
other. However, the results for the entanglement entropy do not agree that well. We 
believe that this disagreement is due to the corrections in FPT which are higher than
first-order. The correction terms have a relatively small effect on the Floquet quasienergies but have a noticeable effect on the Floquet eigenfunctions. Consequently, the entanglement 
entropy deviates significantly from the exact 
numerical values. Nevertheless, we note that the qualitative behavior of 
the entanglement entropy is the same in the two cases. In Figs.~\ref{fig80} (c) and \ref{fig80} (d), we compare the same plots as in
Figs.~\ref{fig80} (a) and \ref{fig80} (b) but for $J=1$, $\mu=10$, $\om=20$, and $V=0.25$,
which is away from a DL point. Since the first-order term dominates over higher order corrections away
from a DL point, we see that the Floquet quasienergy and entanglement entropy agree with each other almost identically.

\subsubsection{Effects of staggered on-site potential}

\begin{figure}[!tbp]
\footnotesize
\begin{center}
\stackunder[5pt]{\includegraphics[width=0.47\hsize]{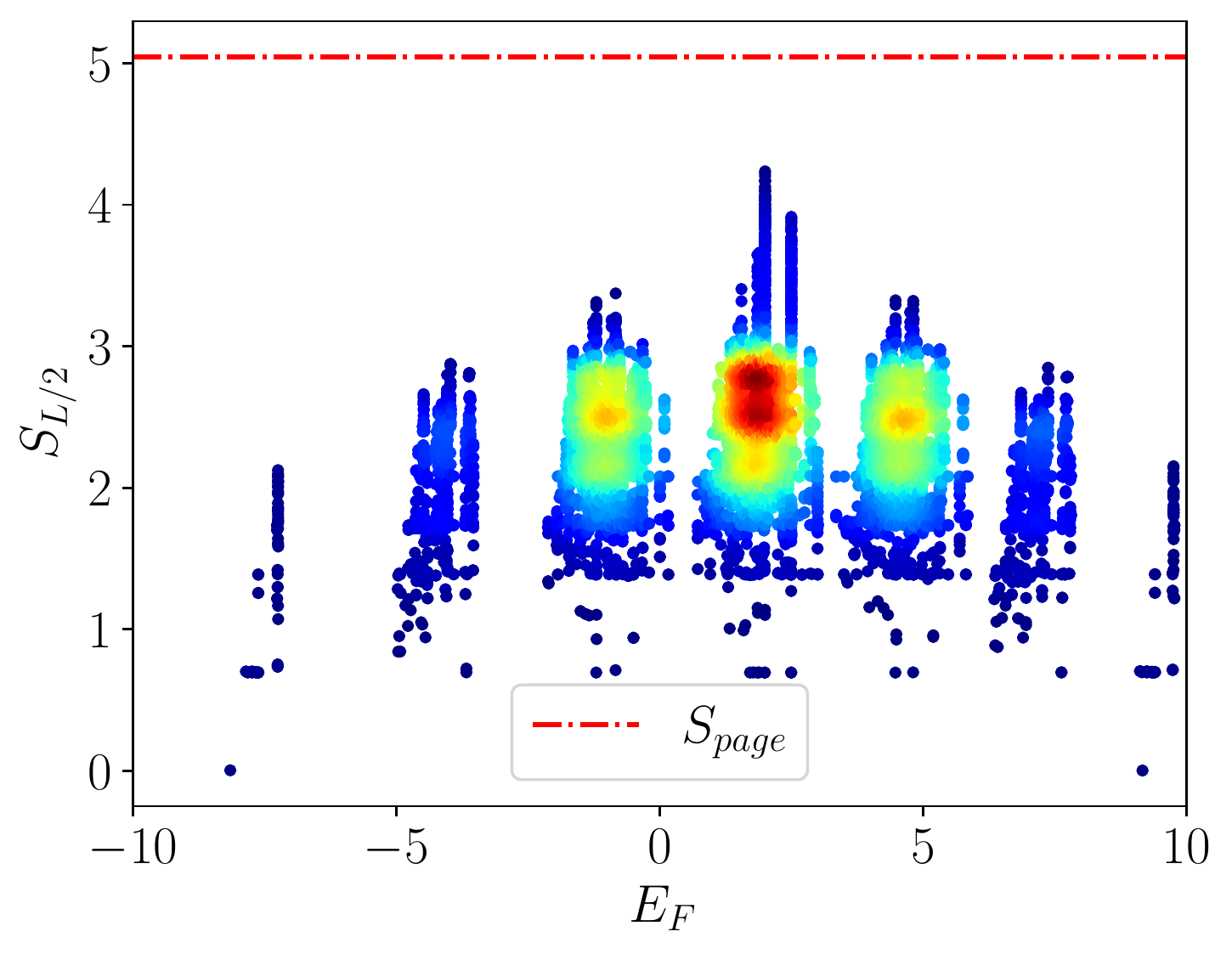}}{\large(a)}
\stackunder[5pt]{\includegraphics[width=0.47\hsize]{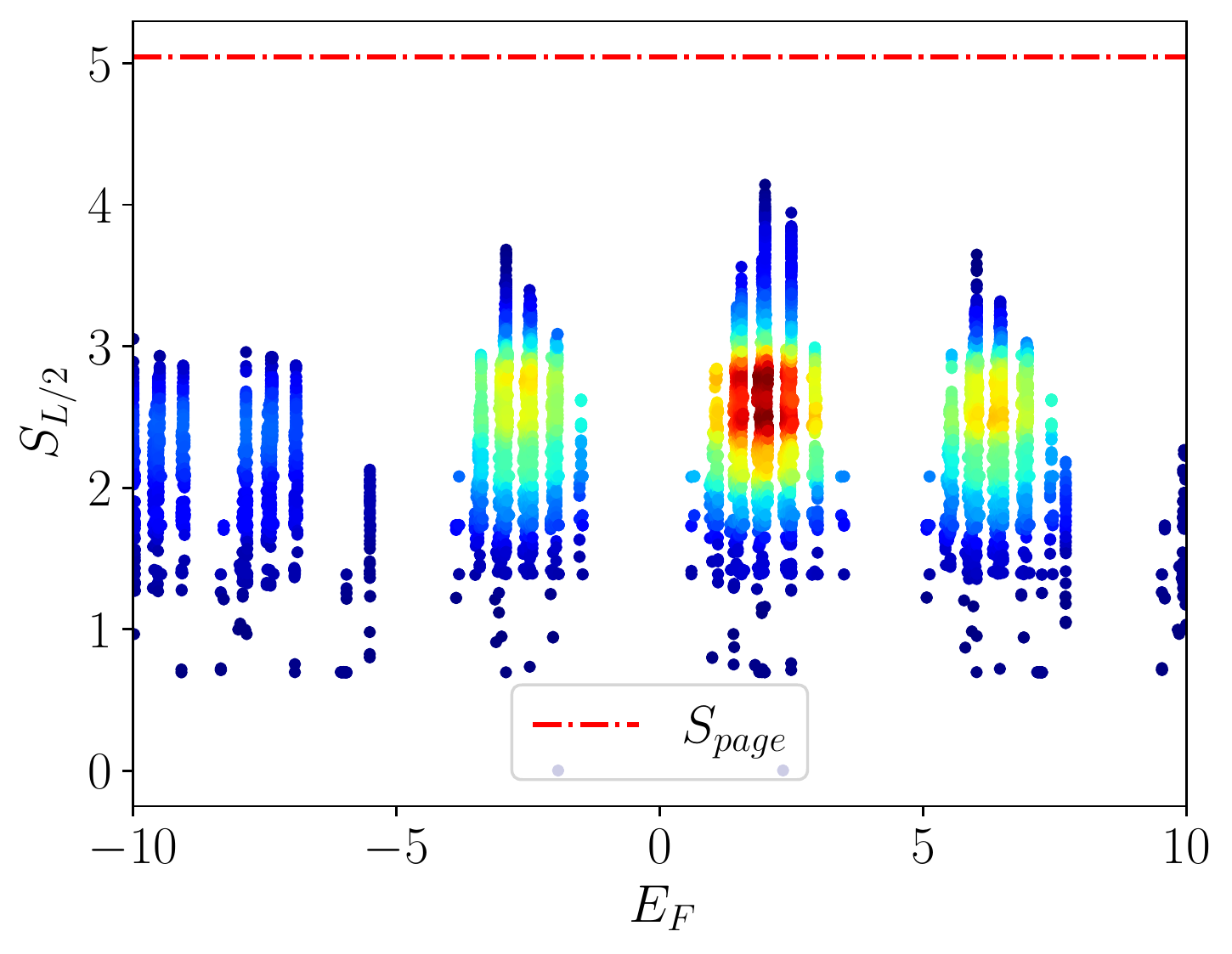}}{\large(b)}
\end{center}
\caption{{\bf Entanglement entropy spectrum with a staggered on-site potential at a DL point:} Plots of $S_{L/2}$ versus $E_{F}$ at a DL point with $J=1$, $\mu= 
\om=20$, $V=0.5$, and (a) $w=1$ and (b) $w=3$. In both cases we see many finger-like structures with multiple secondary fragments, with the effect becoming
clearer with increasing strength of the staggered potential $w$. 
The color intensity indicates the density of states, revealing that the majority of states demonstrate athermal behavior with $S_{L/2}$ being much smaller than the 
thermal value.} \label{fig07} \end{figure}

\begin{figure}[!tbp]
\begin{center}
\includegraphics[width=0.87\hsize]{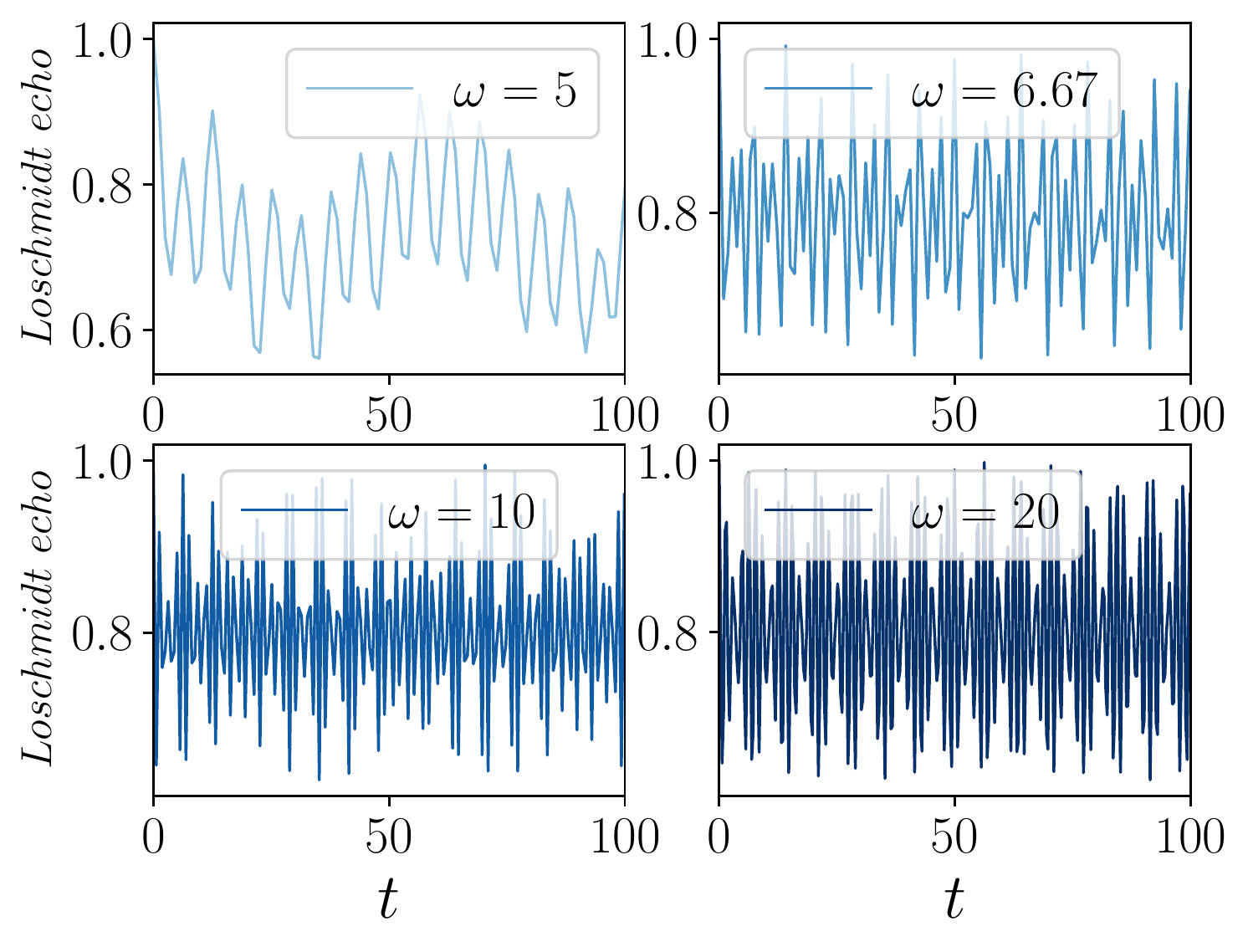}
\end{center}
\caption{{\bf Dynamics of Loschmidt echo in the presence of a staggered potential:} Plots showing the dynamics of Loschmidt echo for the initial state $\ket{21012101}$ 
at four DL points with $J=1$, $V=0.5$, $w=3$, $\mu=20$, and $\om=5, ~6.67, ~10$ and $20$. In all four cases, the Loschmidt echos show an oscillatory behavior with an extremely 
slow decay, suggesting that the system retains the information of the initial state for a long period of time.}\label{fig10} \end{figure}

It is interesting to incorporate the effects of a staggered on-site potential with amplitude $w$ in the period-4 model with $\phi=7\pi/4$ since such a 
potential commutes with the unperturbed Hamiltonian $H_{0}$. Hence, in the presence of such a potential, the first-order FPT Hamiltonian becomes 
$H_{F,\rm{stagg}}^{(1)}=w\sum_{j=1}^{L}(-1)^{j}n_{j}$.
This term can also be incorporated within the effective spin model. Assuming that
all the unit cells are singly occupied (so that there is no boundary field), we obtain
\bea H ~=~ \sum_{j=1}^{L/2} ~\left[\sigma_{j}^{x} ~+~ \frac{V}{4}(1-\sigma_{j}^{z}\sigma_{j+1}^{z}) ~+~ w\sigma_{j}^{z}\right] \label{stagg}. \eea 
(In general, the $\sigma^{z}$ term only appears for unit cells with single occupation, and has no effect on unit cells with $n_{j}=0$ or 2). The other 
sectors where only some of the unit cells have single occupation get modified 
in a similar way. In Fig.~\ref{fig07}, we show the variation of the 
entanglement entropy 
$S_{L/2}$ versus $E_{F}$ for a system with $J=1$, $\mu=\om=20$, $V=0.5$, and (a) $w=1$ and (b) $w=3$. In both cases the spectrum contains multiple finger-like 
primary structures with further secondary fragments~\cite{staggdetu}, and the secondary fragments become more prominent with increasing value of the staggered 
potential. In Fig.~\ref{fig10}, the dynamics of the Loschmidt echo is shown for a system with $w=3$, and $\om=5, ~6.67, ~10$ and $20$, respectively. For all 
four DL points, the Loschmidt echo for the initial state $\ket{21012101}$ 
exhibits an oscillatory behavior for a long period of time, which implies 
that the system thermalizes very slowly.

\subsubsection{Effects of resonances}

We have so far discussed cases with $V \ll \mu$ where the system shows very slow thermalization at a DL point. In this section, we will address the effects of 
resonances at two DL points, i.e., $\mu=\omega=V$ and $\mu=2\omega=V$. For both cases, we note that 
$\mu,~V \gg J$, which is the opposite limit to the previously examined cases, and 
we want to investigate whether this limit can give rise to non-ergodic behavior 
similar to the period-2 model. To do so, we will first derive an effective 
Hamiltonian based on the first-order FPT. Similar to the period-2 case, we will first identify the non-trivial processes for a system with
four sites. Due to the periodic pattern of the on-site potential, we need 
to consider a total of sixteen 
independent processes to construct the time-dependent effective Hamiltonian.
These are shown in Table 6.

\vspace*{.4cm}
\begin{table}[h!]
\begin{center}
\begin{tabular}{|c|c|c|c|} 
\hline
Pattern of periodic potential & Process & Effective time-dependent Hamiltonian\\
\hline
+~+~-~-&1100 $\leftrightarrow$ 1010 & $
H(t)=\left(\begin{array}{cc}
2\mu(t)+V & J\\ J&0\end{array}\right)$\\
\hline
+~+~-~-&0100 $\leftrightarrow$ 0010 & $
H(t)=\left(\begin{array}{cc}
\mu(t) & J\\ J&-\mu(t)\end{array}\right)$\\
\hline
+~+~-~-&0101 $\leftrightarrow$ 0011 & $
H(t)=\left(\begin{array}{cc}
0 & J\\ J&-2\mu(t)+V\end{array}\right)$\\
\hline
+~+~-~-&1101 $\leftrightarrow$ 1011 & $
H(t)=\left(\begin{array}{cc}
V+\mu(t)& J\\ J&V-\mu(t)\end{array}\right)$\\
\hline
+~-~-~+&1100 $\leftrightarrow$ 1010 & $
H(t)=\left(\begin{array}{cc}
V & J\\ J&0\end{array}\right)$\\
\hline
+~-~-~+&0100 $\leftrightarrow$ 0010 & $
H(t)=\left(\begin{array}{cc}
-\mu(t) & J\\ J&-\mu(t)\end{array}\right)$\\
\hline
+~-~-~+&0101 $\leftrightarrow$ 0011 & $
H(t)=\left(\begin{array}{cc}
0 & J\\ J&V\end{array}\right)$\\
\hline
+~-~-~+&1101 $\leftrightarrow$ 1011 & $H(t)=\left(\begin{array}{cc}
\mu(t)+V & J\\ J&V+\mu(t)\end{array}\right)$\\
\hline
-~-~+~+&1100 $\leftrightarrow$ 1010 & $
H(t)=\left(\begin{array}{cc}
-2\mu(t)+V & J\\ J&0\end{array}\right)$\\
\hline
-~-~+~+&0100 $\leftrightarrow$ 0010 & $
H(t)=\left(\begin{array}{cc}
-\mu(t) & J\\ J&\mu(t)\end{array}\right)$\\
\hline
-~-~+~+&0101 $\leftrightarrow$ 0011 & $
H(t)=\left(\begin{array}{cc}
0 & J\\ J&2\mu(t)+V\end{array}\right)$\\
\hline
-~-~+~+&1101 $\leftrightarrow$ 1011 & $
H(t)=\left(\begin{array}{cc}
V-\mu(t)& J\\ J&V+\mu(t)\end{array}\right)$\\
\hline
-~+~+~-&1100 $\leftrightarrow$ 1010 & $
H(t)=\left(\begin{array}{cc}
V & J\\ J&0\end{array}\right)$\\
\hline
-~+~+~-&0100 $\leftrightarrow$ 0010 & $
H(t)=\left(\begin{array}{cc}
\mu(t) & J\\ J&\mu(t)\end{array}\right)$\\
\hline
-~+~+~-&0101 $\leftrightarrow$ 0011 & $
H(t)=\left(\begin{array}{cc}
0 & J\\ J&V\end{array}\right)$\\
\hline
-~+~+~-&1101 $\leftrightarrow$ 1011 & $H(t)=\left(\begin{array}{cc}
-\mu(t)+V & J\\ J&V-\mu(t)\end{array}\right)$\\
\hline
\end{tabular}
\end{center}
\caption{Allowed processes and their corresponding effective time-dependent Hamiltonians for a four-site system with all possible patterns of the
periodic potential in a period-4 model in which both resonances and dynamical localization are present.}
\end{table}
\vspace*{.4cm}

As an example, we will derive the effective time-independent first-order FPT Hamiltonian for the first process shown in Table 6. In this case, the effective 
time-dependent Hamiltonian can be written as
\bea H(t)&=&(\mu(t)+V/2) ~I ~+~ (\mu+V/2) ~\sigma^{z} ~+~ J ~\sigma^{x},\non\\
H_{0}&=&(\mu+V/2) ~\sigma^{z},~~~~ H_{1} ~=~ J ~\sigma^{x}, \eea
where $H_{0}$ and $H_{1}$ are the unperturbed Hamiltonian and the perturbation, respectively, and we will assume that $\mu,V \gg J$. The instantaneous eigenvalues of $H_{0}$ 
are $E_{\pm}=\pm(\mu(t)+V/2)$. The eigenfunctions corresponding to $E_{k}^{\pm}$ are given by $\ket{+}=\left(\begin{array}{cc}
1 \\ 0 \end{array}\right)$ and $\ket{-}=\left(\begin{array}{cc}
0 \\ 1 \end{array}\right)$.
These two eigenvalues again satisfy the condition given in Eq.~\eqref{per26}. Therefore, following the usual steps of degenerate FPT, we obtain
\bea \bra{+}H_{F}^{(1)}\ket{+}&=& 0,~~~~~\bra{-}H_{F}^{(1)}\ket{-}~=~ 0,\non\\
\bra{+}H_{F}^{(1)}\ket{-}&~=~&J ~I(\mu,V,T),~~~~~
\bra{-}H_{F}^{(1)}\ket{+}~=~J ~I^{*}(\mu,V,T),\non\\
I(\mu,V,\om)&=&\frac{e^{i(2\mu+V)T/4}\sin [(2\mu+V)T/4]}{(2\mu+V)T/2} ~+~ \frac{e^{i(2\mu+3V)T/4}\sin [(2\mu-V)T/4]}{(2\mu-V)T/2}. \non\\ \eea
Substituting $\mu=V=\omega$, $I(\mu,V,\om)$ turns out to be $4i/(3\pi)$. Thus, the effective Hamiltonian for a system consisting of four sites with this specific choice of periodic 
potential pattern becomes (setting $J=1$)
\bea H_{F}^{(1)}&=&\frac{4i}{3\pi} ~n_{0} ~c_{2}^{\dagger}c_{1} ~(1-n_{3}) ~+~ 
{\rm H.c.} \eea
Following the same procedure, we can compute the effective Hamiltonian for all the other processes.

\vspace{.4cm}
\begin{table}[h!]
\begin{center}
\begin{tabular}{|c|c|c|c|} 
\hline
Pattern of periodic potential & Process & First-order Floquet Hamiltonian\\
\hline
+~+~-~-&1100 $\leftrightarrow$ 1010 & $
H_{F}^{(1)}=\frac{4i}{3\pi}n_{0}c_{2}^{\dagger}c_{1}(1-n_{3})+ {\rm H.c.}$\\
\hline
+~+~-~-&0100 $\leftrightarrow$ 0010 & $
H_{F}^{(1)}=0$\\
\hline
+~+~-~-&0101 $\leftrightarrow$ 0011 & $
H_{F}^{(1)}=\frac{4i}{3\pi}(1-n_{0})c_{2}^{\dagger}c_{1}n_{3}+ {\rm H.c.}$\\
\hline
+~+~-~-&1101 $\leftrightarrow$ 1011 & $
H_{F}^{(1)}=0$\\
\hline
+~-~-~+&1100 $\leftrightarrow$ 1010 & $
H_{F}^{(1)}=0$\\
\hline
+~-~-~+&0100 $\leftrightarrow$ 0010 & $
H_{F}^{(1)}=(1-n_{0})c_{2}^{\dagger}c_{1}(1-n_{3})+ {\rm H.c}.$\\
\hline
+~-~-~+&0101 $\leftrightarrow$ 0011 & $
H_{F}^{(1)}=0$\\
\hline
+~-~-~+&1101$\leftrightarrow$1011 & $H_{F}^{(1)}=n_{0}c_{2}^{\dagger}c_{1}n_{3}+
{\rm H.c.}$\\
\hline
-~-~+~+&1100 $\leftrightarrow$ 1010 & $
H_{F}^{(1)}=\frac{4i}{3\pi}n_{0}c_{2}^{\dagger}c_{1}(1-n_{3})+ {\rm H.c.}$\\
\hline
-~-~+~+&0100 $\leftrightarrow$ 0010 & $H_{F}^{(1)}=0$\\
\hline
-~-~+~+&0101 $\leftrightarrow$ 0011 & $
H_{F}^{(1)}=-\frac{4i}{3\pi}(1-n_{0})c_{2}^{\dagger}c_{1}n_{3}+ {\rm H.c.}$\\
\hline
-~-~+~+&1101 $\leftrightarrow$ 1011 & $
H_{F}^{(1)}=0$\\
\hline
-~+~+~-&1100 $\leftrightarrow$ 1010 & $
H_{F}^{(1)}=0$\\
\hline
-~+~+~-&0100 $\leftrightarrow$ 0010 & $
H_{F}^{(1)}=(1-n_{0})c_{2}^{\dagger}c_{1}(1-n_{3})+ {\rm H.c.}$\\
\hline
-~+~+~-&0101 $\leftrightarrow$ 0011 & $
H_{F}^{(1)}=0$\\
\hline
-~+~+~-&1101 $\leftrightarrow$ 1011 & $H_{F}^{(1)}=n_{0}c_{2}^{\dagger}c_{1}n_{3}
+ {\rm H.c.}$\\
\hline
\end{tabular}
\end{center}
\caption{First-order effective FPT Hamiltonians corresponding to the allowed correlated processes for a four-site system with all possible patterns of periodic 
potential in the case of dynamical localization and resonance for a period-4 model.}
\end{table}
\vspace{.4cm}

\begin{center}
\begin{figure}[!tbp]
\footnotesize
\stackunder[5pt]{\includegraphics[width=0.47\hsize]{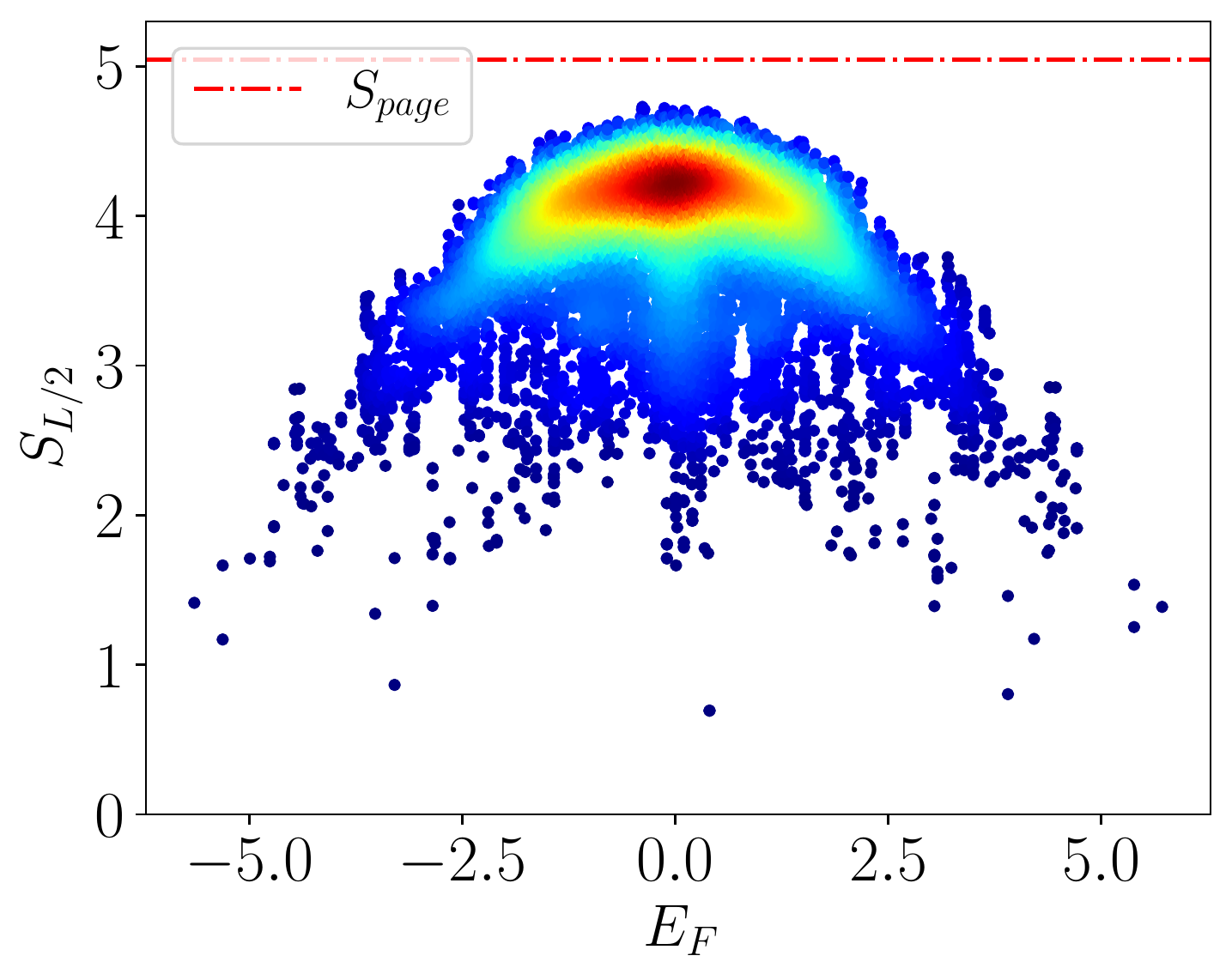}}{\large(a)}
\stackunder[5pt]{\includegraphics[width=0.47\hsize]{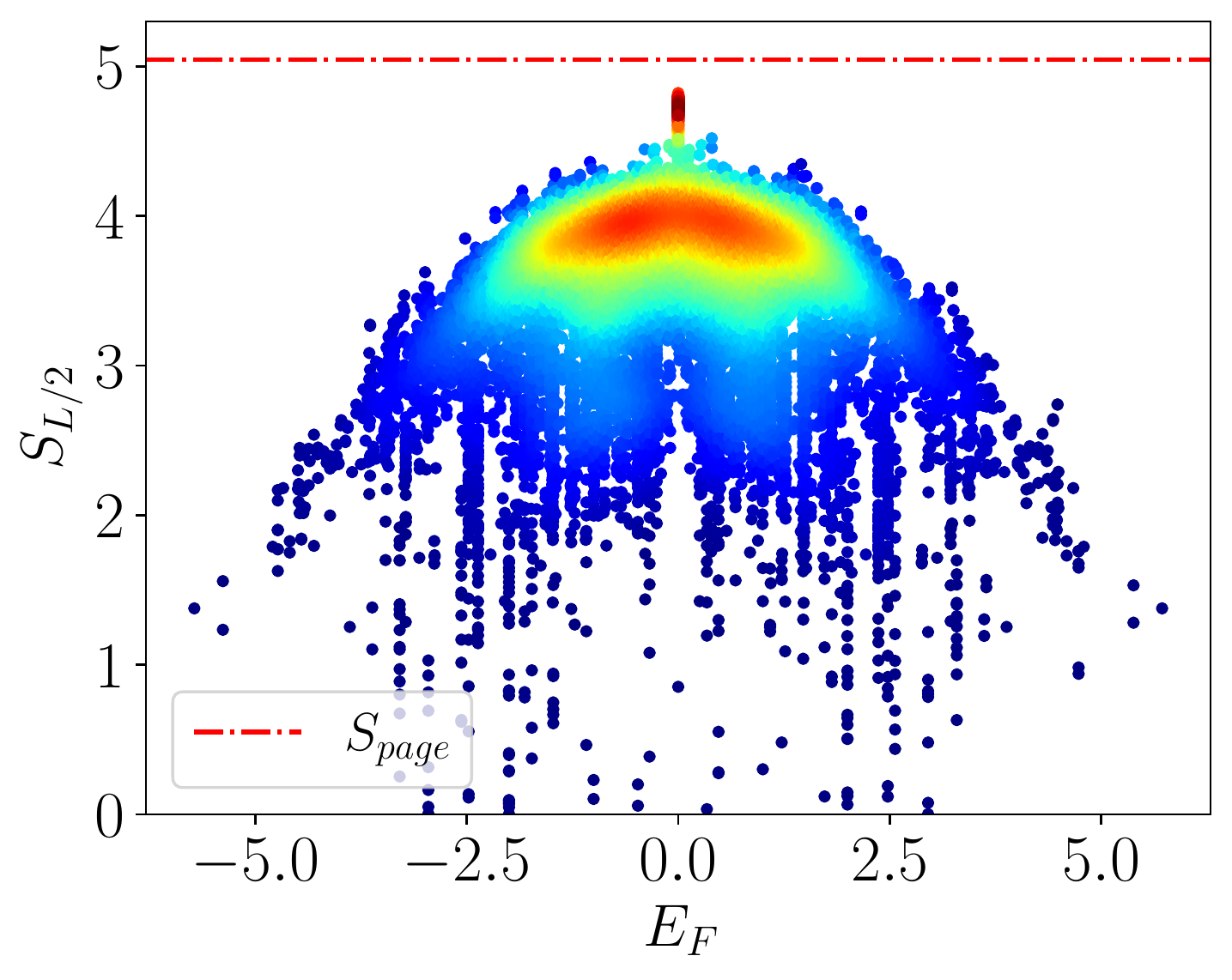}}{\large(b)}
\caption{{\bf Entanglement entropy spectrum for a resonant case at a DL point of the period-4 model:} Plots showing the entanglement entropy $S_{L/2}$ versus the 
quasienergy $E_{F}$ obtained from (a) an exact Floquet calculation and 
(b) the first-order FPT 
Hamiltonian, for a DL point exhibiting a resonance with $J=1$ and $\mu=\om=V=20$. The color intensity indicates the density of states, suggesting that most of the states attain the 
thermal value. However, there are also many low-entanglement states present near the middle of the spectrum.
The quasienergy spectrum obtained from FPT (b) agrees well with the exact numerically computed spectrum (a). However, the entanglement entropy obtained from the first-order 
FPT is much less than the exact numerically obtained values for many of 
the states.} \label{fig08} \end{figure}\end{center}

\begin{figure}[!tbp]
\footnotesize
\begin{center}
\stackunder[5pt]{\includegraphics[width=0.5\hsize]{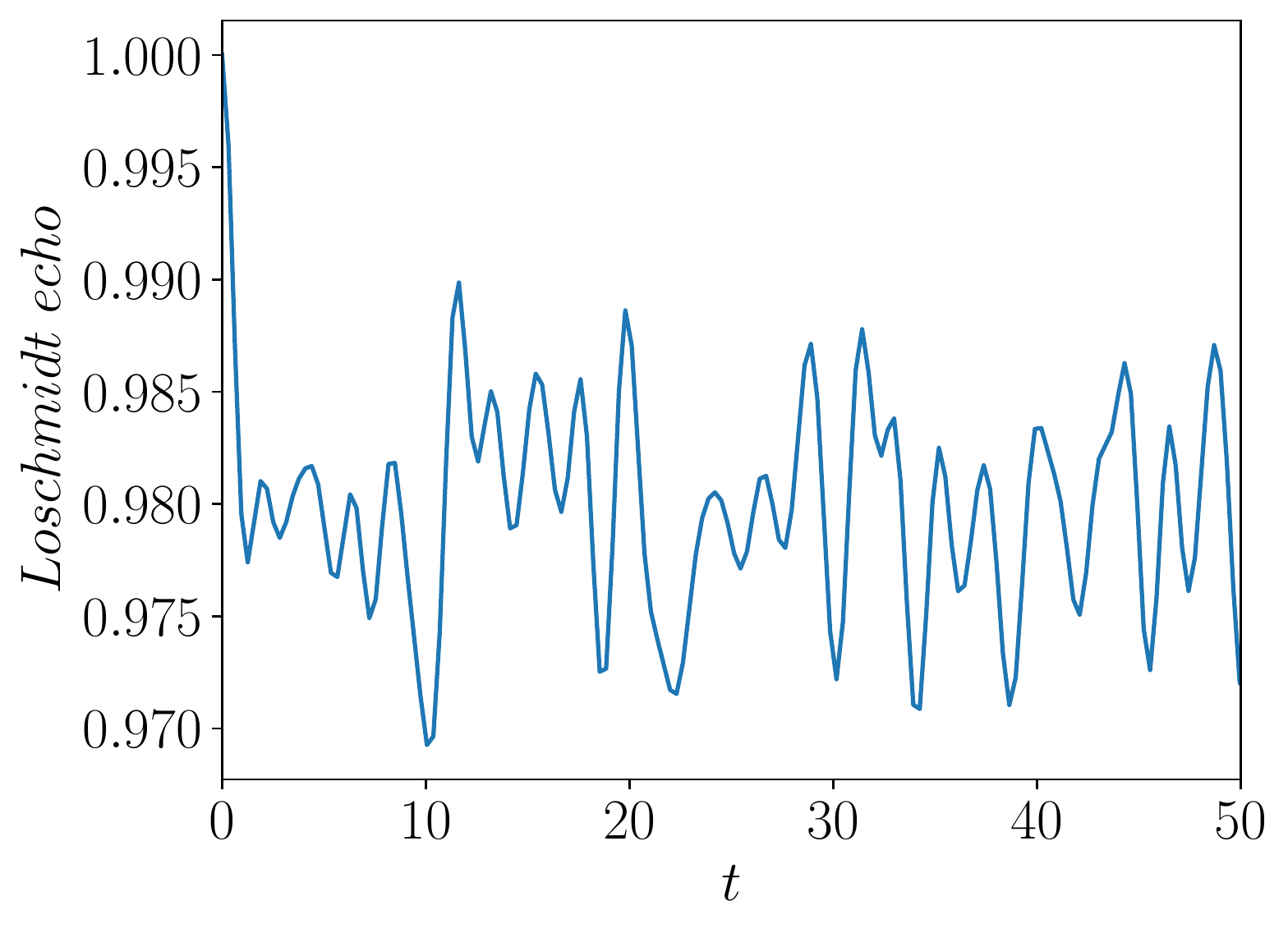}}{\large(a)}
\stackunder[5pt]{\includegraphics[width=0.5\hsize]{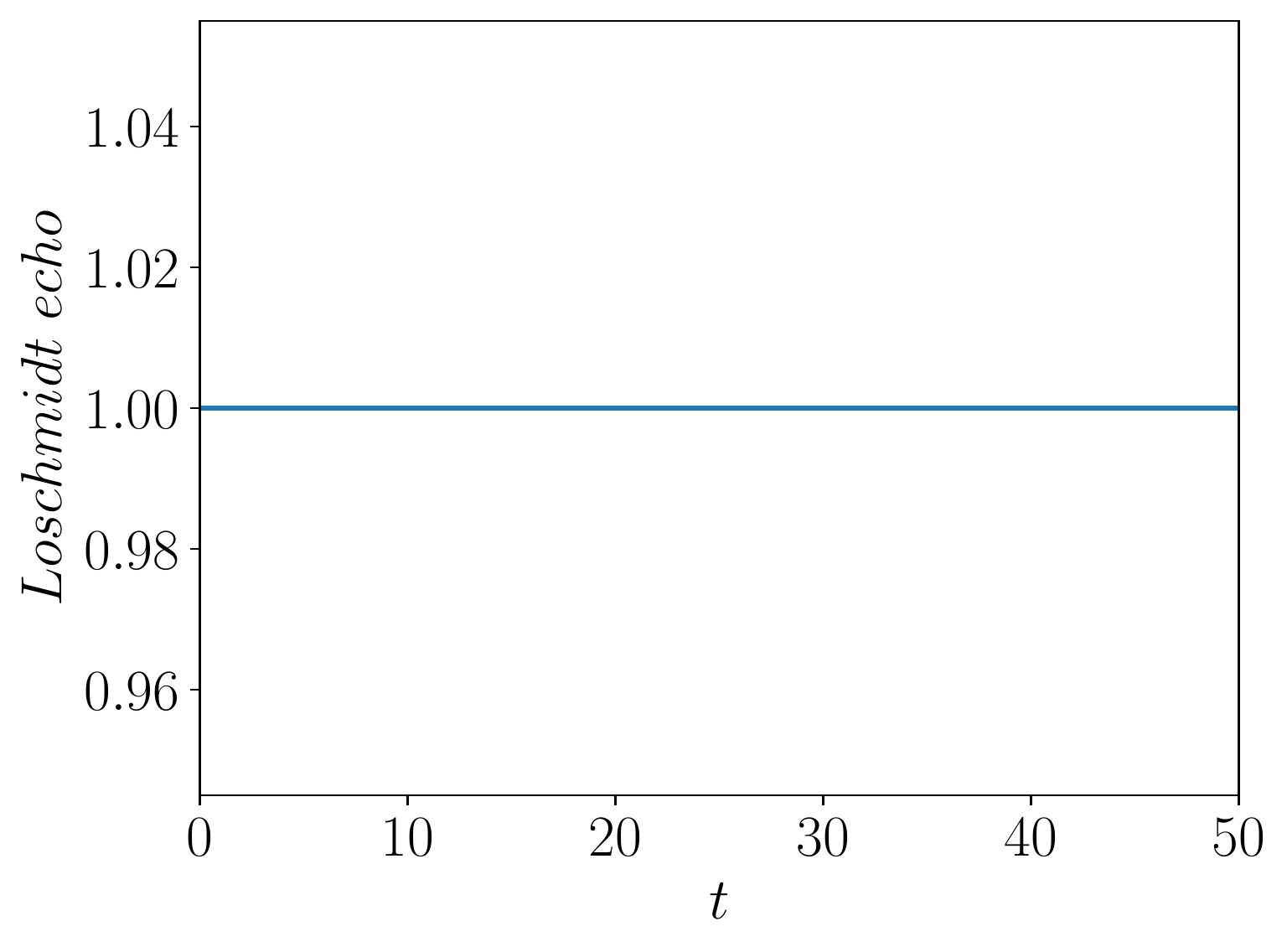}}{\large(b)}\\
\stackunder[5pt]{\includegraphics[width=0.6\hsize]{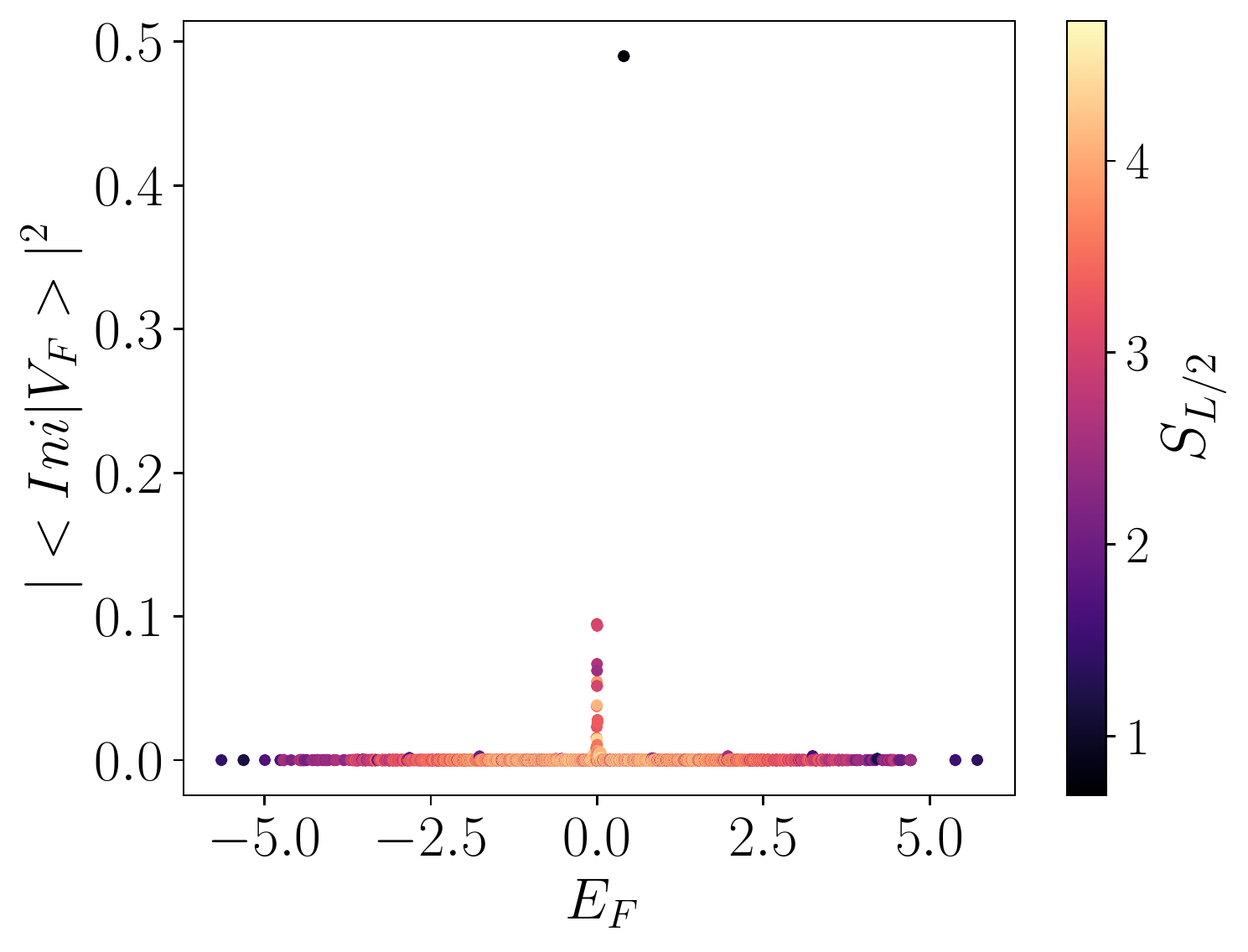}}{\large(c)}
\end{center}
\caption{{\bf Dynamics of Loschmidt echo and the overlap with the Floquet eigenstates for a resonant case at a DL point of the period-4 model:} (a) Dynamics of the
Loschmidt echo for an initial state ($\ket{Ini}$), $\ket{1100110011001100}$, obtained from (a)
an exact numerical calculation and (b) from the first-order effective 
Hamiltonian, for a system with $J=1$, $\mu=\om=20$, $V=20$, and $L=16$. In both cases, the Loschmidt echo is found to stay close to 1, showing that 
the system retains the memory of the initial state for a long period of time. (c) Overlap of the initial state with the Floquet eigenstates for the same 
parameter values with the color bar showing the variation of $S_{L/2}$. The Floquet eigenstates having the highest overlap with this initial state lie exactly 
in the middle of the Floquet eigenvalue spectrum and have extremely low entanglement.} \label{fig13} \end{figure}

These results enable us to deduce the complete first-order effective Hamiltonian for $\mu = \om = V$, namely,
\bea H&=&\frac{4i}{3\pi} ~\sum_{j=1}^{L/2} ~(-1)^{j} ~\left[(1-n_{2j})c_{2j+2}^{\dagger}~c_{2j+1}n_{2j+3}+n_{2j}c_{2j+2}^{\dagger}c_{2j+1}(1-n_{2j+3}) + {\rm H.c.} \right] \non \\
&& ~+ ~\sum_{j=1}^{L/2} ~\left[(1-n_{2j+1})~c_{2j+3}^{\dagger}c_{2j+2}(1 - n_{2j+4}) +n_{2j+1}c_{2j+3}^{\dagger}c_{2j+2}n_{2j+4} + {\rm H.c.}\right]. \label{ham60} \eea
The form of this Hamiltonian suggests that certain nearest-neighbor hoppings are forbidden, as elaborated below, when the DL and resonance condition are simultaneously 
satisfied; in principle this can lead to an anomalous thermalization behavior. In Fig.~\ref{fig08}, we show the variation of $S_{L/2}$ with $E_{F}$ obtained from exact numerical calculations and from
the first-order FPT analysis, for $J=1$, and $\mu=\om=V=20$. As anticipated, the middle of the entanglement spectrum consists of many low-entanglement states 
possibly due a fragmented nature of the Hilbert space as described below. For example, we can see that this effective Hamiltonian supports a simple fragment
which consists of only one state, $\ket{1100110011001100}$, and its 
translated partners. To construct this single fragment, we note the following constraints following from Eq.~\eqref{ham60}. The hopping on the
bonds $(2j+1,2j+2)$ is only 
possible if the neighboring sites have $(n_{2j},n_{2j+3})=(0,1)$ or $(1,0)$. 
However, the hopping on the bonds $(2j+2,2j+3)$ requires the neighboring sites 
to have $(n_{2j+1},n_{2j+4}) = (0,0)$ or $(1,1)$. These two constraints enables us to show that the above state forms a fragment on its own, and it does not 
mix with other states due to the action of the Hamiltonian. In Fig.~\ref{fig13}
(a) and (b), we study the dynamics of the Loschmidt echo for the initial state $\ket{1100110011001100}$ using the exact Floquet dynamics and the first-order
effective Hamiltonian, respectively. In both cases, we find that the Loschmidt echo stays very close to 1, with some small oscillations in (a). In
Fig.~\ref{fig13} (c), we see that these initial states have highest overlap 
with two mid-spectrum Floquet 
eigenstates with $S_{L/2}=\ln 2$. Since $S_{L/2}$ for this initial state is much smaller than $S_{page} (\simeq 5.1$ for $L=16)$, this state is likely to retain 
its initial memory for a long period of time.

\begin{figure}[!tbp]
\footnotesize
\begin{center}
\stackunder[5pt]{\includegraphics[width=0.5\hsize]{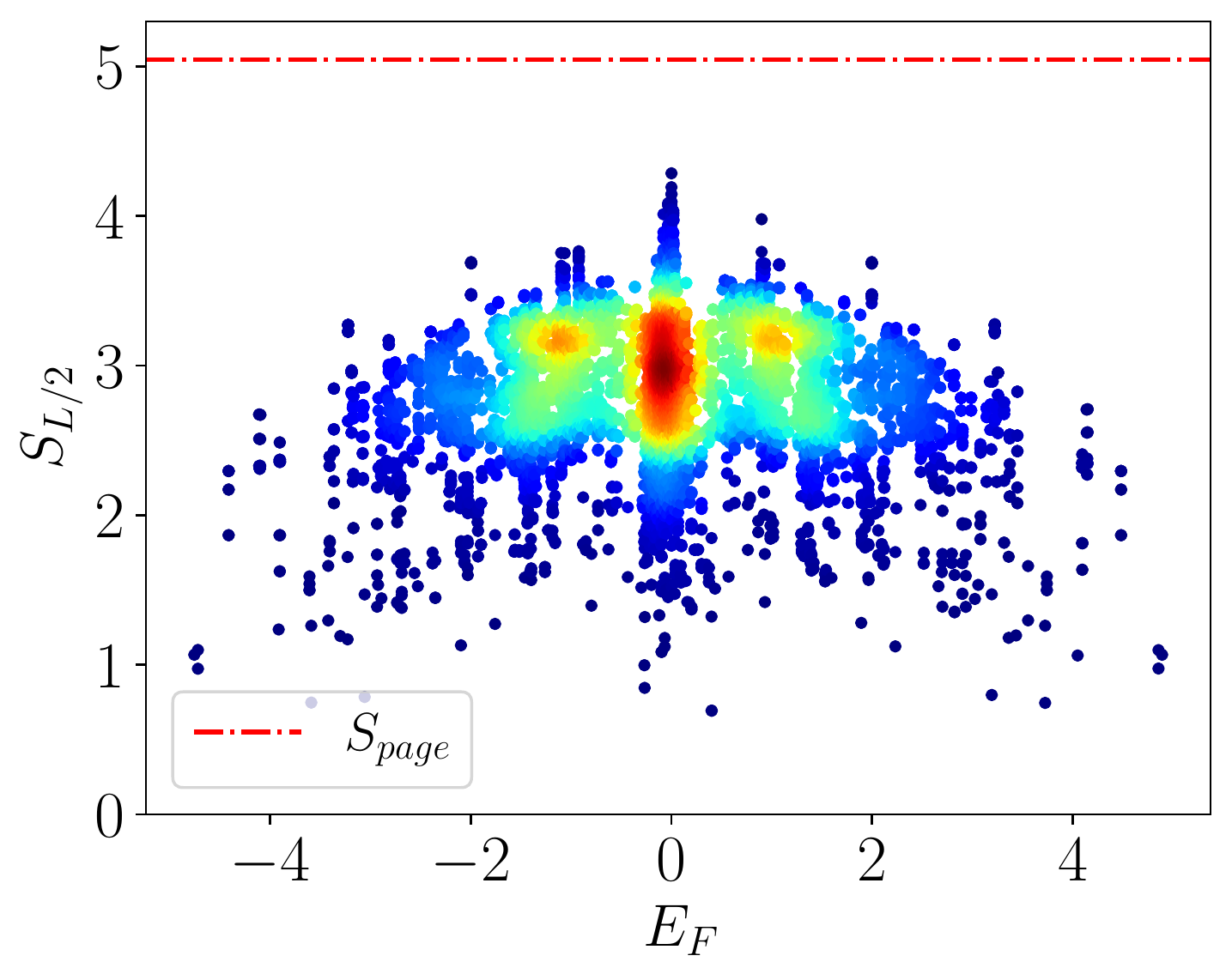}}{\large(a)}
\stackunder[5pt]{\includegraphics[width=0.5\hsize]{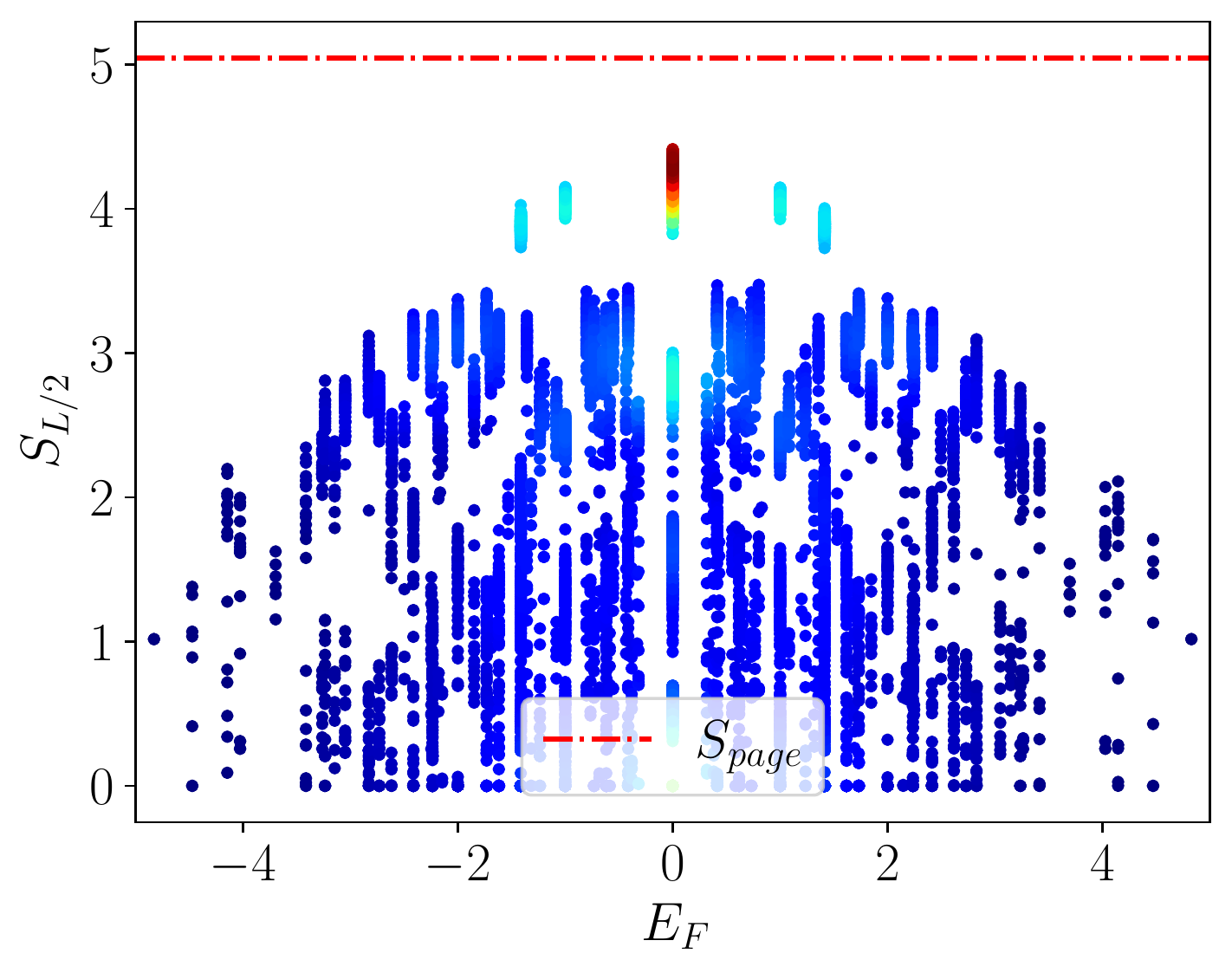}}{\large(b)}
\end{center}
\caption{{\bf Entanglement spectrum at another DL point and at resonance for the 
period-4 model:} Plots showing the entanglement entropy $S_{L/2}$ obtained from 
(a) exact numerical calculations and (b) first-order Floquet Hamiltonian, at a DL 
point with $J=1$, and $\mu=2\om=V=20$. Both calculations show that the quasienergy
spectrum consists of multiple fragments with several low-entanglement states lying near the middle of the spectrum. The color intensity indicates the density of states, 
revealing that the majority of states do not attain the thermal value of te
entanglement entropy which is given by the upper envelop of the plots.}
\label{fig14} \end{figure}

Now we will consider the case $\mu=2\om=V$. Although both $\mu=\om$ and 
$\mu=2\om$ give rise to DL for a non-interacting system, the presence of $V$ 
makes a significant difference in the effective Hamiltonian description. Going through the same procedure as before, we obtain the effective Hamiltonian
\bea H&=&\sum_{j=1}^{L/2} ~[(1-n_{2j+1}) ~(1-n_{2j+4}) ~+~ n_{2j+1} ~n_{2j+4}]
~(c_{2j+2}^{\dagger}c_{2j+3} ~+~ {\rm H.c.}). \eea
Note that there is no hopping on the bonds $(n_{2j+1},n_{2j+2})$.
This implies that the occupation number $n_{2j}+n_{2j+1}$ in the
$j$-th unit cell commutes with the effective Hamiltonian for all values of
$j$. Hence there are $L/2$ 
approximately conserved quantities, which can protect some of the mid-spectrum states from thermalization for a long time. In Fig.~\ref{fig14}, the 
entanglement entropy spectrum obtained by (a) exact 
numerics and (b) first-order FPT are shown for $J=1$, $\mu=2\om=V=20$, and $L=16$. The fragmentation in the spectrum points towards the existence of conserved charges following 
from the first-order effective Hamiltonian.

\section{Experimental accessibility}

Flat-band induced quantum many-body scars and HSF have
been observed in recent years in the context of equilibrium systems. One of the 
common mechanisms for these is compact localization which requires special 
kinds of lattice structures. However, DL-induced flat bands can appear in extremely 
simple lattice models; therefore, this mechanism is experimentally advantageous
compared to the systems demonstrating compact localization. This is one of the 
main reasons that motivated us to pursue this idea. Moreover, 
the intricate interplay between DL and resonances in this class of models induces 
a HSF which is different from the models of HSF discussed in the literature until 
now. A well-studied model of HSF has a conservation of the total particle number 
and the total dipole moment~\cite{tomasi,yang}. However, the class of models 
studied here does not 
conserve the dipole moment but conserves a staggered Ising interaction (see
Appendix B); hence this model merits a more detailed investigation. Further,
our model for HSF does not seem to arise from any limit of an equilibrium model,
unlike an earlier model of HSF which appear in the large $V$ limit of a
model of spinless fermions which have a nearest-neighbor interaction
with strength $V$~\cite{tomasi,yang}. Therefore, periodic driving is necessary to 
realize this new kind of non-equilibrium HSF. Furthermore, the kind 
of driving we chose for our proposed models can be realized in cold-atom 
systems, and therefore opens up new possibilities for further explorations in 
experimental settings.

\section{Discussion}
\label{sec6}

The central results presented in our paper are as follows. We have unraveled an 
intricate dynamical behavior of a class of disorder-free one-dimensional interacting 
spinless fermionic models with a periodically driven on-site potential which is
also periodic in space. In the absence of interactions, this class of models 
exhibits DL for a particular set of parameter values, giving rise to one or more 
flat bands. We have focused in detail on two models, corresponding to potentials
with period-2 and period-4 on the lattice. For the period-2 model, we describe a 
dynamical phase transition which can be observed in the relaxation behavior of correlators in the absence of any interactions. Our investigation shows
that a crossover behavior between different power laws of the decay of
correlations generally occurs away from the DL points. We find that 
in the period-2 model, the flat bands which arise due to DL 
are stable in the presence of a comparatively weak interaction strength
due to an emergent integrability. Further, the spectrum of 
half-chain entanglement entropy as a function of the Floquet quasienergy for a weakly
interacting system reveals that there are many low-entanglement states near the middle 
of the quasienergy spectrum, implying that the system may evade thermalization for a 
long time. The persistent oscillations in the correlation functions and in the Loschmidt echo which survive for a long period of time support the above statement 
about thermalization. However, these oscillations decay rapidly in time when we 
move away from these fine-tuned parameter values.

Remarkably, our model also exhibits Hilbert space fragmentation due to the 
presence of kinetic constraints when the DL and resonance condition are simultaneously satisfied and the interaction is strong. In the case of 
period-4, the behavior appears to be much more intriguing, even in the regime
of weak 
interaction. The period-4 model possesses two mirror-symmetric configurations, corresponding to two values of the phase, 
$\phi=0$ and $\phi=7\pi/4$. Our study reveals that the $\phi=0$ case is identical to the period-2 model at the DL, although the conditions for DL for the two 
cases are slightly different. The $\phi=7\pi/4$ model is much more rich compared 
to the earlier models. In the strong driving limit, the first-order Floquet perturbation theory suggests that the $\phi=\pi/4$ model 
at the DL points reduces to the SSH Hamiltonian with perfect dimerization, with the hopping amplitudes alternating as $\ga_{1}=0$ and 
$\ga_{2}=1$ respectively. Hence the system supports robust zero-energy edge modes, which are topologically protected. The entanglement spectrum in this regime 
demonstrates a finger-like structure with many low-entanglement states near the 
middle of the quasienergy spectrum. We find that there are some initial states 
which either show long-time persistent oscillations or do not participate in the dynamics 
at all. We put forward two possible mechanisms for this ergodicity breaking.

\noi (i) There exists an extensive numbers of conserved quantities at the DL 
points giving rise to sectors which are decoupled from the each other. The 
number of sectors grows exponentially with the system size as $3^{L/2}$, which is
slower than the growth of the Hilbert space dimension $2^{L}$. We would like
to emphasize that these quantities are only approximately conserved quantities; the conservation becomes more and more exact as the driving amplitude is increased. 

\noi (ii) Another possibility is that there are many configurations of frozen 
states which do not evolve with time. However,
these states quickly thermalize as we move away from a DL point. 

In the presence of interactions, the period-4 model with $\phi = 7 \pi/4$ is found
to have a Floquet Hamiltonian which describes the transverse field Ising model with longitudinal fields at the boundaries. We find it surprising and remarkable that 
periodic driving of a period-4 model can be tuned to generate well-known systems 
like the SSH model and the transverse field Ising model which have been extensively 
studied for many years.

We also discuss the effects of a staggered on-site potential on the DL. In this case, we 
find that the finger-like structure in the entanglement spectrum further breaks up into secondary fragments, and the Loschmidt echo produces long-time coherent oscillations 
at the same fine-tuned parameter values. Next, we examine the stability of this 
non-ergodic behavior whenever the DL and resonance condition are simultaneously 
satisfied. To study this regime, we choose two different sets of parameter values, $\mu=\om=V \gg J$, and $\mu=2\om=V \gg J$. In both cases, the non-interacting part of the 
effective Hamiltonian supports DL. For $\mu=\om=V \gg J$, the entanglement spectrum 
again demonstrates many low-entanglement states and slow thermalization of the system. In this regime, the effective Floquet Hamiltonian found using first-order perturbation
theory shows that DL and resonances together put strict restrictions on the allowed hopping processes, and these restrictions protect some of the mid-spectrum states from 
thermalization. These kinetic constraints on the dynamics generate dynamically disconnected sectors, a phenomenon called Hilbert space fragmentation. For $\mu=2
\om=V \gg J$, we see a fractured entanglement spectrum with many segments and with many 
low-entanglement states near the middle of the quasienergy spectrum. The first-order effective Hamiltonian for this case shows that there are an extensive numbers of conserved 
quantities. Furthermore, as in the previous case, some processes are again strictly forbidden due to the combination of DL and resonance. These two mechanisms can, in principle, lead the 
system towards non-ergodic behavior. 

We would like to emphasize that the results obtained from the first-order Floquet
Hamiltonian (such as the appearance of a large number of conserved quantities) agree 
well with the results from an exact numerical calculation of the Floquet operator
only when (i) the driving amplitude and frequency are much larger than all the other
parameters of the system, and (ii) the time scale of observation
of correlation functions and Loschmidt echos is not very large. The two sets 
of results are expected to deviate from each other at very long times because the 
effects of higher-order terms in the FPT then become important.

In summary, we have presented a number of models in this
paper which can be tailored by Floquet engineering to exhibit rich topological and 
dynamical phase diagrams which have no counterparts in a time-independent 
(undriven) model.

\section{Acknowledgements}

S.A. thanks Sumilan Banerjee and Nilanjan Roy for useful discussions.
S.A. thanks MHRD, India for financial support through a PMRF.
D.S. thanks SERB, India for funding through Project No. JBR/2020/000043.

\section{Appendix A}

In this appendix we will consider the model described in Eq.~\eqref{effp2},
\beq H ~=~ \sum_{j=1}^L ~(n_j - n_{j+3})^2 ~\left(c_{j+2}^{\dagger} c_{j+1} ~+~ 
{\rm H.c.} \right), \label{a1} \eeq
where we have set the coefficient in front to be equal to 1 for simplicity, and we
assume $L$ to be even. This model has three conserved quantities given by the
total particle number and the total staggered Ising interactions ($\sigma_i^z 
\sigma_{i+1}^z$ where $\sigma_i^z = 2 n_i - 1$) on odd- and even-numbered bonds,
\bea C_1 &=& \sum_{j=1}^L ~n_j, \non \\
C_2 &=& \sum_{j=1}^{L/2} ~(-1)^j ~(2 n_{2j-1} - 1)~(2 n_{2j} -1), \non \\
C_3 &=& \sum_{j=1}^{L/2} ~(-1)^j ~(2 n_{2j} - 1)~(2 n_{2j+1} -1). \eea
We will use a transfer matrix method to determine the number of
zero-energy states which consist of a single state in the number
basis for the Hamiltonian given in Eq.~\eqref{a1}. We see from
that Hamiltonian that there cannot be any hopping between sites $j+1$ and $j+2$
if either \\
(i) the occupation numbers at sites $(n_j,n_{j+3})$ are either $(0,0)$ or $(1,1)$,
or \\
(ii) the occupation numbers at sites $(n_{j+1},n_{j+2})$ are either $(0,0)$ or $(1,1)$. \\
Hence, any configuration which satisfies any of the above conditions for {\it all}
values of $j$ must necessarily be a zero-energy state. This leads us to define an 
$8 \times 8$ transfer matrix $T_1$ whose rows correspond to the eight possible occupation
numbers $(n_j,n_{j+1},n_{j+2})$, i.e., $(111)$, $(110)$, $(101)$, $(100)$, $(011)$,
$(010)$, $(001)$ and $(000)$, and the columns correspond in a similar way to the
eight possible occupation numbers $(n_{j+1},n_{j+2},n_{j+3})$. The conditions
given above imply that $T_1$ must have the form
\beq T_1 ~=~ \left( \begin{array}{cccccccc}
1 & 1 & 0 & 0 & 0 & 0 & 0 & 0 \\ 
0 & 0 & 1 & 0 & 0 & 0 & 0 & 0 \\ 
0 & 0 & 0 & 0 & 1 & 0 & 0 & 0 \\ 
0 & 0 & 0 & 0 & 0 & 0 & 1 & 1 \\ 
1 & 1 & 0 & 0 & 0 & 0 & 0 & 0 \\ 
0 & 0 & 0 & 1 & 0 & 0 & 0 & 0 \\ 
0 & 0 & 0 & 0 & 0 & 1 & 0 & 0 \\ 
0 & 0 & 0 & 0 & 0 & 0 & 1 & 1 \end{array} \right). \eeq
We now have to find the eigenvalues of this matrix. We first note that 
the matrix has two $4 \times 4$ blocks which are not coupled to each other: 
the blocks consist of the rows and columns numbered (1235) and (4678), and
the two blocks have identical eigenvalues. We can
therefore look at either of the two blocks and find the four eigenvalues;
the eigenvalues of $T_1$ will then be given by these four eigenvalues, each repeated
twice. The block corresponding to (1235) takes the form
\beq T'_1 ~=~ \left( \begin{array}{cccc}
1 & 1 & 0 & 0 \\
0 & 0 & 1 & 0 \\
0 & 0 & 0 & 1 \\
1 & 1 & 0 & 0 \end{array} \right). \eeq
We find that one of the eigenvalues of $T'_1$ is zero. The other three eigenvalues
must therefore be solutions of a cubic equation which turns out to be 
\beq \lambda^3 - \lambda^2 - 1 = 0. \eeq
The solutions of this equation are found to be
\beq \lambda ~=~ \frac{1}{3} ~+~ \frac{2}{3} ~\cos ~\left[ \frac{1}{3} 
\cos^{-1} \left( \frac{29}{2} \right) ~-~ \frac{2 \pi k}{3} \right], \eeq
where $k$ can take the values $0, ~1, ~2$. We then find that the three 
eigenvalues are $1.466$ and $-0.233 \pm 0.793 i$ approximately.
The eigenvalues of the original transfer matrix $T_1$ are therefore given by
$1.466$, $-0.233 \pm 0.793 i$ and 0, each repeated twice. Hence, for
a system with a large number of sites $L$, the number of zero-energy states
grows exponentially as $1.466^L$ (compared to the total number of states
which grows as $2^L$).

It is interesting to compare the number of zero-energy states in this model with the
number of such states in a different model which also has a kinetic constraint on 
the nearest-neighbor hoppings~\cite{tomasi,yang}. The Hamiltonian of that model 
is given by
\bea H ~= ~\sum_{j=1}^L ~[1 ~-~ (n_j - n_{j+3})^2] ~\left(c_{j+2}^{\dagger} 
c_{j+1} ~+~ {\rm H.c.} \right). \eea
This model is known to have three conserved quantities given by the total
particle number and the total dipole numbers ($n_i n_{i+1}$) on odd- and 
even-numbered bonds,
\bea C_4 &=& \sum_{j=1}^L ~n_j, \non \\
C_5 &=& \sum_{j=1}^{L/2} ~n_{2j-1} ~n_{2j}, \non \\
C_6 &=& \sum_{j=1}^{L/2} ~n_{2j} ~n_{2j+1}. \eea
It has been shown recently that this model can appear as the effective Hamiltonian
of a periodically driven system for some special values of the driving 
parameters~\cite{prefrag}. In this model, we see that there cannot be any
hopping between sites $j+1$ and $j+2$ if either \\
(i) the occupation numbers at sites $(n_j,n_{j+3})$ are either $(0,1)$ or $(1,0)$,
or \\
(ii) the occupation numbers at sites $(n_{j+1},n_{j+2})$ are either $(0,0)$ or $(1,1)$. \\
To find the number of zero-energy states in this model, we again construct an
$8 \times 8$ transfer matrix which now takes the form 
\beq T_2 ~=~ \left( \begin{array}{cccccccc}
1 & 1 & 0 & 0 & 0 & 0 & 0 & 0 \\ 
0 & 0 & 0 & 1 & 0 & 0 & 0 & 0 \\ 
0 & 0 & 0 & 0 & 0 & 1 & 0 & 0 \\ 
0 & 0 & 0 & 0 & 0 & 0 & 1 & 1 \\ 
1 & 1 & 0 & 0 & 0 & 0 & 0 & 0 \\ 
0 & 0 & 1 & 0 & 0 & 0 & 0 & 0 \\ 
0 & 0 & 0 & 0 & 1 & 0 & 0 & 0 \\ 
0 & 0 & 0 & 0 & 0 & 0 & 1 & 1 \end{array} \right). \eeq
This matrix has a $6 \times 6$ block consisting of the rows and columns numbered
(124578) and a $2 \times 2$ block consisting of the rows and columns numbered
(36) which are not coupled to each other. The $6 \times 6$ block is found to
have eigenvalues $(1/2) (1 \pm \sqrt{5})$, $(1/2) (1 \pm i \sqrt{3})$, 0 and
0, while the $2 \times 2$ block has eigenvalues $\pm 1$. Hence the number
of zero-energy states in this model grows with the system size as $\tau^L$,
where $\tau = (1 + \sqrt{5})/2 \simeq 1.618$ is the golden ratio.

\section{Appendix B}

In this appendix, we will derive the first-order FPT for a two-site model with nearest-neighbor hopping $J$, and on-site potentials $\mu_{1}$ and $\mu_{2}$, which can then be 
generalized to a system with larger system sizes. We will further assume that the on-site potential is periodically driven in time with a square pulse protocol. Assuming $\mu_{1},~\mu_{2} \gg J$, we can recast the Hamiltonian as
\bea H&=&H_{0}+H_{1},~~~~{\rm where}~~~~
H_{0}=\left(\begin{array}{cc}
\mu_{1} f(t) & 0\\ 0 &\mu_{2} f(t)\end{array}\right)~~~{\rm and}~~~
H_{1}=\left(\begin{array}{cc}
0 & J\\ J &0\end{array}\right). \eea 
The eigenfunctions corresponding to $E_{1,2} = \mu_1 f(t)$ and $\mu_{2}f(t)$ are given by
$\ket{+}=\left(\begin{array}{cc}
1 \\ 0 \end{array}\right)$ and $\ket{-}=\left(\begin{array}{cc}
0 \\ 1 \end{array}\right)$
respectively.
The instantaneous eigenvalues $E_{1,2}$ satisfy the degeneracy condition in Eq.~\eqref{per26}. Following the procedure for degenerate FPT outlined in Eqs.~(\ref{per28}-\ref{per30}), we find that 
\beq \bra{+}H_{F}^{(1)}\ket{+} ~=~
\bra{-}H_{F}^{(1)}\ket{-} ~=~ 0,~~~~~
\bra{+}H_{F}^{(1)}\ket{-}~=~\bra{-}H_{F}^{(1)}\ket{+}^*=J e^{iB}\left(\frac{\sin B}{B}\right), \eeq
where $B=(V_{1}-V_{2})T/4$. Hence $H_{F}^{(1)}$ is given by 
\bea H_{F}^{(1)}=J e^{iB} \left(\frac{\sin B}{B}\right) ~
(c_{1}^{\dagger}c_{2} ~+~ c_2^\dagger c_1). \eea

\section{Appendix C}

In this section we will derive the third-order effective Hamiltonian using Floquet perturbation theory for the period-2 model for $V=0$ at a dynamical 
localization point obtained from the first-order effective Hamiltonian. Following 
the usual steps of perturbation theory, we find that the third-order effective 
Floquet Hamiltonian is
\bea
\bra{m}H_{F}^{(3)}T\ket{n}~&=&- ~\bra{m}M^{(3)}\ket{n} ~+~ \frac{1}{3}\bra{m}
(M^{(1)})^3\ket{n},\non\\
\bra{m}M^{(3)}\ket{n}&=&\sum_{p,q} ~\bra{m}H_1\ket{p}\bra{p}H_1\ket{q}\bra{q}H_1\ket{n}\int_{0}^{T}e^{i\int_{0}^{t}(E_m(t_1)-E_p(t_1)) ~dt_1} ~dt \non \\
&& \times \int_{0}^{t}e^{i\int_{0}^{t'}(E_p(t_2)-E_q(t_2)) ~dt_2} ~dt' 
\int_{0}^{t'}e^{i\int_{0}^{t''}(E_q(t_3)-E_n(t_3)) ~dt_3} ~dt'',\non\\
\bra{m}(M^{(1)})^3\ket{n}&=&\sum_{p,q} ~\bra{m}H_1\ket{p}\bra{p}H_1\ket{q}\bra{q}H_1\ket{n}\int_{0}^{T}e^{i\int_{0}^{t}(E_m(t_1)-E_p(t_1))~ dt_1} ~dt \non \\
&& \times \int_{0}^{T}e^{i\int_{0}^{t'}(E_p(t_2)-E_q(t_2)) ~dt_2} ~dt' 
\int_{0}^{T}e^{i\int_{0}^{t''}(E_q(t_3)-E_n(t_3)) ~dt_3} ~dt''.\label{third} \eea
While writing Eq. \eqref{third}, we use the fact that the second-order term $M^{(2)}$ 
is zero for our particular model with our choice of driving protocol due to the symmetry discussed in Eq. \eqref{per33}. Since the perturbation part of the 
Hamiltonian for our model is off-diagonal in the basis, $\ket{m}=\ket{\pm}$, the only non-zero matrix elements for the third-order effective Hamiltonian are
\bea \bra{+}H_{F}^{(3)}\ket{-} ~=~ - ~\bra{+}M^{(3)}\ket{-} ~+~ \frac{1}{3}\bra{+}
(M^{(1)})^3\ket{-},~~~~ \bra{-}H_{F}^{(3)}\ket{+} ~=~ \bra{+}H_{F}^{(3)}
\ket{-}^{*},\label{third1} \eea
where
\bea \bra{+}M^{(3)}\ket{-}&=&\bra{+}H_1\ket{-}\bra{-}H_1\ket{+}\bra{+}H_1\ket{-}\int_{0}^{T}e^{2i\int_{0}^{t}\mu(t_1) ~dt_1} ~dt \non \\
&& \times \int_{0}^{t}e^{-2i\int_{0}^{t'}\mu(t_2) ~dt_2} ~dt' \int_{0}^{t'}e^{2i\int_{0}^{t''}\mu(t_4) ~dt_4} ~dt'', \label{m3} \\
\bra{+}(M^{(1)})^3\ket{-}&=&\bra{+}H_1\ket{-}\bra{-}H_1\ket{+}\bra{+}H_1\ket{-}\int_{0}^{T}e^{2i\int_{0}^{t}\mu(t_1) ~dt_1} ~dt \non \\
&& \times \int_{0}^{T}e^{-2i\int_{0}^{t'}\mu(t_2) ~dt_2} ~dt' ~\int_{0}^{T}e^{2i\int_{0}^{t''}\mu(t_4) ~dt_4} ~dt''. \label{third2} \eea
For our model, we find using Eq. \eqref{matper2} that
\bea \bra{+}H_1\ket{-}\bra{-}H_1\ket{+}\bra{+}H_1\ket{-}=8 ~e^{-ik}\cos^{3} k.
\eea
Fo our choice of driving protocol, the integral in Eq. \eqref{m3} can be written as
\bea &&\int_{0}^{T}e^{2i\int_{0}^{t}\mu(t_1) ~dt_1} ~dt \int_{0}^{t}e^{-2i\int_{0}^{t'}
\mu(t_2) ~dt_2} ~dt' ~\int_{0}^{t'}e^{2i\int_{0}^{t''}\mu(t_4) ~dt_4} ~dt''\non\\
&&=\int_{0}^{T/2}e^{2i\mu t} ~dt \int_{0}^{t}e^{-2i\mu t'} ~dt' 
\int_{0}^{t'}e^{2i\mu t''} ~dt''\non\\
&& ~~~+ \int_{T/2}^{T}e^{-2i\mu (t-T)} ~dt \int_{0}^{t}e^{-2i\int_{0}^{t'}\mu(t_1)
~dt_1}~dt' \int_{0}^{t'}e^{2i\int_{0}^{t''}\mu(t_3) ~dt_3} ~dt''.\non\\ \label{third3}
\eea
At the dynamical localization points $\mu = n \om$ obtained from the first-order FPT Hamiltonian, the first integral in Eq. \eqref{third3} is given by
\beq \int_{0}^{T/2}e^{2i\mu t} ~dt \int_{0}^{t}e^{-2i\mu t'} ~dt' \int_{0}^{t'}
e^{2i\mu t''}dt''
~=~ - ~\frac{T}{4\mu^2} ~~~~{\rm when}~~~\mu=n\om. \eeq
The second integral in Eq. {\eqref{third3}} can be written as a sum of three integrals,
\bea
&&\int_{T/2}^{T}e^{-2i\mu (t-T)} ~dt \int_{0}^{t}e^{-2i\int_{0}^{t'}\mu(t_1) ~dt_1}
~dt'\int_{0}^{t'}e^{2i\int_{0}^{t''}\mu(t_3) ~dt_3} ~dt''\non\\
&& = ~~\int_{T/2}^{T}e^{-2i\mu(t-T)} ~dt \int_{0}^{T/2} ~e^{-2i\mu t'} ~dt' \int_{0}^{t'}e^{2i\mu t''} ~dt''\non\\
&&~~+ ~\int_{T/2}^{T}e^{-2i\mu(t-T)} ~dt \int_{T/2}^{t}e^{2i\mu(t'-T)} ~ dt' \left(\int_{0}^{T/2} e^{2i\mu t''} ~dt'' +\int_{T/2}^{t'}e^{-2i\mu(t''-T)}~ 
dt''\right).\non\\\label{third4} \eea
The three integrals in Eq. \eqref{third4} at the dynamical localization
points reduce to
\bea
&&{\bf 1.}~~~ \int_{T/2}^{T}e^{-2i\mu(t-T)} ~dt \int_{0}^{T/2} e^{-2i\mu t'} ~dt' \int_{0}^{t'}e^{2i\mu t''} ~dt'' ~=~ 0, \non \\
&&{\bf 2.}~~~ \int_{T/2}^{T}e^{-2i\mu(t-T)} ~dt \int_{T/2}^{t}e^{2i\mu(t'-T)} ~\int_{0}^{T/2} e^{2i\mu t''} ~dt'' ~= ~0, \non \\
&& {\bf 3.} ~~~\int_{T/2}^{T}e^{-2i\mu(t-T)} ~dt \int_{T/2}^{t}e^{2i\mu(t'-T)} ~dt' \int_{T/2}^{t'}e^{-2i\mu(t''-T)}~ dt'' 
~=~ -\frac{T}{4\mu^2} \eea
when $\mu = n \om$.
Similarly, we can show that the integral related to $(M^{(1)})^3$ in 
Eq. \eqref{third2} is
\bea \int_{0}^{T}e^{2i\int_{0}^{t}\mu(t_1) ~dt_1} ~dt \int_{0}^{T}e^{-2i\int_{0}^{t'}\mu(t_2)~ dt_2} ~dt' \int_{0}^{T}e^{2i\int_{0}^{t''}\mu(t_4) ~dt_4} ~dt'' 
~=~ 0~~~~{\rm when}~~\mu=n\om. \eea
Therefore, the third-order effective Hamiltonian for $V=0$ at the dynamical localization points $\mu = n \om$ is given by
\bea H^{(3)}_{F} ~=~ \frac{4J^3}{\mu^2} ~\sum_{k} ~\cos^{3} k ~\left(e^{-ik}a_{k}^{\dagger}b_{k}+\rm{H.c.}\right). \eea
Interestingly, we see that $H_{F}^{(3)}$ scales as $J^3/\mu^2$ at the dynamical localization points $\mu= n\om$, and it does not explicitly depend on $\om$ at 
these special points.

\bibliographystyle{SciPost_bibstyle}

\end{document}